\documentclass[manuscript, screen]{acmart}

\usepackage{graphicx}
\usepackage{float} 
\usepackage{subfig} 
\usepackage{makecell}
\usepackage{booktabs}
\usepackage{longtable}
\usepackage{multirow}
\usepackage{pifont}
\usepackage{caption}
\usepackage{listings} 
\usepackage{xcolor}
\usepackage{epstopdf}
\usepackage{adjustbox}
\usepackage{url}
\usepackage{tabularx}
\usepackage{amsmath}

\lstdefinelanguage{Alloy}{
  keywords={sig, abstract, extends, pred, fun, fact, assert, run, check, for, but, let, all, no, lone, some, one, set, sum, seq, int, disj, not, iff, implies, else, in, univ, iden, seq},
  keywordstyle=\bfseries\color{blue},
  sensitive=false,
  morecomment=[l]{//},
  morecomment=[s]{/*}{*/},
  morestring=[b]",
  basicstyle=\small\ttfamily,
}

\newcommand{\eat}[1]{}

\AtBeginDocument{%
  }

\setcopyright{acmcopyright}
\copyrightyear{2018}
\acmYear{2018}
\acmDOI{XXXXXXX.XXXXXXX}

\acmPrice{15.00}
\acmISBN{978-1-4503-XXXX-X/18/06}

\begin{document}
\title{A Comprehensive Survey on Database Management System Fuzzing: Techniques, Taxonomy and Experimental  Comparison}

\author{Xiyue Gao}
\email{xygao@xidian.edu.cn}
\author{Zhuang Liu}
\email{zliu_01@stu.xidian.edu.cn}
\author{Jiangtao Cui}
\email{cuijt@xidian.edu.cn}
\affiliation{%
  \institution{Xidian University}
  \department{School of Computer Science and Technology}
  \streetaddress{266 XiFeng Road, Chang'an District}
  \city{Xi'an}
  \country{China}}
\author{Hui Li}
\email{hli@xidian.edu.cn}
\authornote{Corresponding author}
\affiliation{%
  \institution{Xidian University}
  \department{School of Computer Science and Technology}
  \streetaddress{266 XiFeng Road, Chang'an District}
  \city{Xi'an}
  \country{China}}
\affiliation{
 \institution{Shanghai Yunxi Technology Co., Ltd.}
 \city{Shanghai}
 \country{China}
}
\author{Hui Zhang}
\author{Kewei Wei}
\author{Kankan Zhao}
\affiliation{%
  \institution{Inspur}
  \department{Shandong Inspur Database Technology Co., Ltd}
  \city{Jinan}
  \country{China}
}

\renewcommand{\shortauthors}{AUTHOR et al.}


\begin{abstract}

Database Management System (DBMS) fuzzing is an automated testing technique aimed at detecting errors and vulnerabilities in DBMSs by generating, mutating, and executing test cases. It not only reduces the time and cost of manual testing but also enhances detection coverage, providing valuable assistance in developing commercial DBMSs. Existing fuzzing surveys mainly focus on general-purpose software. However, DBMSs are different from them in terms of internal structure, input/output, and test objectives, requiring specialized fuzzing strategies. Therefore, this paper focuses on DBMS fuzzing and provides a comprehensive review and comparison of the methods in this field. We first introduce the fundamental concepts. Then, we systematically define a general fuzzing procedure and decompose and categorize existing methods. Furthermore, we classify existing methods from the testing objective perspective, covering various components in DBMSs. For representative works, more detailed descriptions are provided to analyze their strengths and limitations. To objectively evaluate the performance of each method, we present an open-source DBMS fuzzing toolkit, OpenDBFuzz. Based on this toolkit, we conduct a detailed experimental comparative analysis of existing methods and finally discuss future research directions.

\end{abstract}

\ccsdesc[500]{Information systems~Database performance evaluation}
\ccsdesc[300]{Software and its engineering~Software testing and debugging}
\ccsdesc[500]{Information systems~Database performance evaluation}

\keywords{Automated database testing, fuzzing, DBMS fuzzing, DBMS fuzz testing, Experimental comparison. }

\received{20 February 2007}
\received[revised]{12 March 2009}
\received[accepted]{5 June 2009}

\maketitle

\section{Introduction}
Database Management System (DBMS), as the fundamental software for managing and organizing data, is widely used in almost all software applications~\cite{fayyad2011data}. Its reliability and security are of paramount importance. DBMS testing aims to verify its functionality, performance, and security to ensure that it meets the expected requirements and standards~\cite{lo2010framework,lee2011performance,bati2007genetic}. After discovering potential errors and vulnerabilities, developers can quickly address and optimize them, which is crucial for the development of commercial DBMSs~\cite{c2021,p2021,db733,sc2019}.

Traditional DBMS testing heavily relies on manual testing, which is not only time-consuming and laborious but also suffers from a limited testing scope. To improve testing efficiency and accuracy, automated testing techniques have gradually emerged~\cite{Jayakumar_2013,software1992,Th_venod_Fosse_1993}. One such technique is DBMS fuzzing, which discovers DBMS bugs by automatically generating, mutating, and executing test cases~\cite{howden1978theoretical}. The goal of fuzzing is to introduce a significant amount of invalid, abnormal, or random data to test the robustness and security of DBMSs against adverse inputs~\cite{afl,oss,g2016,llvm,go}. This approach significantly reduces the time and cost associated with manual testing while improving the coverage of bug detection~\cite{fb}.

\begin{figure}[tp]
    \centering
    \includegraphics[width=1\textwidth]{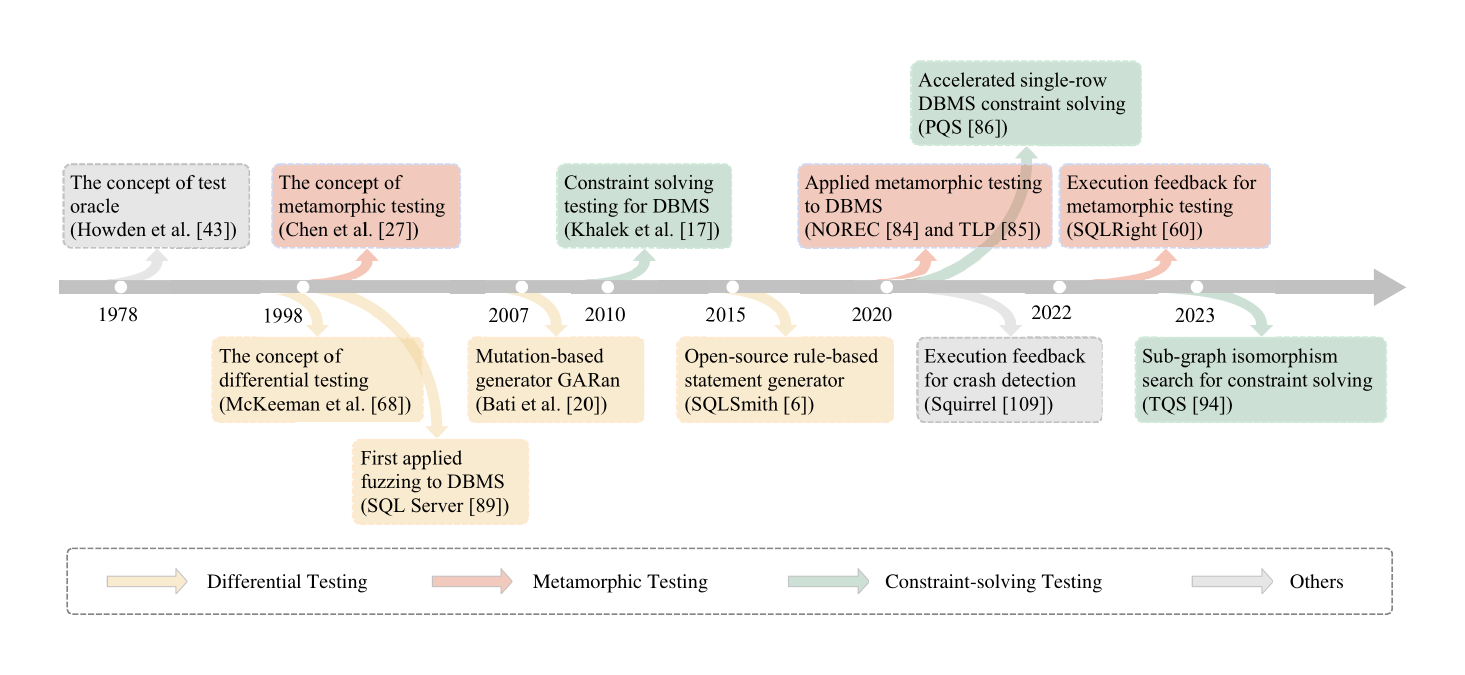}
    \caption{Main milestones of DBMS fuzzing.}
    \label{fig:Milestones}
\end{figure}


As shown in Figure~\ref{fig:Milestones}, the study of DBMS fuzzing has a long history. In 1978, Howden et al. first introduced the concept of test oracle, which is a source of information about whether the output of a system is correct or not~\cite{howden1978theoretical}. Since then, many types of methods have been proposed to construct a test oracle. In 1998, McKeeman introduced the concept of differential testing~\cite{mckeeman1998differential}, and Chen et al. proposed metamorphic testing~\cite{chen1998metamorphic}. In the same year, SQL Server first applied fuzzing to the database field~\cite{slutz1998massive}, which generated a large number of statements based on the rules and conducted differential tests on different DBMSs. In 2007, the DBMS differential testing method proposed by Bati et al. adopted a mutation-based generator named GARan to generate test cases~\cite{bati2007genetic}. In 2010, constraint-solving testing~\cite{abdul2010automated} was proposed as another way to obtain the test oracle in DBMS fuzzing, which provides the ground truth for the test cases. In 2015, SQLSmith~\cite{git2023sqlsmith} was released, which still generates statements based on predefined rules, and can also connect to reference databases to serve as test oracles. Thanks to the open-source factor, compared to the SQL Server fuzzing tool, SQLSmith has rapidly attracted attention from both academic and industrial. In 2020, Squirrel~\cite{zhong2020squirrel} first introduced execution feedback to improve code coverage in crash detection fuzzing. In the same year, Rigger et al. proposed several new fuzzing methods, including NoREC~\cite{Rigger_2020}, TLP~\cite{rigger2020finding} and PQS~\cite{rigger2020testing}. NoREC and TLP apply metamorphic testing to the database domain, enabling fuzzing on a single DBMS. PQS is a constraint-solving testing method, and its oracle only requires that the execution result contain one row of the solving outcome, thereby accelerating the solving process compared to the former approach, which requires matching all the rows. In 2022, Liang et al. introduced SQLRight~\cite{liang2022detecting}, a metamorphic testing method combined with execution feedback, and is scalable to allow new oracles. In 2023, Tang et al. presented a novel constraint-solving testing method, TQS~\cite{tang2023detecting}, which incorporated subgraph isomorphism search into DBMS fuzzing, significantly enhancing bug detection efficiency.

DBMS fuzzing techniques can be classified from different perspectives. 
\begin{itemize}
    \item One classification approach is based on the way expected results are obtained, as shown in Figure~\ref{fig:Milestones}: differential testing, metamorphic testing, and constraint-solving. Differential testing compares output differences across databases~\cite{Marinov_2010}, mutation testing modifies input samples to assess the tolerance of the DBMS to exceptional inputs~\cite{chen1998metamorphic,chu2017cosette,Lemieux_2018}, and constraint-solving utilizes solvers for benchmark ground truths~\cite{Mcminn_2015,Shah_2011,Abdul_Khalek_2008,Bruno_2006}.
    \item Fuzzing techniques can also be categorized by the transparency of the DBMS at test time: black-box, gray-box, and white-box~\cite{blazytko2019grimoire}. Black-box testing assesses the behavior and functionality of DBMS solely through input and output observations~\cite{mohan2018finding}. Gray-box testing collects state information to generate targeted test cases~\cite{Wang_2019,B_hme_2017,B_hme_2016}, while white-box testing involves a deep understanding of the internal structure and emphasizes code coverage and quality assessment~\cite{Godefroid_2008}.
\end{itemize} 

Existing fuzzing surveys focus mainly on \textit{general-purpose} software. However, as the structures and testing objectives for different types of software can vary significantly, specific fuzzing methods~\cite{blazytko2019grimoire} have to be carefully designed in order to tailor for each particular type of software system, e.g. Database. Compared to other types of software system, DBMS contains some unique modules and functionalities, including SQL optimizer, executor, transaction engine, etc~\cite{graefe1995cascades,li2016optmark,leis2015good,Osman_2012}. This leads to significant differences between DBMS and other software, e.g., the way how to generate test cases and how to verify the correctness of execution results. However, there does NOT exist any systematic survey that provides an in-depth study and comparison of DBMS fuzzing approaches. To fill this gap, in this paper, we will thoroughly review existing work on DBMS fuzzing, including basic definitions, horizontal and vertical taxonomies, experimental comparisons, and future directions.

Specifically, the basic concepts and notations of DBMS fuzzing are introduced first, laying the foundation for further exploration. Then, we systematically define a general fuzzing procedure, we also decompose and categorize existing methods. The advantages, disadvantages, and contributions of the methods in each stage are also analyzed. In addition, we further classify existing methods from the testing objective perspective, covering various components in DBMSs. In order to objectively evaluate the performance of each method, we present a comprehensive open-source toolkit, OpenDBFuzz. Based on this toolkit, we test and analyze the corresponding syntax validity, semantic validity, code coverage, and the number of unique bugs on PostgreSQL, MySQL, and SQLite databases, respectively. Finally, future research directions for DBMS fuzzing are discussed, identifying areas for further exploration and development. The contributions of this paper are as follows.
\begin{enumerate}
    \item We systematically define the general DBMS fuzzing procedure, decompose and categorize existing methods. Furthermore, the advantages, disadvantages, and contributions of the methods in each stage are concluded. 
    \item We classify existing DBMS fuzzing methods based on testing components, encompassing various components in DBMSs, and provide detailed descriptions of representative works.
    \item We present OpenDBFuzz, the most comprehensive open-source toolkit for DBMS fuzzing, and conduct fair experimental comparisons to evaluate state-of-the-art methods using the same experimental configurations. 
    \item We thoroughly explore the issues and challenges of DBMS fuzzing and forecast future research directions in this field.
\end{enumerate}

The remainder of this paper is organized as follows: in Section~\ref{sec:preli}, we introduce the preliminaries. Considering that the DBMS fuzzing process is complex and that the best method may not perform optimally at every step, we comprehensively disassemble, classify and compare the existing methods from the horizontal perspective in Section~\ref{sec:hor}. In addition, some fuzzing methods only target internal components in the DBMS. Therefore, in Section~\ref{sec:ver}, we classify and describe representative fuzzing works from a vertical perspective. In Section~\ref{sec:exp}, we present OpenDBFuzzy and conduct a comprehensive experimental evaluation. In Section~\ref{sec:cha}, we summarize the existing challenges. Section~\ref{sec:con} concludes the paper.

\section{Preliminary}\label{sec:preli}
This section provides an introduction to the notations, definitions, and backgrounds necessary for describing DBMS fuzzing techniques.

\subsection{Notations}
The notations and their explanations used in this paper are summarized in Table~\ref{tab:Summary of notations}. These notations will be used throughout the following sections. It is worth noting that adding an `s' to any notation indicates multiple instances of that item. For example, `Ss' denotes multiple statements.
\begin{table}[ht]
\centering
\small
\caption{Summary of Commonly Used Notations in This Paper}
\scalebox{1}{
\label{tab:Summary of notations}
    \begin{tabular}{|m{1.5cm}<{\centering}|m{12cm}|}
      \Xhline{0.5pt}
      \textbf{Notation} & \textbf{\quad Definition} \\
      \Xhline{0.5pt}
      $Q$ & \quad A SQL query, specifically refers to a SELECT statement \\
      \Xhline{0.5pt}
      $D$ & \quad The target DBMS\\
      \Xhline{0.5pt}
      $S$ & \quad A statement, which can be any SQL sentence supported by the target DBMS \\
      \Xhline{0.5pt}
      $p$ & \quad A predicate in the statement \\
      \Xhline{0.5pt}
      $TX$ & \quad A transaction,
      which begins with `BEGIN' and ends with `COMMIT' and contains some statements\\
      \Xhline{0.5pt}
      $TC$ & \quad A test case, which contains some statements or some transactions\\
      \Xhline{0.5pt}
      $t$ & \quad A table in the target DBMS\\
      \Xhline{0.5pt}
      $D(TC)$ & \quad  The execution result of the test case $TC$ in the DBMS $D$ \\
      \Xhline{0.5pt} 
      $G(TC,D)$ & \quad The ground truth of the execution result for the test case $TC$ in the DBMS $D$\\
      \Xhline{0.5pt}
    \end{tabular}}
\end{table}

\subsection{Basic Definitions}

Database bugs encompass three main types: crashes, logic bugs, and performance bugs~\cite{Miller_1990,lo2010framework}. Crashes occur when program errors or memory leaks interrupt the database while queries are executed. Compared to other types of bugs, crashes are relatively easier to identify. Logic bugs occur when the execution results deviate from the expected ones or violate common sense in the DBMS. Such bugs are usually hidden and difficult to detect. Performance bugs manifest themselves as significant differences in the execution time of the same query in different versions of DBMS, or large performance gaps between a query and its equivalent~\cite{Graefe_1993,Gu_2012,giakoumakis2008testing,rigger2020testing,Galindo_Legaria_2004}. However, it is important to note that not all of these differences are considered to be actual bugs. Because determining the optimal execution plan is inherently an NP-hard problem~\cite{ibaraki1984optimal}, some databases practically choose suboptimal plans to speed up the optimization process. We consider performance bugs to have less impact on normal database usage, and therefore have a lower priority than the other two types of bugs.

DBMS fuzzing detects crashes, logic bugs, and performance bugs by generating a large number of SQL test statements in the DBMS~\cite{git2023sqlsmith,Poess_2004,Mishra_2008,Yan_2018,abdul2010automated}. According to the rules of statement generation, DBMS fuzzing methods can be divided into generation-based and mutation-based fuzzing. Generation-based fuzzing employs deterministic rules to randomly generate statements, while mutation-based fuzz testing expands statements by performing mutations on seed queries.

\begin{figure}[t]
    \centering
    \includegraphics[width=13cm]{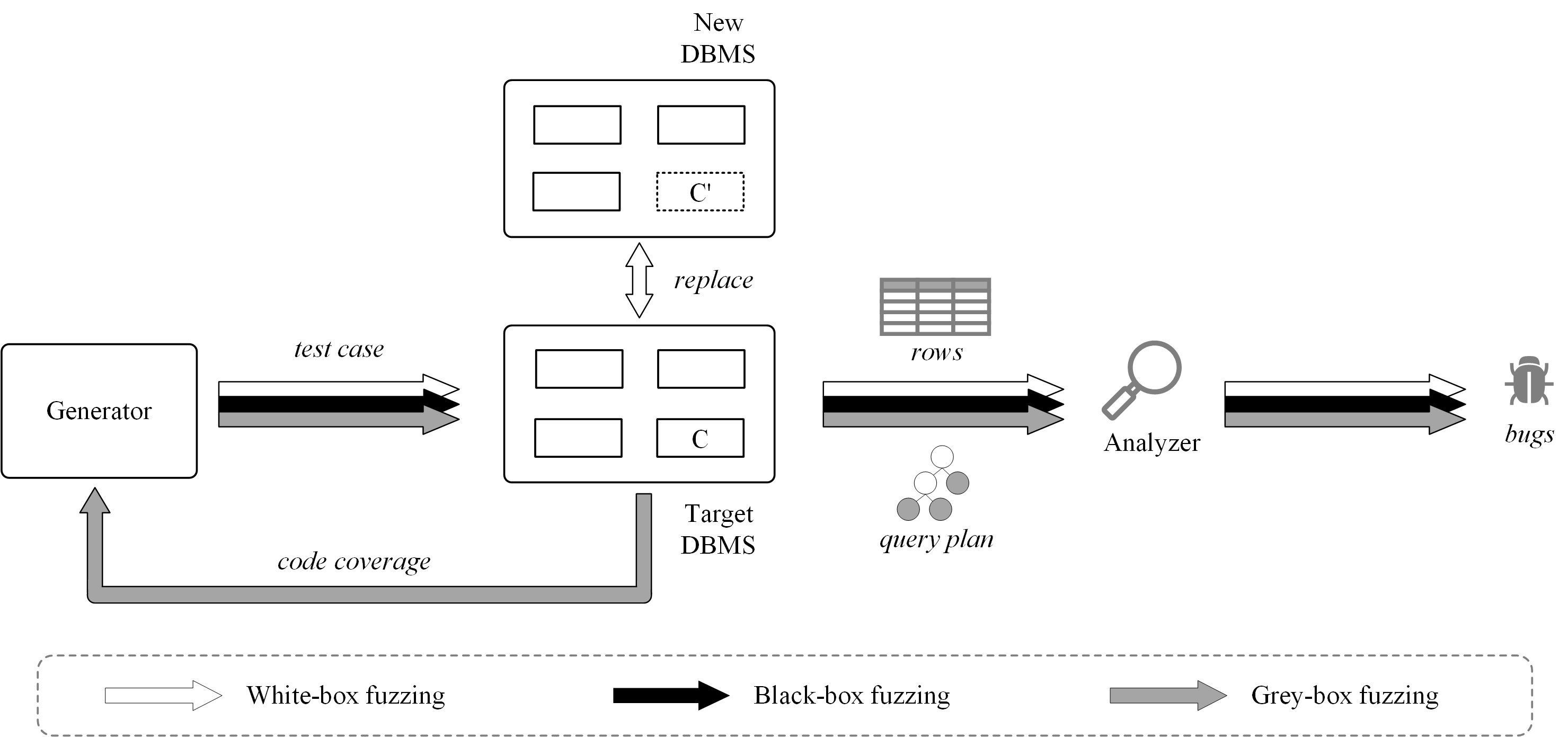}
    \caption{Illustration of the differences between black-box fuzzing, grey-box fuzzing and white-box fuzzing. The black arrow part represents the process of black-box fuzzing, which is also the basis of white-box and gray-box fuzzing. The gray and white arrows represent gray-box and white-box fuzzing, respectively.}
    \label{fig:black and grey box}
\end{figure}

DBMS fuzzing can also be divided into black-box fuzzing, grey-box fuzzing, and white-box fuzzing according to the transparency of the target DBMS, as shown in Figure~\ref{fig:black and grey box}. Black-box fuzzing treats the DBMS as a black box, where a series of SQL statements or transactions are inputted, and a series of execution results or execution plans are obtained accordingly. The bugs are then determined by validating the results and plans. Grey-box fuzzing collects state information from the DBMS, such as code branch coverage and status, to guide statement generation and improve the efficiency of DBMS testing. By comparison, white-box fuzzing requires testing with an understanding of the internal structure and implementation mechanisms of the DBMS. For example, intramorphic testing~\cite{rigger2022intramorphic} works by replacing a specific component in the system to get another version and then testing between the two versions to detect bugs in that component. However, white-box fuzzing is currently limited to theoretical research on general systems, and there is no white-box fuzzing tool for DBMS, which is also the focus of future research.

\begin{figure}[t]
\centering
\subfloat[Differential Testing]{
\includegraphics[width=0.55\textwidth]{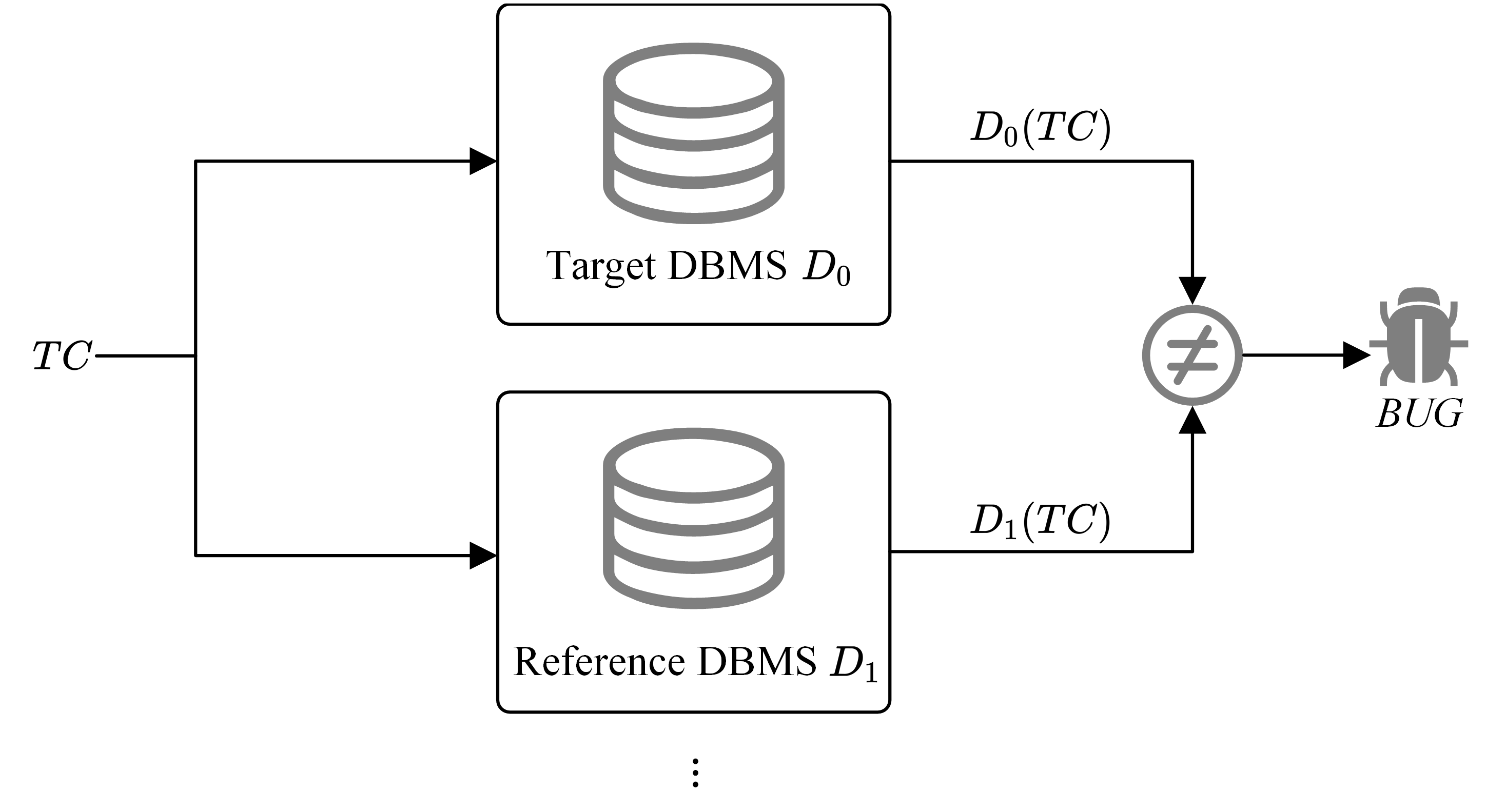}} \\
\subfloat[Metamorphic Testing]{
\includegraphics[width=0.55\textwidth]{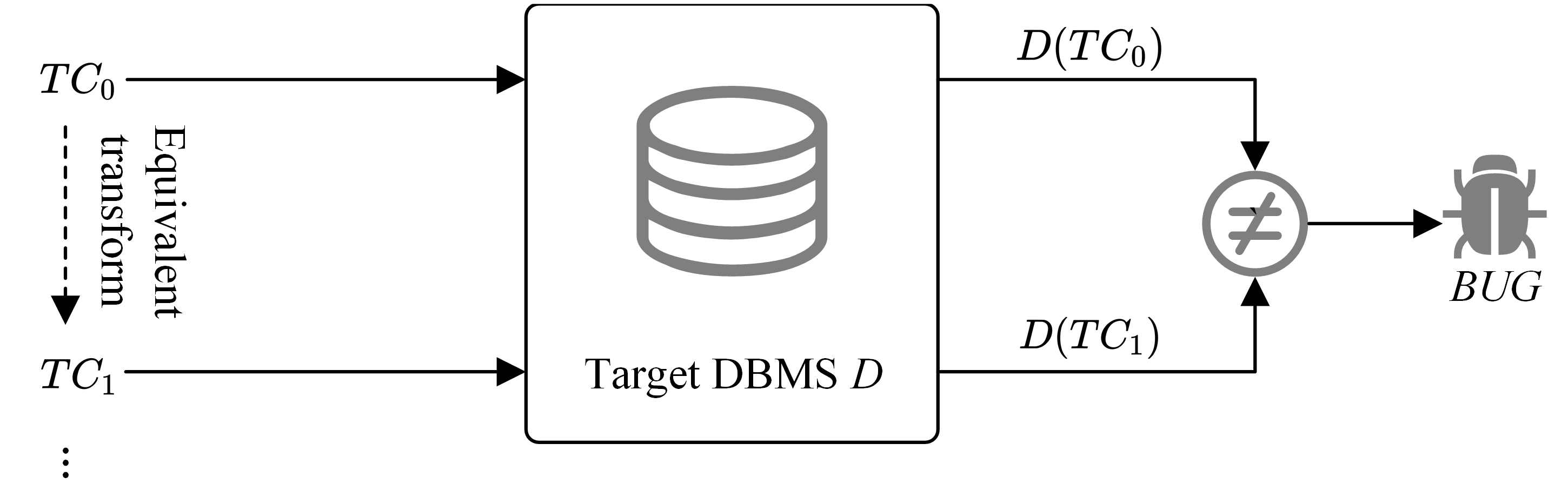}} \\
\subfloat[Constraint-solving Testing]{
\includegraphics[width=0.55\textwidth]{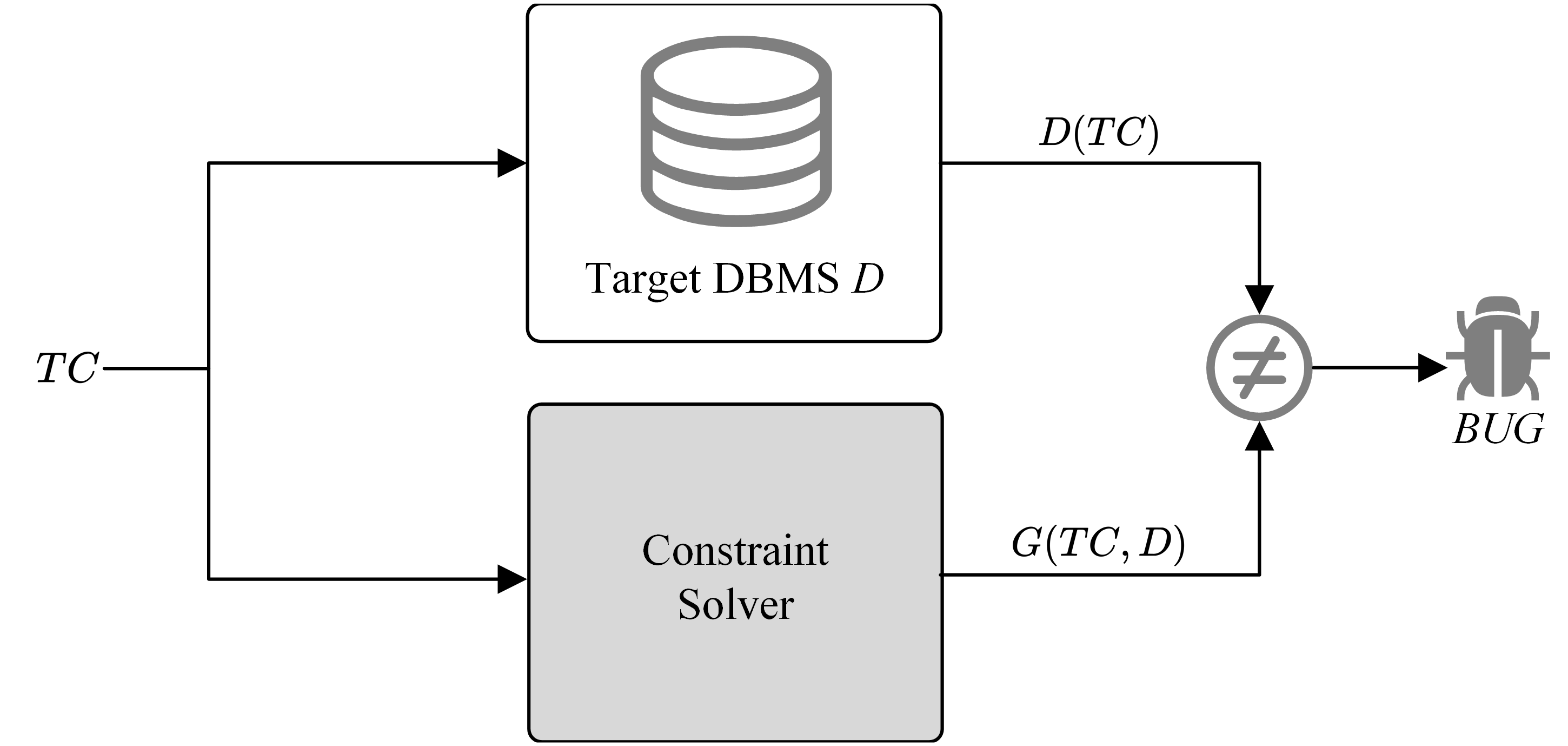}}
\caption{Illustration of the differences between differential testing, metamorphic testing, and constraint-solving testing.}
\label{fig:Classification by Oracle}
\end{figure}

A testing oracle refers to a mechanism to obtain the expected results of test cases. DBMS fuzzing can also be divided into differential testing, metamorphic testing, and constraint-solving testing according to different oracles. Figure~\ref{fig:Classification by Oracle} shows the difference between them. Their formal definitions are shown in Definition~\ref{def:differential testing},~\ref{def:metamorphic testing} and~\ref{def:constraint-solving testing}.

\begin{definition} [Differential Testing] Differential testing detects bugs by comparing the results of the same test case between different databases or versions.  
For a given test case $TC$, a target DBMS $D_0$ and some reference databases $D_1$, $D_2$, ... , $D_n$, the differential test verifies that $ {\forall}$ $j$ : $D_{0}(TC) = D_{j}(TC)$.
\label{def:differential testing}
\end{definition}

\begin{definition} [Metamorphic Testing] Metamorphic testing performs equivalent transformations on original statements to construct equivalence oracles, ensuring that the execution results remain unchanged. For a given test case $TC_0$ and a test database $D$, some equivalent statements $TC_1$, $TC_2$, ... , $TC_n$ can be obtained through transformation. Metamorphic testing validates that $ {\forall}$ $i$, $j$ : $D(TC_i) = D(TC_j)$.
\label{def:metamorphic testing}
\end{definition}

\begin{definition} [Constraint-solving Testing] Constraint-solving testing utilizes constraint solvers to obtain the ground truth of query results and finds the DBMS bugs by comparing the ground truth with the eventual execution results.
For a given test case $TC$, and a test database $D$, constraint-solving testing first uses the constraint solver to obtain the ground truth $G(TC,D)$, and then verifies that $D(TC) = G(TC, D)$.
\label{def:constraint-solving testing}
\end{definition}

\section{Fuzzing Step-based Taxonomy}\label{sec:hor}

\begin{figure}[htp]
    \centering
    \includegraphics[width=\textwidth]{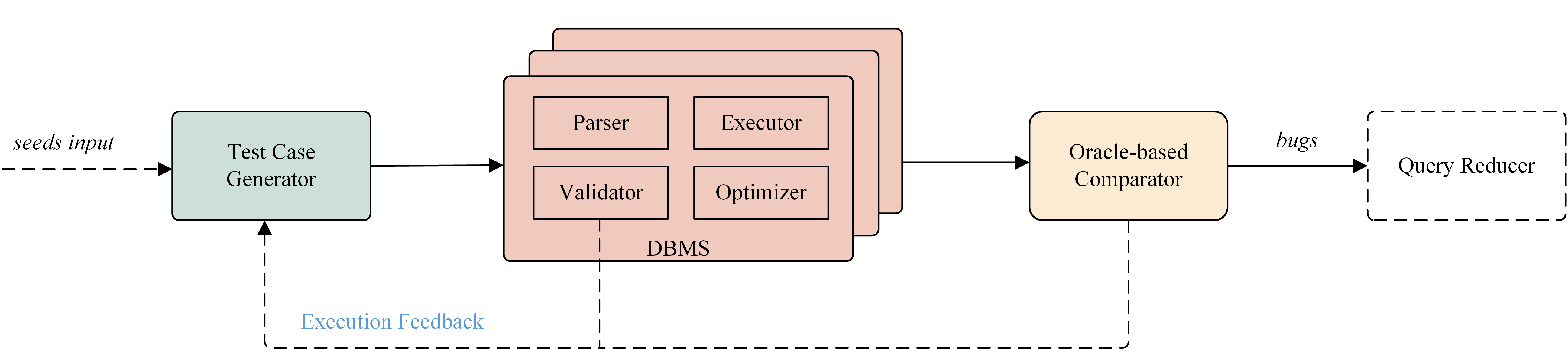}
    \caption{General procedure of DBMS fuzzing. The green part represents the generator, the red part represents the DBMS, the yellow part signifies the oracle-based comparator, while the blue part corresponds feedback. These rules shall be applicable to this figure and the subsequent pipelines.}
    \label{fig:fuzzing horizontal}
\end{figure}

The general (horizontal) procedure for DBMS fuzzing is illustrated in Figure~\ref{fig:fuzzing horizontal}. Basically, all fuzzers contain a test case generator and an oracle-based comparator. The former generates diverse test cases using different strategies to cover various functionalities of the DBMS as comprehensively as possible. The latter utilizes various types of oracles to analyze the execution results. If the execution results do not match the expected ones from the oracle, a bug is identified~\cite{Wang_2021}.

In addition to the aforementioned common modules, some fuzzers may also include the seed input, feedback, and query reducer modules. Seed inputs are typically combined with mutation-based approaches to enrich the state space of DBMS fuzzing. Feedback provides guidance for generating subsequent test cases based on execution information from the current test case~\cite{Pacheco_2007}. The query reducer module is used to reduce the difficulty of triggering test cases to facilitate the localization and resolution of bugs by database developers~\cite{Stobie_2005}.

\subsection{Test Case Generator}

\begin{table}[t]
\caption{Comparison of Test Case Generators}
\label{tab:generator comparation}
\begin{tabular}{|c|c|c|c|c|c|}
\hline
\textbf{Fuzzer} & \textbf{Year} & \textbf{\makecell[c]{Generator\\Type}}         & \textbf{\makecell[c]{Generator\\Strategy}} & \textbf{Feedback}          & \textbf{\makecell[c]{Database\\Instance}}                              \\ \hline
RAGS\cite{slutz1998massive}            & 1998 & \multirow{13}{*}{Generation-based} & \multirow{2}{*}{\makecell[c]{AST Model\\(Static Configuration)}}   & \multirow{13}{*}{\makecell[c]{No Feedback\\(Black-Box)}} & \multirow{3}{*}{Existing Databases}           \\ \cline{1-2}
SQLsmith\cite{git2023sqlsmith}   & 2015     &  &                            &                                    &                                        \\ \cline{1-2} \cline{4-4}
APOLLO\cite{jung2019apollo}     & 2019 &  &     \multirow{2}{*}{\makecell[c]{AST Model\\(Dynamic Configuration)}}  &                          &                                                                        \\ \cline{1-2} \cline{6-6}
AMOEBA\cite{liu2022automatic}     & 2022 &  &                            &                                    &                                        \\ \cline{1-2}
\cline{4-4}
Go-Randgen\cite{git2023gorandgen} & 2019     & &  \multirow{6}{*}{\makecell[c]{AST Model\\(Static Configuration)}} &                            & \multirow{9}{*}{Random Databases}                                       \\ \cline{1-2}
SQLancer\cite{rigger2020testing}   & 2020     &  &                            &                                    &                                        \\ \cline{1-2}
Artemis\cite{mi2021artemis}    & 2021     &  &                            &                                    &                                        \\ \cline{1-2}

DT2\cite{cui2022differentially}         & 2022      &  &                            &                                    &                                        \\ \cline{1-2}
DQE\cite{song2023testing}        &  2023     &  &                            &                                    &                                        \\ \cline{1-2}
Troc\cite{dou2023detecting}       &  2023    &  &                            &                                    &                                        \\ \cline{1-2} \cline{4-4}
TQS\cite{tang2023detecting}        &  2023    &  &        \makecell[c]{AST Model\\(Dynamic Configuration)} &                            &                                                                   \\ \cline{1-2} \cline{4-4} 
ADUSA\cite{abdul2010automated}      &  2010    & & Alloy  Model &                            &                                                               \\ \hline
GARan\cite{bati2007genetic}           & 2007     & \multirow{15}{*}{Mutation-based}  & \multirow{7}{*}{\makecell[c]{SQL Structure\\Mutation}} & \multirow{5}{*}{\makecell[c]{Feedback from\\internal status\\(Grey-Box)}}   & Existing Databases                   \\ \cline{1-2} \cline{6-6}
Squirrel\cite{zhong2020squirrel}    &  2020    &  & &                            & \multirow{14}{*}{Random Databases}                                         \\ \cline{1-2}
Squill\cite{wen2023squill}     & 2023     &  &                            &                                    &                                        \\ \cline{1-2}
SQLRight\cite{liang2022detecting}   &  2022    &  &                            &                                    &                                        \\ \cline{1-2}
DynSQL\cite{jiang2023dynsql}     &  2023    &  &                            &                                    &                                        \\ \cline{1-2} \cline{5-5}
Eqsql\cite{zhang2021duplicate}      &  2021    & & & \multirow{2}{*}{\makecell[c]{No Feedback\\(Black-Box)}} &                                                                            \\ \cline{1-2}
MutaSQL\cite{chen2020testing}    & 2020     &  &                            &                                    &                                        \\ \cline{1-2} \cline{5-5} \cline{4-4} 
LEGO\cite{liang2023sequence}       &  2023    & & \multirow{3}{*}{\makecell[c]{SQL Sequence\\Mutation}} & \makecell[c]{Feedback from\\ internal status\\(Grey-Box)}                   &                                     \\ \cline{1-2} \cline{5-5}
Griffin\cite{fu2022griffin}    &  2022    & & & \makecell[c]{No Feedback\\(Black-Box)} &                                                                            \\ \cline{1-2} \cline{5-5} \cline{4-4} 
QPG\cite{ba2023testing}        &  2023    & & \makecell[c]{DBMS State\\ Mutation} &                         \makecell[c]{Feedback from\\ external interface\\(Black-Box)}   &                                                        \\ \hline
\end{tabular}
\end{table}



The test case generator is the first step in DBMS fuzzing and is used to automatically generate test cases. There are primarily two types of implementation: generation-based and mutation-based. Generation-based methods start from scratch to create test cases without any initial input. It typically relies on a specific SQL syntax model~\cite{Valduriez_1992} to randomly generate constants and expressions, aiming to enumerate potential test cases that satisfy specific criteria or cover specific aspects of the DBMS. On the other hand, mutation-based methods begin with existing seed queries and apply various modifications or mutations to generate new test cases~\cite{Zhou_2021,Shah_2011}. In Table~\ref{tab:generator comparation}, we gain insight into the generation strategies of existing fuzzers and categorize them. Specifically, we compare them in terms of database instance, generation type, strategy, and feedback.

\subsubsection{Database instance}
Before generating test cases, a database instance must be available. Depending on their sources, they fall into two categories: existing databases and random databases. Existing databases come from benchmark tests such as TPC-H~\cite{tpc} or real datasets such as SCOTT~\cite{scott}. Fuzzers based on existing databases~\cite{slutz1998massive, git2023sqlsmith, jung2019apollo, bati2007genetic} first acquire the schema of the target database and then generate syntactically correct statements based on that schema. However, as both the schema and the data are fixed, the generated statements cannot effectively explore the entire query space, leading to incomplete detection. On the other hand, fuzzers based on random databases~\cite{git2023gorandgen, liu2022automatic, rigger2020testing, rigger2020finding, cui2022differentially, song2023testing, dou2023detecting, abdul2010automated, mi2021artemis, tang2023detecting, zhong2020squirrel, wen2023squill, fu2022griffin} create random database instances from scratch. The random generation of databases has been widely explored~\cite{gray1994quickly, bruno2005flexible, houkjaer2006simple, neufeld1993generating}. A common practice is to first create tables, indexes and views randomly and then populate them with data using INSERT, UPDATE, and DELETE statements~\cite{Rigger_2020}. The advantage of using randomly generated databases is their potential to thoroughly explore the query space. When coupled with execution feedback, these solutions can effectively explore the search space.

\subsubsection{Generator type and strategy}
After acquiring the database instance, we can proceed to generate test cases using either a generation-based or a mutation-based approach. Among them, the generation-based approach consists primarily of two strategies: Abstract Syntax Tree (AST) model and Alloy model~\cite{ALLoy01}. 

\begin{figure}[tp]
    \centering
    \includegraphics[width=0.98\textwidth]{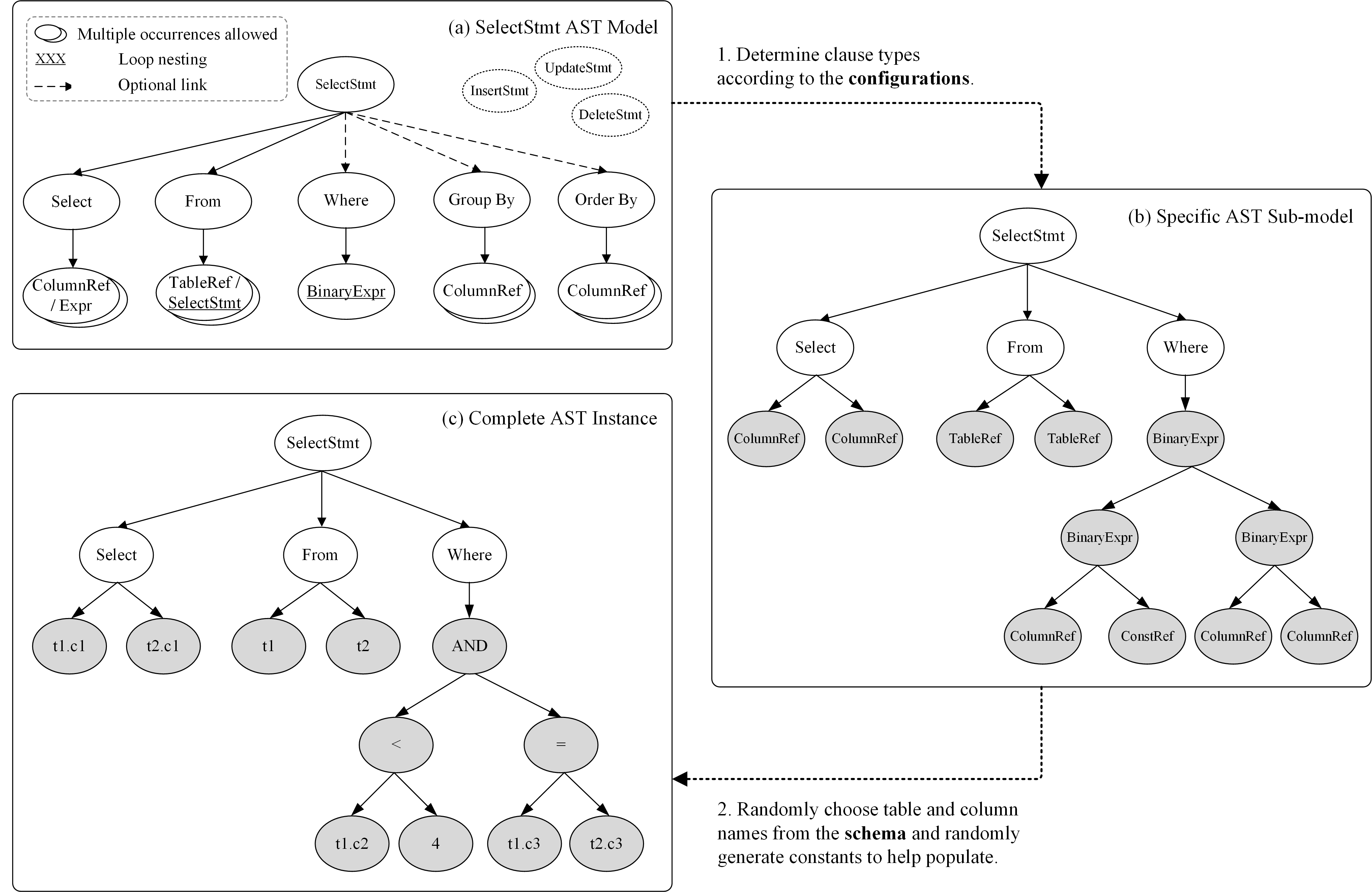}
    \caption{Working process of generation-based generators based on AST model.}
    \label{fig:ast model generator}
\end{figure}

The general process of the AST model is depicted in Figure~\ref{fig:ast model generator}. First, we need an AST model. AST itself is a tree-like data structure, which can represent the various components of SQL query statements (such as \textsf{SELECT, FROM, WHERE, JOIN}, etc.) and their relationships as nodes and branches for subsequent parsing and processing. Moreover, an AST model is a meta-model that covers the shared syntax structure of all SQL statements' AST instances. On top of that, the clause types are determined according to the configurations to obtain a specific AST sub-model. Then, table and column names are randomly selected from the schema of a specific database instance and combined with randomly generated constants to populate the AST sub-model, resulting in a complete AST instance. From the process, it can be observed that the AST model can only guarantee syntactic correctness, rather than semantic correctness.

In the AST model-based generator, the aforementioned configurations can be both static and dynamic. The static configurations refer to the static parameters manually set before generation, which can guide the generation of test cases. Common static configurations include:
\begin{itemize}
    \item \textbf{Customizing the Probability of SQL Statements}: This parameter allows for customizing the distribution of test case types, such as the probability of \textsf{SELECT, INSERT, UPDATE, DELETE}, etc.~\cite{slutz1998massive, git2023gorandgen}.
    \item \textbf{Customizing the Probability of SQL Clauses}: This parameter allows for customizing the distribution of different clauses, such as \textsf{WHERE, GROUP BY, ORDER BY}, etc.~\cite{slutz1998massive, git2023gorandgen}.
    \item \textbf{Limiting the Maximum number of Joined Tables}: This parameter helps constrain the complexity of join relationships, preventing the generation of large-scale multi-table join statements and enhancing the efficiency of statement generation~\cite{slutz1998massive, git2023gorandgen}.
    \item \textbf{Limiting the Maximum Number of Rows Returned}: This parameter restricts the execution time and resource consumption of the test cases by adding a \textsf{LIMIT} clause to the original statement~\cite{slutz1998massive, git2023gorandgen}.
    \item \textbf{Limiting the Maximum Depth of the AST}: This parameter is set to limit the generation time of test cases, as the generation time tends to increase exponentially with the increase in AST depth. Without this limitation, numerous complex SQL statements would be generated, which may expand the search space but reduce the overall efficiency of bug detection~\cite{rigger2020testing, Rigger_2020, rigger2020finding, cui2022differentially, song2023testing, dou2023detecting}.
\end{itemize}


However, static configurations require manual adjustments based on specific requirements. This process can be time-consuming, and manually setting parameters may not sufficiently explore the state space in a targeted manner. Dynamic configurations of the AST model adjust generator parameters based on previous execution results, which include: 
\begin{itemize}
    \item \textbf{Dynamic Probability Table}: The dynamic probability table stores the dynamic probability of various SQL clauses. If an SQL query triggers a bug, the probability of the clauses appearing in that SQL query increases. This ensures that the generator explores the input space with maximum efficiency in generating test cases~\cite{jung2019apollo, liu2022automatic}.
    \item \textbf{Schema Graph}:  The Schema Graph describes the connection relationships between tables, and by randomly traversing the schema graph, a join pattern can be obtained. It can adjust the selection probability of each table of the schema. This dynamic adaptation prevents the generation of duplicate test cases and significantly improves the efficiency of fuzzing~\cite{tang2023detecting}.
\end{itemize}

\begin{figure}

\begin{minipage}{\textwidth}
\centering
\begin{minipage}{.4\textwidth}
\begin{lstlisting}[language=Alloy, caption=Alloy Syntax Model,frame=tlrb,label=lst:alloy syntax model,aboveskip=5pt]{Name}
abstract sig FieldName {}
abstract sig TableName {}
abstract sig Value {}
abstract sig Table {
    name : one TableName,
    fields : some FieldName
}
sig Term {
    field : one FieldName,
    agg : lone AggregateName
}
sig Select {
    fields : some Term
}
sig From {
    tables: some Table
}
sig Where {
    condition : one Condition
}
sig Condition {
    leftExpr: one Expr,
    extendCond: some ExtendCondition
}
sig ExtendCondition{
    logicalOp: one LogicalOperator,
    rightCond: one Condition
}
sig Expr{
    leftOperand: one FieldOrValue,
    cmpOp: one ComparisionOperator,
    rightOperand: one FieldOrValue
}
sig FieldOrValue{
    field: lone FieldName,
    value: lone Value
}
sig SelectStatement{
    selectClause: one Select, 
    fromClause: one From, 
    whereClause: one Where
}
\end{lstlisting}
\end{minipage} 
\hspace{3em}
\begin{minipage}{.4\textwidth}
\begin{lstlisting}[language=Alloy, caption=Alloy Data Model,frame=tlrb,label=lst:alloy data model]{Name}
sig id, name, studentID,
    courseID, grade extends 
    FieldName {}
    
sig students, grades extends
    TableName {}

fact field_value_not_empty {
    all f: FieldOrValue |
    (
        f.field in FieldName
        and no f.value
    ) or
    (
        f.value in Values and
        no f.field
    )
}

fact select_field_in_table {
    all f: Select.fields.field | 
    some t: From.tables |
    f in t.fields
}

fact where_field_in_table{
    all f: Where.(...).field |
    (
        some t: From.tables
    ) |
    (
        f in t.fields
    )
}

fact unique_select_terms {
    all a, b : SELECT.fields |
    (
        a.field = b.field and 
        a.agg = b.agg 
    ) => a=b
}
\end{lstlisting}
\end{minipage}
\end{minipage}
\end{figure}

Alloy~\cite{jackson2002alloy} is a language for describing structured properties. When applied to describe SQL generation rules, it can specify not only SQL syntax rules, but also table names, column names, constant values in SQL statements, and constraints between table names and column names. As shown in Listing~\ref{lst:alloy syntax model} and Listing~\ref{lst:alloy data model}, we provide a simplified example of using the Alloy model to generate SQL statements. The Alloy syntax model first defines four signatures: `FieldName', `TableName', `Value' and `Table'. Then, `Term' is defined to represent an item of Select clause, it consists of one FieldName and at most one AggregateName. The `Select' signature contains multiple terms, while the `From' signature contains multiple tables. Since the conditions in the `Where' clause may be nested, `Condition' and `ExtendCondition' are defined to describe this possible relationship. A `Condition' consists of one `Expr' and some  `ExtendCondition', where `Expr' consists of a left operand, an operator, and a right operand. Importantly, the left or right operand can be a field or a value. Therefore, a signature called `FieldOrValue' is introduced to represent this operand. It consists of a `FieldName' and `Value' with keyword `lone', where `lone' signifies the semantics of at most one. Finally, by combining `Select' clause, `From' clause, and `Where' clause, we construct a simple Select syntax model. 

\begin{figure}[t]
\centering
\subfloat[SQL Structure Mutation]{
\includegraphics[width=0.8\textwidth]{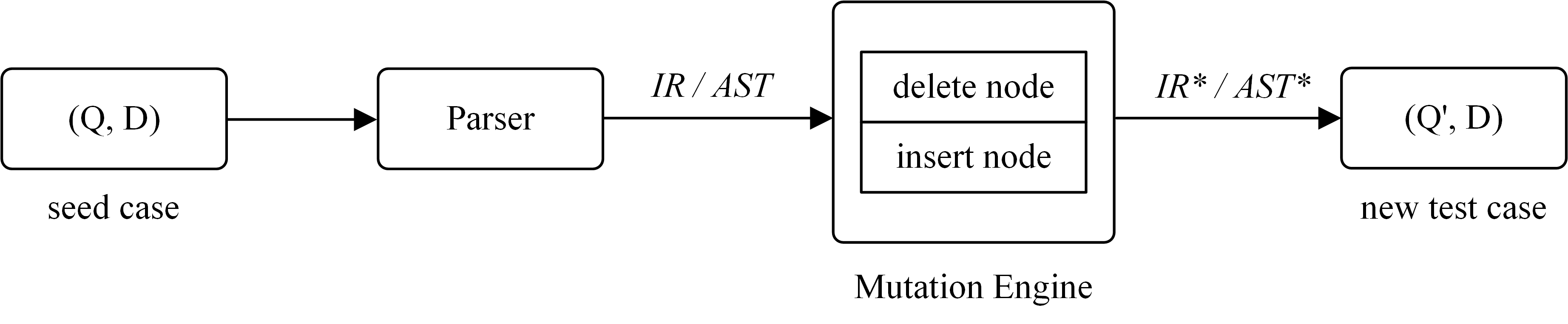}} \\
\subfloat[SQL Sequence Mutation]{
\includegraphics[width=0.8\textwidth]{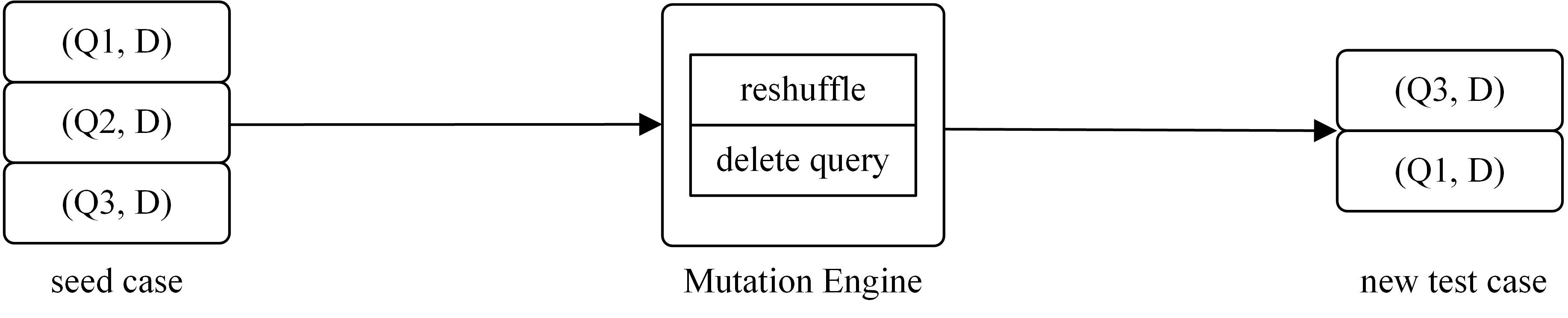}} \\
\subfloat[DBMS State Mutation]{
\includegraphics[width=0.8\textwidth]{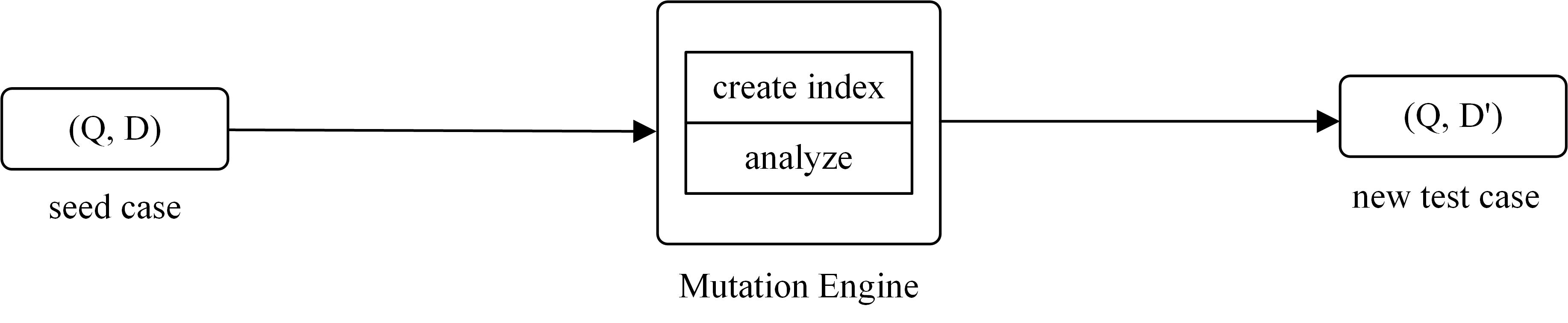}}
\caption{Working process of mutation-based generators.}
\label{fig:mutation-based generator}
\end{figure}

The Alloy data model describes the specific values for `FieldNames' and `TableNames'. Additionally, it introduces a fact named `field\_value\_not\_empty', which ensures that for `FieldOrValue', either the field or the value must be non-empty. The facts `select\_field\_in\_table' and `where\_field\_in\_table' place constraints on the column names appearing in the select and where clauses, requiring them to originate from tables stipulated in the from clause. The fact `unique\_select\_terms' guarantees that there are no duplicate fields in the select clause. The Alloy syntax model is similar to the AST model, while the former guarantees the semantic accuracy of generated statements by constraining column names in select clause and where clause. Therefore, a generator based on the Alloy model~\cite{abdul2010automated} can achieve a higher level of semantic correctness compared to the AST model.

Mutation-based generators can be divided into three major categories: SQL structure mutation~\cite{bati2007genetic, zhong2020squirrel, wen2023squill, liang2022detecting, jiang2023dynsql, zhang2021duplicate, chen2020testing}, SQL sequence mutation~\cite{liang2023sequence, fu2022griffin}, and DBMS state mutation~\cite{ba2023testing}. Figure~\ref{fig:mutation-based generator} illustrates the differences between three mutation strategies. SQL structure mutation generates new test cases by modifying the structure of statement, and it is the most common type of mutation. SQL sequence mutation generates new test cases by reshuffling or deleting queries from the original test cases. Its advantage is that it does not rely on SQL syntax, which reduces the adaptation cost for new DBMSs. DBMS state mutation uses DDL operations to create indexes or modify table definitions to generate new database schemas. In addition, it combines old SQL statements with the newly generated schemas to form new test cases. Since the same statement usually has different execution plans under different database schemas, and these diverse execution plans trigger distinct code branches, state mutations contribute to an increased degree of testing for the DBMS source code.

\begin{figure}

\begin{minipage}{\textwidth}
\centering
\begin{minipage}{.6\textwidth}
\centerline{\includegraphics[width=\textwidth]{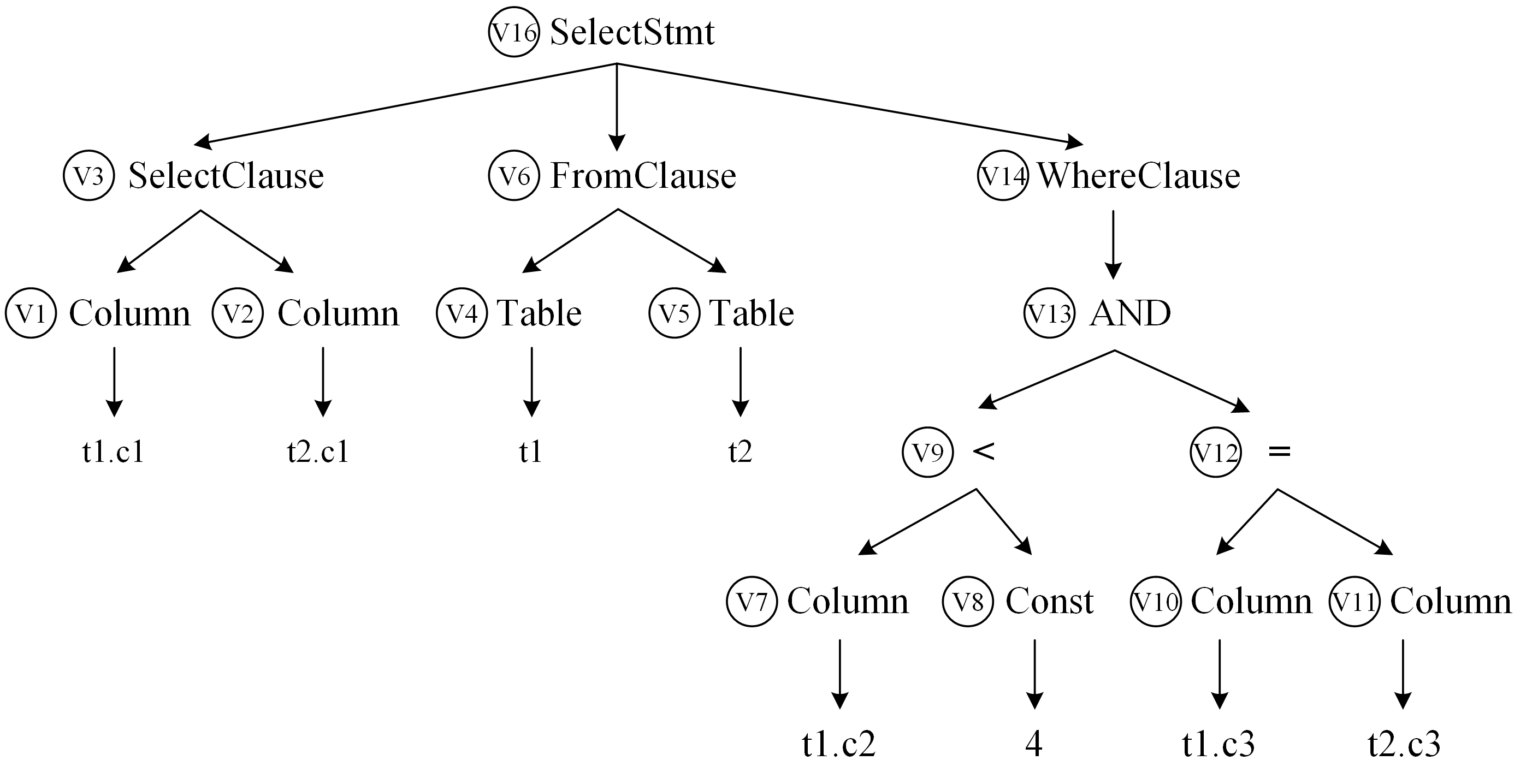}}

\end{minipage} 
\hspace{1em}
\begin{minipage}{.36\textwidth}
\begin{lstlisting}[language=, label=label=lst:ir, basicstyle=\small]{Name}
V1=(Column, l=null, r=null)
V2=(Column, l=null, r=null)
V3=(SelectClause, l=V1, r=V2)
V4=(Table, l=null, r=null)
V5=(Table, l=null, r=null)
V6=(FromClause, l=V4, r=V5)
...
V13=(AndExpr, l=V9, r=V12)
V14=(WhereClause, l=V13, r=null)
V15=(unknown, l=V6, r=V14)
V16=(SelectStmt, l=V3, r=V15)
\end{lstlisting}

\end{minipage}
\end{minipage}
\caption{An example of AST and its corresponding IR, on the left, the AST contains a series of tree nodes, and on the right, V1-V16 represent a series of IR nodes corresponding to each tree node. V15 does not correspond to any tree node in the AST, its introduction serves the purpose of conveniently representing the SelectStmt.}
\label{figure:An example of IR}
\end{figure}

One of the implementation methods for the SQL structure mutation, known as the AST mutation, suffers from various limitations. Due to strict type constraints and complex mutation operations, altering the AST corresponding to an SQL statement is as challenging as modifying the SQL statement itself~\cite{zhong2020squirrel}. Therefore, the IR mutation was proposed to simplify mutation operations. As shown in Figure~\ref{figure:An example of IR}, IR nodes are actually AST tree nodes with constant values removed. Each IR node has a specific type and can have a maximum of two child nodes. Each mutation operation only needs to deal with the two child nodes of an IR node. There are three mutation operations applied to IR nodes: 
\begin{itemize}
    \item \textbf{Insertion}: Insertion involves creating a new IR node and inserting it into an existing IR node. For example, a `Table' can be inserted into a `FromClause'.
    \item \textbf{Replacement}: Replacement can be achieved by modifying the type of an IR node, such as replacing column `x' with `count(x)'.
    \item \textbf{Deletion}: Deletion means removing at least one child node from an IR node.
\end{itemize}

\subsubsection{Execution feedback}
Most generators do not take advantage of feedback information and treat the target DBMS as a black box, which accepts SQL statements or transactions as input and output execution results or execution plans. Some generators of black-box fuzzers~\cite{ba2023testing} obtain feedback such as the validity of query plans or test cases from the query or query plan interface provided by the DBMS, guiding the subsequent generation process. This kind of feedback is essentially different from the feedback from gray-box fuzzers. The gray-box fuzzer generators~\cite{liang2023sequence,bati2007genetic,zhong2020squirrel,wen2023squill,liang2022detecting,jiang2023dynsql} collect the internal status of the DBMS, such as collecting code coverage to guide the generation of test cases that maximize code coverage. In contrast, the feedback of the black-box fuzzers relies only upon the external interface of the DBMS without any knowledge of the source code. Black-box fuzzers are easier to implement and can be used to test any commercial DBMSs. However, it is essentially a random exploration of the entire input space. Gray-box fuzzers guide generation based on the DBMS's internal state and can usually achieve higher code coverage, but it can only be used for open source DBMSs. More types of feedback and information are elaborated on in Section~\ref{sec:feedback}.

\subsection{Oracle-based Comparator}

\begin{table}[t]
\caption{Comparison of Oracles}
\label{tab:oracle comparison}
\begin{tabular}{|c|c|c|c|}
\hline
\textbf{Fuzzer} & \textbf{Oracle Type}                & \textbf{Feature}                         & \textbf{Test Scope}               \\ \hline
Squirrel\cite{zhong2020squirrel}        & \multirow{5}{*}{Crash}              & \multirow{5}{*}{Database Crashes}                       & \multirow{5}{*}{Crash Bugs}       \\ \cline{1-1}
Squill\cite{wen2023squill}         &                                     &                                          &                                   \\ \cline{1-1}
Griffin\cite{fu2022griffin}         &                                     &                                          &                                   \\ \cline{1-1}
LEGO\cite{liang2023sequence}            &                                     &                                          &                                   \\ \cline{1-1}
DynSQL\cite{jiang2023dynsql}          &                                     &                                          &                                   \\ \hline
RAGS\cite{slutz1998massive}            & \multirow{6}{*}{Differential}       & \multirow{5}{*}{Different Databases}     & \multirow{5}{*}{Logic Bugs}       \\ \cline{1-1}
SQLsmith\cite{git2023sqlsmith}        &                                     &                                          &                                   \\ \cline{1-1}
Go-Randgen\cite{git2023gorandgen}      &                                     &                                          &                                   \\ \cline{1-1}
GARan\cite{bati2007genetic}           &                                     &                                          &                                   \\ \cline{1-1}
DT2\cite{cui2022differentially}             &                                     &                                          &                                   \\ \cline{1-1} \cline{3-4} 
APOLLO\cite{jung2019apollo}          &                                     & Different Versions of A Same Database       & \multirow{2}{*}{Performance Bugs} \\ \cline{1-3}
AMOEBA\cite{liu2022automatic}          & \multirow{8}{*}{Metamorphic}        & \multirow{4}{*}{Statement Rewriting} &                                   \\ \cline{1-1} \cline{4-4} 
MutaSQL\cite{chen2020testing}         &                                     &                                          & \multirow{11}{*}{Logic Bugs}      \\ \cline{1-1}
Eqsql\cite{zhang2021duplicate}           &                                     &                                          &                                   \\ \cline{1-1} 
NoREC\cite{Rigger_2020}           &                                     &                     &                                   \\ \cline{1-1}
SQLRight\cite{liang2022detecting}        &                                     &        &                                   \\ \cline{1-1} \cline{3-3}
DQE\cite{song2023testing}             &                                     & Statement Type Transformation         &                                   \\ \cline{1-1} \cline{3-3}
TLP\cite{rigger2020finding}             &                                     & Query Partition                          &                                   \\ \cline{1-1} \cline{3-3}
Troc\cite{dou2023detecting}            &                                     & Transaction Splitting       &                                   \\ \cline{1-3}
ADUSA\cite{abdul2010automated}           & \multirow{4}{*}{Constraint-solving} & \multirow{2}{*}{Forward Solving(SAT Solver)}              &                                   \\ \cline{1-1}
Artemis\cite{mi2021artemis}         &                                     &                                          &                                   \\ \cline{1-1} \cline{3-3}
PQS\cite{rigger2020testing}             &                                     & Backward Solving                       &                                   \\ \cline{1-1} \cline{3-3}
TQS\cite{tang2023detecting}             &                                     & Forward Solving(Logical Solver)                 &                                   \\ \hline
\end{tabular}
\end{table}

In manual testing, expected results can be calculated manually, and potential errors or vulnerabilities can be detected by comparing expected results with the eventual execution results on the target DBMS. However, for fuzzing, it is not a trivial task to obtain the expected results of random test cases. Oracle refers to the means to obtain the expected results~\cite{howden1978theoretical}. Existing oracles can be classified into four types: crash oracle, differential oracle, metamorphic oracle, and constraint-solving oracle. Table~\ref{tab:oracle comparison} presents the classification of fuzzers along with their respective oracle implementation methods and applicable domains.


\subsubsection{Crash oracle}

The crash oracle, also called a no-oracle, determines the occurrence of a crash based on the running state of the target database. Fuzzing methods based on crash oracle~\cite{zhong2020squirrel, wen2023squill, liang2023sequence, fu2022griffin, jiang2023dynsql} can only detect crashes. After executing test cases, bugs are determined based on whether the DBMS stops running. 

\subsubsection{Differential oracle}
Differential oracles~\cite{slutz1998massive, git2023sqlsmith, git2023gorandgen, cui2022differentially, bati2007genetic} detect logic bugs or performance bugs by comparing the execution results of different DBMSs (or different versions of the same DBMS). It tries to get expected results from another DBMS. The advantage of differential oracle testing is that it is relatively simple. Nevertheless, it cannot detect common bugs between the target DBMS and the referenced one.

\subsubsection{Metamorphic oracle}

To avoid the drawbacks of the differential oracle, many fuzzing methods~\cite{liu2022automatic, chen2020testing, zhang2021duplicate, Rigger_2020, rigger2020finding, song2023testing, liang2022detecting} use the metamorphic oracle to detect logic bugs or performance bugs. Metamorphic oracle applies equivalent transformations on the original statements to construct equivalent statements, ensuring that the execution results remain consistent. There are several ways to construct metamorphic oracles:

\begin{figure}[t]
\centering
\subfloat[Constraint Rewriting]{
\includegraphics[width=0.9\textwidth]{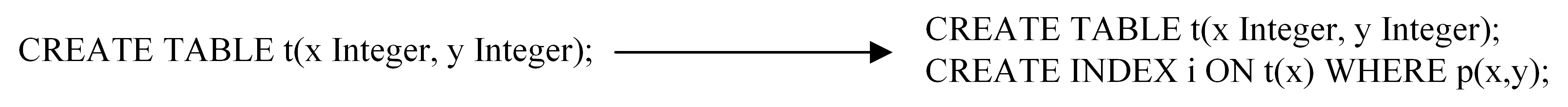}} \\
\subfloat[Structure Rewriting]{
\includegraphics[width=0.91\textwidth]{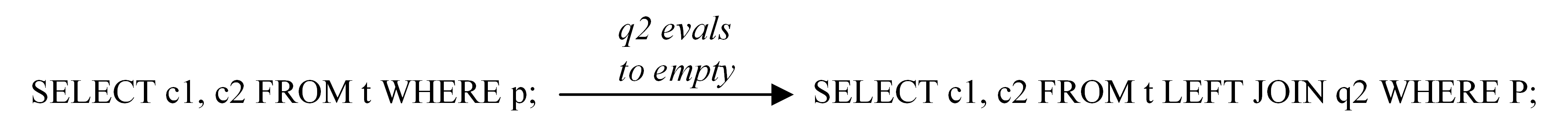}} \\
\subfloat[Expression Rewriting]{
\includegraphics[width=0.9\textwidth]{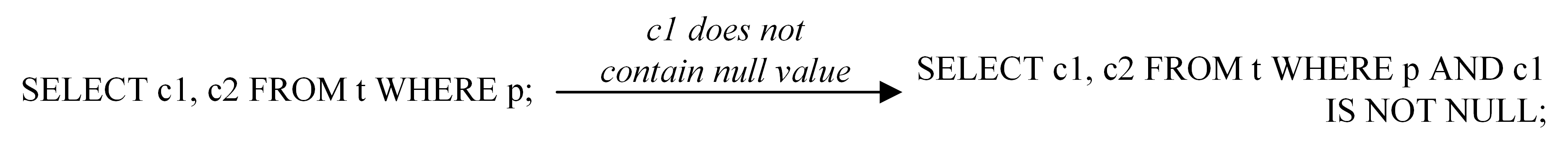}}
\caption{Example of statement rewriting strategies.}
\label{fig:statement rewriting}
\end{figure}

\begin{figure}[tp]
    \centering
    \includegraphics[width=0.7\textwidth]{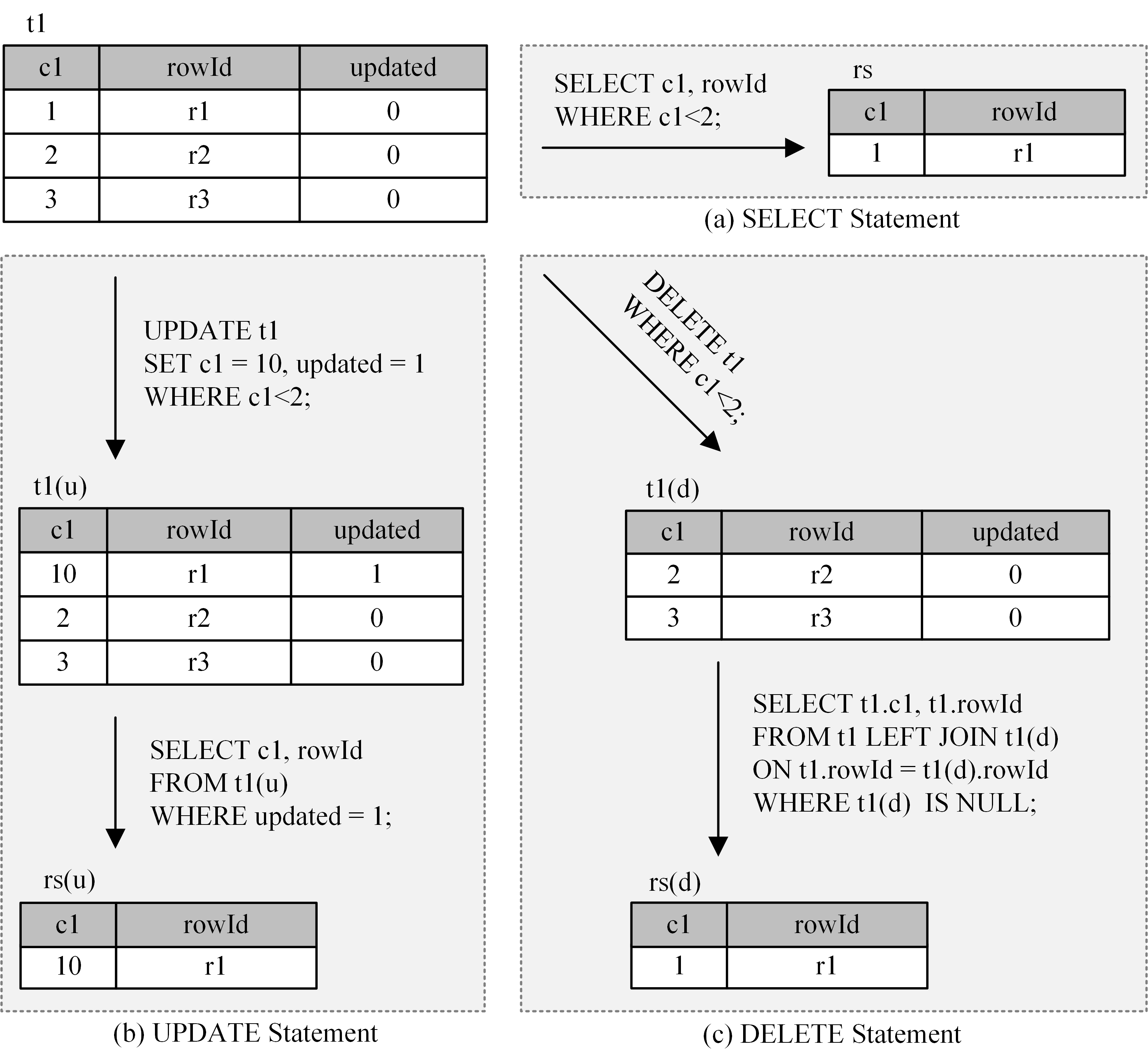}
    \caption{Example of transformation of statement type}
    \label{fig:transformation of statement Type}
\end{figure}

\begin{itemize}
\item \textbf{Statement Rewriting}: Statement Rewriting refers to the process of modifying a statement without changing its semantics or results. Common methods of rewriting statements include constraint rewriting, structure rewriting, and expression rewriting, as shown in  Figure~\ref{fig:statement rewriting}. Constraint rewriting refers to modifying the constraints of a table, such as creating an index. Obviously, the execution result of the same \textsf{SELECT} statement should remain unchanged after creating the index. Structure rewriting refers to adding new clauses such as \textsf{JOIN, DISTINCT} and \textsf{LIMIT} based on the original statement. Expression Rewriting refers to the process of rewriting logical or arithmetic expressions, such as adding predicates that always hold true to the \textsf{WHERE} clause or changing the string comparison operator from `=' to `LIKE', which will not affect the result of the expression~\cite{liu2022automatic, chen2020testing, zhang2021duplicate, Rigger_2020}. 
\item \textbf{Statement Type Transformation}: Statement type transformation involves constructing \textsf{UPDATE} and \textsf{DELETE} statements that correspond to a given \textsf{SELECT} statement, where the three types of statements should affect the same tuples~\cite{song2023testing}. By adding tagging columns 'rowid' and 'updated' to the table, the \textsf{SELECT} statements can return 'rowid' of the affected rows, as shown in Figure~\ref{fig:transformation of statement Type}. For \textsf{DELETE} statements, the affected rows can be determined by calculating the difference between the set of 'rowid' before and after deletion. Similarly, for \textsf{UPDATE} statements, the affected rows can be obtained by querying the 'rowid' where the 'updated' column is marked as true.
\item \textbf{Query Partition}: Query partition refers to dividing an original SQL query into multiple partitions, and the merged results of the partitioned queries should be consistent with the results of the original query~\cite{rigger2020finding}. For example, a query `\textsf{SELECT * FROM t1}' can be partitioned into three queries: `\textsf{SELECT * FROM t1 WHERE c1<10}', `\textsf{SELECT * FROM t1 WHERE c1>=10}', and `\textsf{SELECT * FROM t1 WHERE c1 IS NULL}';
\item \textbf{Transaction Splitting}: Transaction splitting refers to splitting transactions into statements for execution in conjunction with external version concurrency control. In other words, by building a multiple version chain of data outside the database, statements within transactions can run in non-transaction mode but read data visible in transaction mode. By comparing the execution results before and after transaction splitting, it is possible to determine whether there exist any logic bugs in the transaction management module of the DBMS~\cite{dou2023detecting}.
\end{itemize}

\subsubsection{Constraint-solving oracle}
Due to the fact that not all test queries can obtain test oracles by constructing equivalent queries, some fuzzers use constraint-solving to obtain expected result. Constraint-solving fuzzing methods~\cite{abdul2010automated, mi2021artemis, rigger2020testing, tang2023detecting} generally relies on forward or backward solving to derive the ground truth of the execution results. 
\begin{itemize}
    \item \textbf{Forward Solving}: Forward solving involves using an external solver such as an SAT solver to evaluate each tuple against predicate constraints in order to obtain the ground truth for the results of the statement execution~\cite{khurshid2004testera, goldberg2007berkmin}. In this process, a logical solver can also be used as a substitute for the SAT solver, translating the join predicates into logical operations to accelerate the solving speed of the join predicates~\cite{tang2023detecting}.
    \item \textbf{Backward Solving}: Backward solving entails initially selecting some tuples as ground truth at random and then using an SAT solver to work backward and obtain a statement whose execution results include these tuples~\cite{rigger2020testing}.
\end{itemize}

In all, each oracle has its own applicable scenarios. Fuzzing based on crash oracles can only detect crash-related issues such as program errors and memory leaks, which limits their application scenarios. Fuzzing based on differential oracles can only detect common functional modules between different databases and cannot identify logic or performance bugs shared by the referenced database and the test one. Fuzzing based on metamorphic oracles does not require a reference database, but suffers from a limited test range. Although constrained solvers are not subject to testing scope limitations, they exhibit lower solving efficiency, resulting in lower testing efficiency. Although the AST interpreter and logical operation solver improve the efficiency of constraint solving, they also suffer from the limitation of the test scope. For instance, the logical operation solver can only be applied to test logic bugs in join optimizations.

\subsection{Execution Feedback}
\label{sec:feedback}

\begin{table}[tp]
\centering
\small
\caption{Comparison of Fuzzer Execution Feedbacks}
    \label{tab:feedback comparation}
    \begin{tabular}{cccccc}
      \toprule[1pt]
      \textbf{Fuzzer} & \textbf{\makecell[c]{Metamorphic}} & \textbf{\makecell[c]{Validation}} & \textbf{\makecell[c]{Coverage}} & \textbf{\makecell[c]{Query Plan}} & \textbf{\makecell[c]{Syntax \& Semantics Error}} \\
      \midrule[0.5pt]
      APOLLO\cite{jung2019apollo} &  \ding{55} & \ding{51} & \ding{55} & \ding{55} & \ding{55} \\
      Squirrel\cite{zhong2020squirrel} &  \ding{55} & \ding{55} & \ding{51} & \ding{55} & \ding{55} \\
      LEGO\cite{liang2023sequence} &  \ding{55} & \ding{55} & \ding{51} & \ding{55} & \ding{55} \\
      SQLRight\cite{liang2022detecting} &  \ding{55} & \ding{55} & \ding{51} & \ding{55} & \ding{55} \\
      QPG\cite{ba2023testing} &  \ding{55} & \ding{55} & \ding{55} & \ding{51} & \ding{55} \\
      AMOEBA\cite{liu2022automatic} &  \ding{51} & \ding{51} & \ding{55} & \ding{55} & \ding{55} \\
      GARan\cite{bati2007genetic} &  \ding{55} & \ding{55} & \ding{51} & \ding{55} & \ding{55} \\
      DynSQL\cite{jiang2023dynsql} &  \ding{55} & \ding{55} & \ding{51} & \ding{55} & \ding{51} \\
      Squill\cite{wen2023squill} &  \ding{55} & \ding{51} & \ding{51} & \ding{55} & \ding{51} \\
      \bottomrule[1pt]
    \end{tabular}
\end{table}

Execution feedback refers to the process of guiding test case generation based on execution results or database status. Execution feedback is not limited to dynamic configuration of generators, but can also adjust mutation strategies and search directions, typically improving the efficiency of fuzz testing. Table~\ref{tab:feedback comparation} summarizes the feedback mechanisms used by existing fuzzers, which usually include seven types: metamorphic feedback, validation feedback, coverage feedback, query plan feedback, syntax error feedback, semantics error feedback, and fitness feedback.

\subsubsection{Metamorphic feedback}
Metamorphic feedback~\cite{liu2022automatic} refers to the use of feedback information to generate more statements that satisfy the conditions for the use of metamorphic oracles. This feedback is often combined with generation-based generators, where the generation process is influenced by the success of equivalent transformations. Specifically, in fuzzers using metamorphic feedback, statements are generated based on a probability table. If a SQL query can employ some of the statement rewriting strategies shown in Figure~\ref{fig:statement rewriting}, the subsequent occurrence probabilities of \textsf{LIMIT, GROUP BY}, and \textsf{WHERE} clauses will increase. Metamorphic feedback improves the applicability between generated statements and the metamorphic oracle.

\subsubsection{Validation feedback}
Validation feedback refers to the process of increasing the probability of generating test cases that can trigger bugs based on the feedback from the comparator. For generation-based generators~\cite{jung2019apollo, liu2022automatic}, if a SQL query triggers a bug, the probability of subsequent occurrence of different SQL clauses within the statement will increase. For mutation-based generators~\cite{wen2023squill}, if a mutation statement derived from a seed statement triggers a bug, validation feedback will add the mutation statement to the seed queue for selection by the generator.

\subsubsection{Coverage feedback}
Coverage feedback aims to generate statements that maximize the coverage of the DBMS code. This feedback is commonly used in conjunction with mutation-based generators~\cite{zhong2020squirrel, wen2023squill, liang2022detecting, jiang2023dynsql, bati2007genetic, liang2023sequence}. By modifying the binary code of the target DBMS during execution to inject additional code that tracks and records execution paths, these fuzzers can collect code coverage information before and after executing each test case. If a mutation query leads to an increase in code coverage, it will be added to the seed queue. Additionally, LEGO~\cite{liang2023sequence} analyzes the type sequences of statements that increase the coverage of the code, and then the mutation processes will generate test cases containing the target type sequences.

\subsubsection{Query plan feedback}
The goal of query plan feedback is for the generator to maximize the number of query plans. A query plan represents the execution path of an SQL query in a DBMS, with different query plans corresponding to distinct code branches. Query plan feedback~\cite{ba2023testing} uses the emergence of new query plans to guide subsequent query generation. Specifically, it models the objective of maximizing query plan quantity as a Multi-Armed Bandit (MAB) problem~\cite{berry1985bandit}. The generation of new query plans corresponds to exploitation, while exploring new mutation operations corresponds to exploration. When considering both exploitation and exploration, more unique query plans can be generated, increasing the possibility of discovering logic bugs.

\subsubsection{Syntax error and semantics error feedback}
Some generators~\cite{wen2023squill} consider statements with semantic and syntactic errors to be valuable because they may activate certain unexplored SQL features. Therefore, these statements are fed back into the seed queue, but given lower priority, rather than being discarded entirely. Some other generators~\cite{jiang2023dynsql} advocate that although syntax errors and semantic errors can improve code coverage, they do not help detect logic bugs. This is because, even though statements with syntax and semantic errors can trigger boundary branches in the code, they will fail during the parsing stage. This failure prevents them from exploring the underlying logic of the DBMS, providing limited help in detecting bugs. Therefore, feedback is utilized to filter out these invalid test cases from the seed queue.


Most fuzzers use coverage feedback to guide the generation of test cases. Intuitively, coverage feedback can help generate statements that cover as many code branches as possible, thus discovering various bugs. Validation feedback captures the characteristics of test cases that trigger bugs, effectively uncovering specific bugs, but it may not increase the variety of bugs discovered. Query plan feedback essentially generates statements with diverse query plans to trigger as many code branches as possible. Metamorphic feedback can guide the generation of test cases that are more suitable for oracles. Syntax and semantic error feedback play a minor role because test cases with syntax errors do not help detect the underlying bugs.

\subsection{Query Reducer}
Query reducers are designed to simplify and summarize complex queries, providing developers with minimal human-readable queries for debugging. Query reduction strategies can be classified into four types: reduce expression, delete clause, delete subquery, and delete IR node. Table~\ref{tab:query reduction comparation} shows the query reduction strategies used by each fuzzer and also indicates whether the reducer preserves semantics.

\begin{table*}[t]
\centering
\small
\caption{Comparison of Fuzzer Query Reducers}
    \label{tab:query reduction comparation}
    \begin{tabular}{ccccc|c}
      \Xhline{1pt}
      \textbf{Fuzzer} & \textbf{\makecell[c]{Reduce  Expression}} & \textbf{\makecell[c]{Delete Clause}} & \textbf{\makecell[c]{Delete Subquery}} & \textbf{\makecell[c]{Delete IR Node}} & \textbf{\makecell[c]{Semantics Preservation}} \\
      \Xhline{0.5pt}
      RAGS\cite{slutz1998massive} &  \ding{51} & \ding{51} & \ding{55} & \ding{55} & \ding{55}  \\
      SQLancer\cite{rigger2020testing} &  \ding{51} & \ding{51} & \ding{55} & \ding{55} & \ding{55}  \\
      MutaSQL\cite{chen2020testing} &  \ding{51} & \ding{51} & \ding{55} & \ding{55} & \ding{55}  \\
      GARan\cite{bati2007genetic} &  \ding{51} & \ding{51} & \ding{55} & \ding{55} & \ding{55}  \\
      SQLRight\cite{liang2022detecting} &  \ding{55} & \ding{55} & \ding{55} & \ding{51} & \ding{51}  \\
      APOLLO\cite{jung2019apollo} &  \ding{51} & \ding{51} & \ding{51} & \ding{55} & \ding{51}  \\
      DynSQL\cite{jiang2023dynsql} &  \ding{51} & \ding{51} & \ding{51} & \ding{55} & \ding{51}  \\
      
      \Xhline{1pt}
    \end{tabular}
\end{table*}

Reducing expressions~\cite{slutz1998massive,bati2007genetic,rigger2020testing,chen2020testing} and deleting clauses refer to the process of reducing queries by simplifying \textsf{WHERE} clauses, arithmetic expressions, or logical expressions. These methods are relatively easy to implement, but cannot guarantee semantic correctness after simplification, resulting in a decrease in reduction efficiency. The removal of subquery operations~\cite{jung2019apollo, jiang2023dynsql} extends the scope of simplification to subqueries, leading to a substantial improvement in the efficiency of simplification. Delete IR node operations are used in mutation-based fuzzers based on  IR~\cite{liang2022detecting} mutation. They treat query simplification as a mutation operation on SQL queries, ensuring semantic correctness by deleting specific IR nodes. When considering the dependency between deleted expressions and remaining expressions, some fuzzers~\cite{liang2022detecting,jung2019apollo, jiang2023dynsql} have achieved semantic preservation.


\section{DBMS Component-based Taxonomy}\label{sec:ver}
\begin{figure}[htp]
    \centering
    \includegraphics[width=0.92\textwidth]{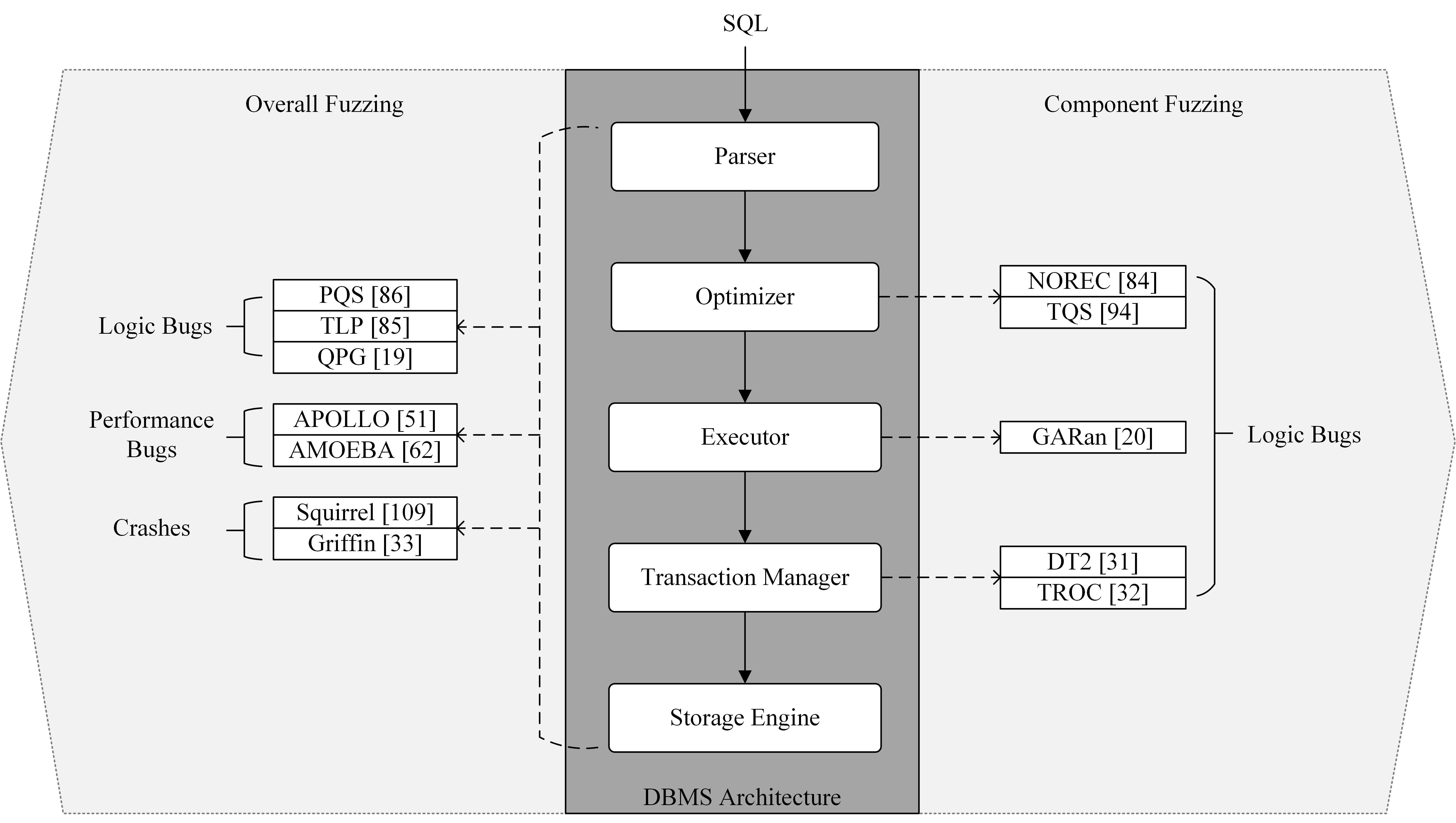}
    \caption{Classification by DBMS components.}
    \label{fig:Classification by database components}
\end{figure}

In this section, we review existing representative fuzzers from the perspective of the DBMS component (vertical). As illustrated in Figure~\ref{fig:Classification by database components}, existing DBMS fuzzers can be divided into overall fuzzing and component fuzzing. Overall fuzzing focuses on the entire DBMS, capable of detecting logic bugs, performance issues, or system crashes. Component fuzzing focuses on testing logic bugs in specific components of the DBMS, such as the optimizer, executor, and transaction manager.

\subsection{Overall Fuzzing}
Overall Fuzzing is necessary for the initial development of DBMS, as it can comprehensively detect logic bugs, performance bugs, and crashes in the entire system. In terms of logic bugs, overall testing not only detects bugs within individual components of the DBMS but also identifies bugs that occur during the interaction between different components. For performance bugs, conducting component fuzzing alone on a single component is often insufficient for detection, as performance disparities typically result from the cumulative impact of multiple components. For crashes, fuzzers basically run against the entire DBMS. While it is possible to perform performance bug and crash detection targeting specific components, such testing is not only time-consuming but also provides only localized results, failing to capture the overall state of the entire DBMS. Therefore, there is little research effort in this area.

\subsubsection{Overall fuzzing on logic bugs}
\begin{figure}[htp]
    \centering
    \includegraphics[width=\textwidth]{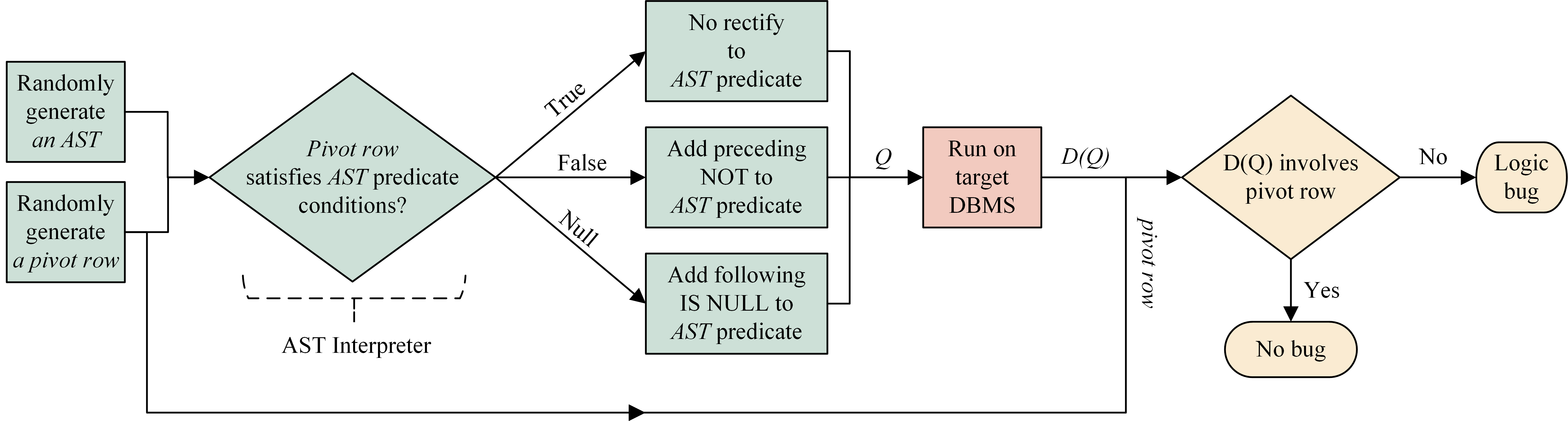}
    \caption{Pipeline of PQS.}
    \label{fig:Pipeline of PQS}
\end{figure}

PQS~\cite{rigger2020testing} begins by randomly generating a pivot row as the ground truth and then constructs an SQL query that includes this pivot row in its result. By checking whether the result of the executed query in the test database contains the pivot row, logic bugs can be detected. Figure~\ref{fig:Pipeline of PQS} illustrates the entire PQS pipeline. The main challenge of PQS is how to generate an SQL query whose execution result contains a known pivot row. The solution to PQS is to first create predicates randomly and then use the AST interpreter to evaluate whether the pivot row satisfies the predicate conditions. The evaluation result can be True, False, or NULL. True indicates that the pivot row can be queried through the predicate, requiring no adjustment. False indicates that the pivot row cannot be queried, so adding \textsf{NOT} before the predicate rectifies it. NULL indicates that the result of the predicate query is unknown, so adding \textsf{IS NULL} after the predicate can help. By employing the above method, PQS ensures that any random predicate can produce a satisfactory SQL query after being queried.

\begin{figure}[htp]
    \centering
    \includegraphics[width=0.95\textwidth]{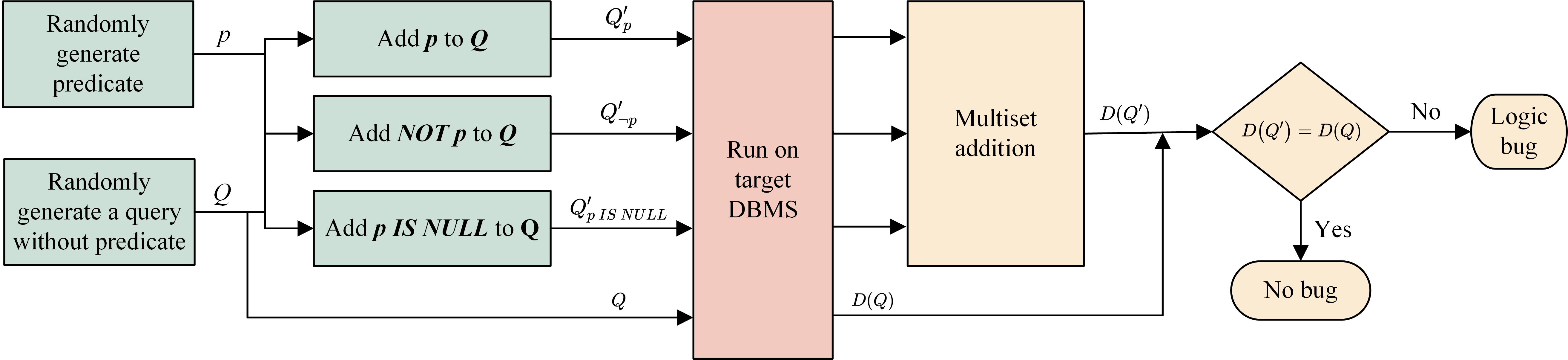}
    \caption{Pipeline of TLP.}
    \label{fig:Pipeline of TLP}
\end{figure}

The core idea of TLP~\cite{rigger2020finding} is that the result of the predicate evaluation always falls within the values of True, False and NULL. Therefore, an original query $Q$ can be decomposed into three partitioned queries: $Q_p^{'}$, $Q_{\neg p}^{'}$, $Q_{p\;IS\,NULL}^{'}$. The expected result of the original query should be equal to the union of the results of the three partitioned queries. Figure~\ref{fig:Pipeline of TLP} illustrates the main process of TLP. Using a generator similar to PQS, the original query is split into three equivalent partitioned queries, and metamorphic oracles are employed for result verification to detect logic bugs. Neither PQS nor TLP adopts feedback but performs a random search throughout the state space.

\begin{figure}[htp]
    \centering
    \includegraphics[width=0.83\textwidth]{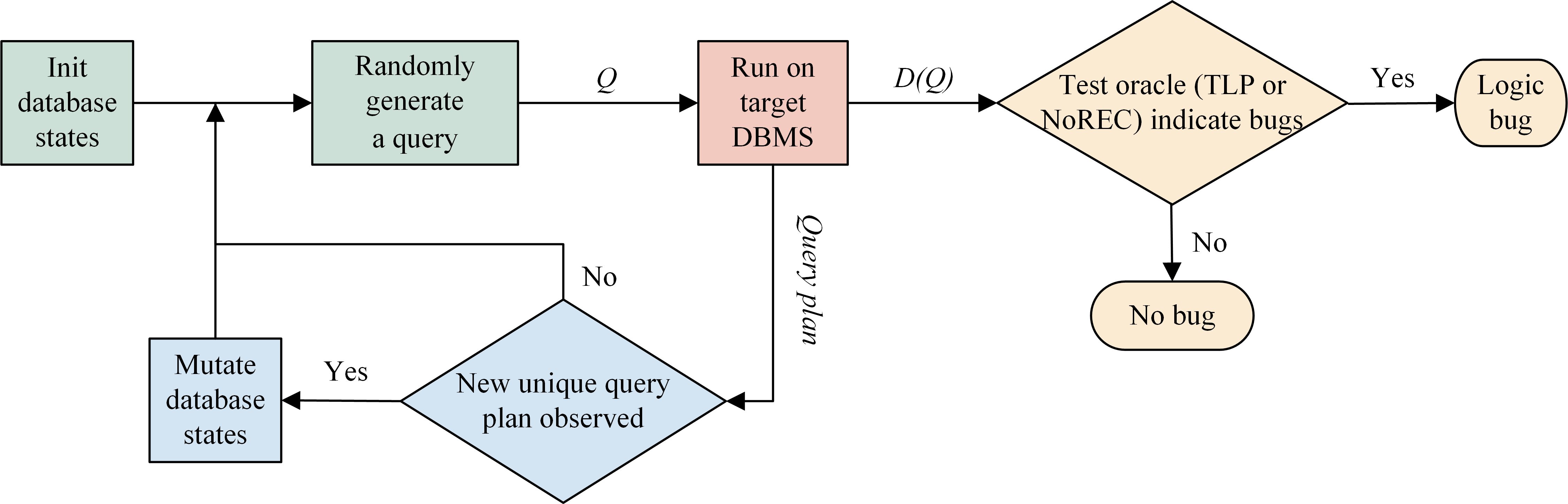}
    \caption{Pipeline of QPG.}
    \label{fig:Pipeline of QPG}
\end{figure}

QPG~\cite{ba2023testing} mutates the state of the database to generate more unique query plans, as shown in Figure~\ref{fig:Pipeline of QPG}. QPG implements a generation-based generator based on PQS, TLP, and NoREC. QPG also utilizes TLP and NoREC to validate query execution results. DDL and DML statements are used to alter the state of the database and obtain different query plans for the same statement. Because different mutation operations contribute differently to generating new query plans, a decision among all mutation operators is necessary. The decision-making process for mutation operators is modeled as a Multi-Armed Bandit~\cite{berry1985bandit} problem. As shown in Equation~\ref{eq:QPG}, $\mu_{i}(t)$ represents the benefit of the $i$-th mutation operator at time $t$, there is a probability of $p$ for random selection, and a probability of $1-p$ to opt for the mutation operator with the highest benefit. Since different query plans represent different execution paths, QPG can enhance the code coverage of DBMS.

\begin{equation}
operator(t) = \left\{
\begin{aligned}
arg max_{i=1...k}(\mu_{i}(t))  \qquad &(1-p)\\
random(k)        \qquad      \qquad        \qquad & (p)
\end{aligned}
\right.
\label{eq:QPG}
\end{equation}

\subsubsection{Overall fuzzing on performance bugs}
APOLLO\cite{jung2019apollo} uses differential testing to detect performance regression bugs. It employs a generation-based generator with a dynamic probability table to generate many random statements. These statements are executed on different versions of the same database program, and performance bugs are detected using a differential oracle. APOLLO incorporates execution feedback to update the probability table, thereby improving the efficiency of the testing process. 
Additionally, Apollo uses query reduction that preserves semantic correctness to quickly obtain minimal test cases that developers can easily debug and fix.

AMOEBA~\cite{liu2022automatic} is based on metamorphic testing to detect performance bugs by comparing the execution time of equivalent queries. It utilizes a generation-based generator to generate a base query, on which it applies structure mutation and expression mutation operations to obtain mutant queries that are equivalent to the base query. Thanks to the feedback from the mutator and validator components, AMOEBA can effectively explore the query space and generate queries that can undergo equivalent mutations and are likely to trigger performance bugs.

\subsubsection{Overall fuzzing on crashes}
Squirrel~\cite{zhong2020squirrel} utilizes a mutation-based generator to mutate seed queries and detect crashes. It begins by randomly selecting a seed query from the seed queue and parsing it into an Abstract Syntax Tree (AST). The AST is then transformed into an IR form. SQL structure mutations are applied to the IR to generate new queries. By analyzing the logical dependencies between the arguments, the semantic validity of the statement after mutation is ensured. Squirrel also utilizes coverage feedback to guide the generator towards exploring the query space with maximum code coverage. 

\begin{figure}[htp]
    \centering
    \includegraphics[width=\textwidth]{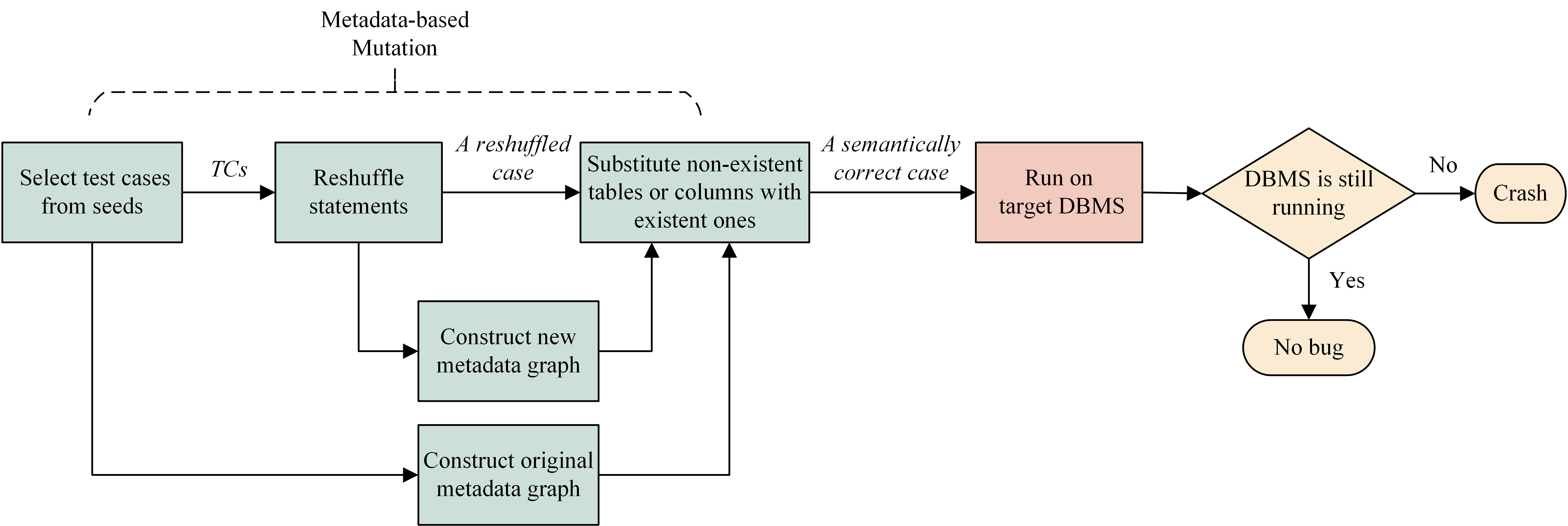}
    \caption{Pipeline of Griffin.}
    \label{fig:Pipeline of Griffin}
\end{figure}

Griffin~\cite{fu2022griffin} utilizes SQL sequence mutation to generate a large number of test cases to detect crashes. As shown in Figure~\ref{fig:Pipeline of Griffin}, Griffin first selects some test cases from the seed input and constructs a metadata graph for them, which describes the relationships between the statements and the tables or columns they create or access. A reshuffled case is then generated by deleting some statements and reordering the SQL sequence, but this may introduce semantic errors. By substituting non-existent tables or columns with existing ones, a semantically correct case can be obtained. Since mutations are performed at the statement sequence level, Griffin does not require dedicated parsers implemented for each DBMS, thus reducing adaptation costs.

Overall fuzzing tests the DBMS as a whole. The advantage is that the test coverage is wide and can detect potential bugs that may be difficult to notice. However, it has the disadvantages of low bug detection efficiency and difficulty in pinpointing the root cause of bugs. In some scenarios, it may be sufficient to just test a specific module to check whether its modifications introduce new bugs, while conducting overall testing may appear redundant and unnecessary.

\subsection{Transaction Testing}
Transactions are used to maintain the consistency and integrity of the data in the DBMS. Logic bugs in transaction implementation can lead to serious consequences, such as incorrect query results and unexpected DBMS blockage. Some testing tools~\cite{dou2023detecting, cui2022differentially} have already confirmed the presence of transaction isolation vulnerabilities in widely used DBMSs, where they fail to achieve transaction isolation levels as claimed. Therefore, conducting fuzzy tests for transactions is crucial.

\begin{figure}[htp]
    \centering
    \includegraphics[width=0.85\textwidth]{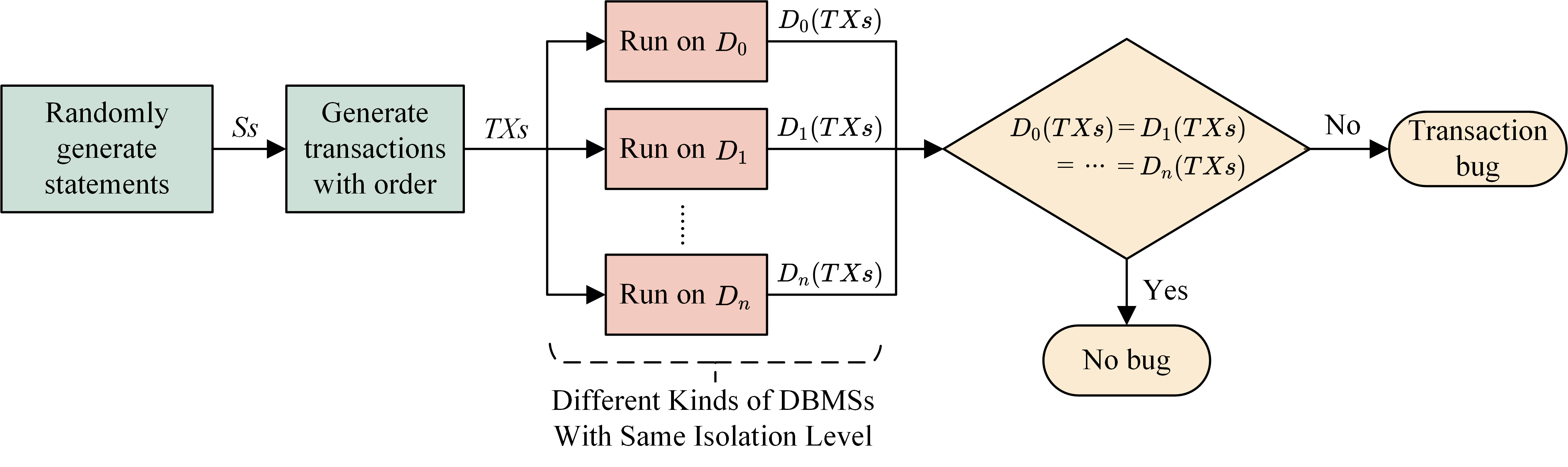}
    \caption{Pipeline of DT2.}
    \label{fig:Pipeline of DT2}
\end{figure}

As illustrated in Figure~\ref{fig:Pipeline of DT2}, DT2~\cite{cui2022differentially} randomly generates statements and combines a series of statements with \textsf{BEGIN} and \textsf{COMMIT} or \textsf{ROLLBACK} to form a transaction. DT2 then randomizes the commit order of statements in all transactions. Finally, transaction bugs are detected by comparing the execution results of transactions on different DBMSs.

\begin{figure}[htp]
    \centering
    \includegraphics[width=\textwidth]{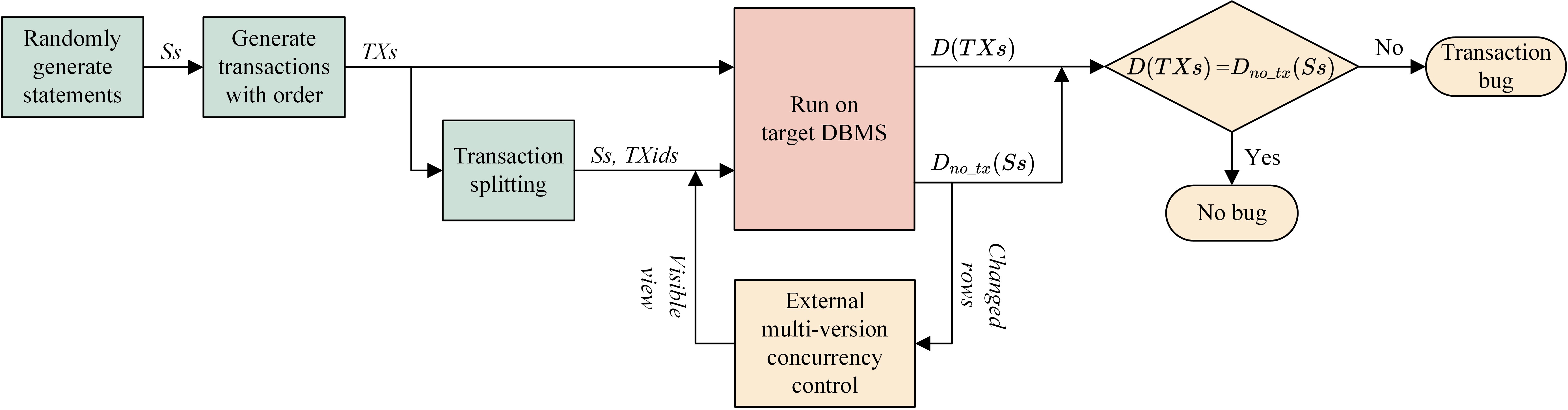}
    \caption{Pipeline of Troc.}
    \label{fig:Pipeline of Troc}
\end{figure}

Troc~\cite{dou2023detecting} utilizes metamorphic testing to detect logic bugs in transactions. The Troc pipeline is illustrated in Figure~\ref{fig:Pipeline of Troc}. Firstly, Troc generates random transactions and random submission orders. It is then tested in two equivalent scenarios: transaction and non-transaction modes. Transaction mode refers to submitting multiple transactions directly to the database. In non-transaction mode, transactions are split into separate statements with transaction ids. Troc  constructs an external multi-version concurrency control system based on the changed rows in each execution. This ensures that every statement executed in non-transaction mode can obtain the same visible view as in transaction mode. Transaction bugs can be detected by comparing the execution results in transaction and non-transaction modes.

\subsection{Optimizer Testing}\label{sec:opt_test}
The optimizer is one of the most complex components of a DBMS. It analyzes different execution plans of a query and selects the best query plan by considering indexes, association conditions, data statistics, and other factors to minimize query time and resource consumption. However, implementation errors in the optimizer can lead to serious logic bugs.


\begin{figure}[htp]
    \centering
    \includegraphics[width=0.9\textwidth]{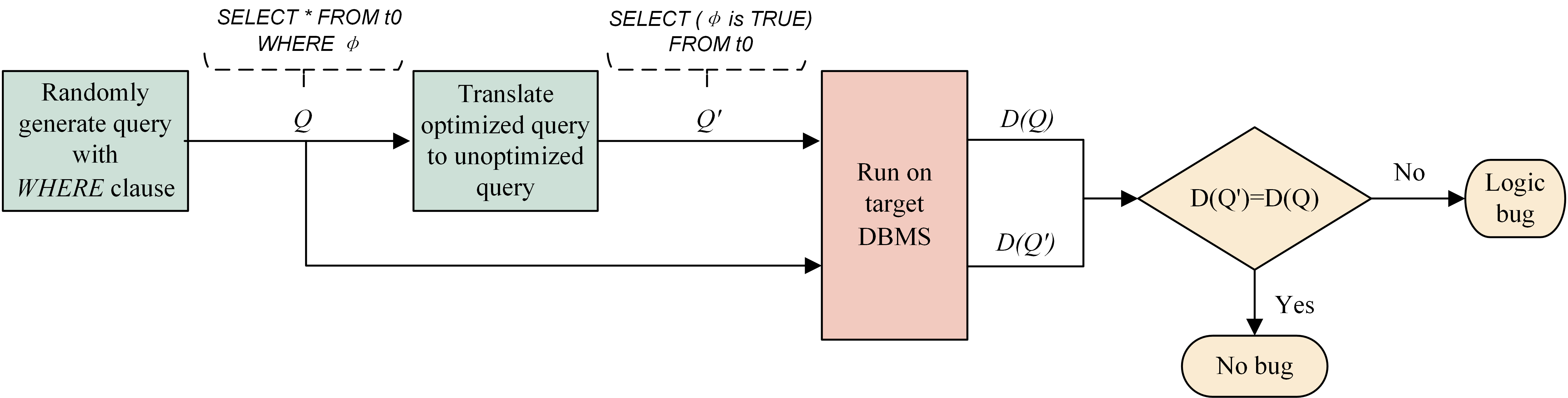}
    \caption{Pipeline of NoREC.}
    \label{fig:Pipeline of NoREC}
\end{figure}

The NoREC pipeline~\cite{Rigger_2020} is illustrated in Figure~\ref{fig:Pipeline of NoREC}. NoREC uses an AST model-based generator to generate queries with \textsf{WHERE} clauses and obtains the unoptimized query by moving the conditions from the \textsf{WHERE} clause to the \textsf{SELECT} clause. The presence of optimizer logic bugs is determined by comparing the number of rows returned by the original query and the number of `True' returned by the unoptimized query.

\begin{figure}[htp]
    \centering
    \includegraphics[width=\textwidth]{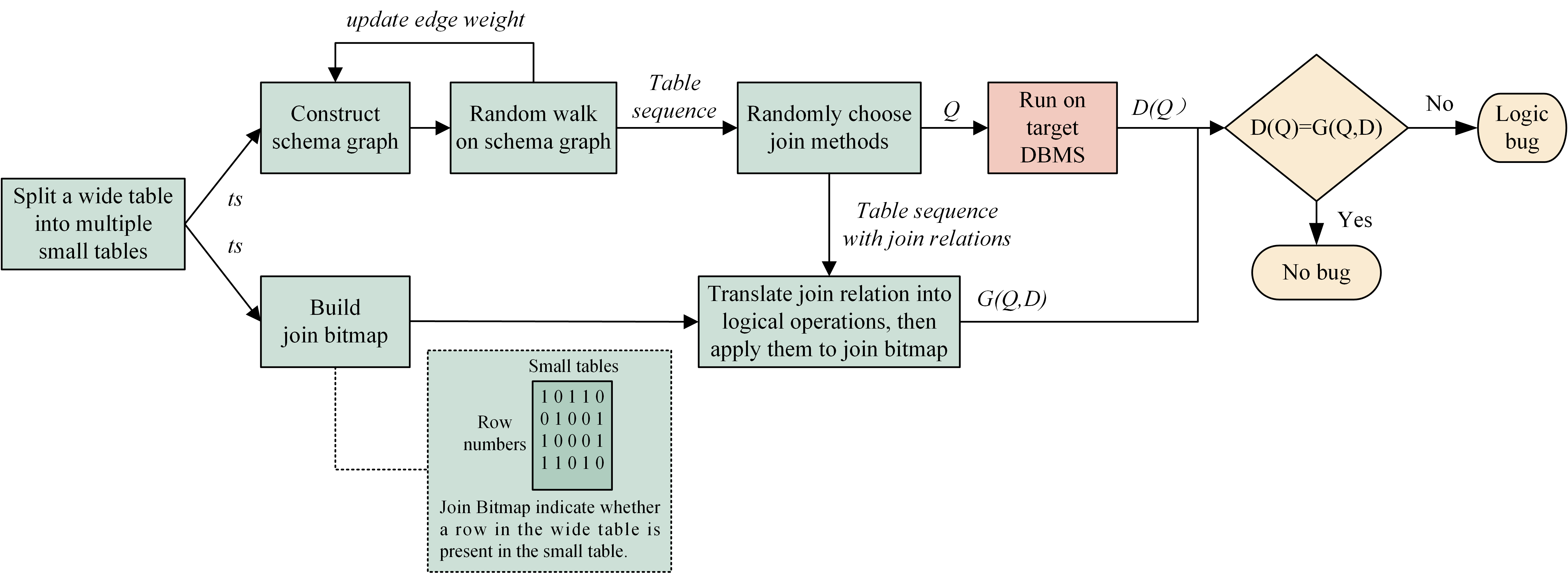}
    \caption{Pipeline of TQS.}
    \label{fig:Pipeline of TQS}
\end{figure}

The pipeline of TQS~\cite{tang2023detecting} is illustrated in Figure~\ref{fig:Pipeline of TQS}. Firstly, TQS splits a wide table into multiple small ones. On the one hand, it constructs a schema graph based on the foreign key relationship. By performing random walks on the schema graph, table sequences can be obtained. Combining these table sequences with randomly selected join methods, a multi-table SQL query `Q' can be generated that does not include any filter predicate. On the other hand, it creates a join bitmap that indicates whether a row in the wide table exists in the small tables. The join relations of an SQL query can be translated into logical operations on the join bitmap. For example, `t1 inner join t2' can be translated into the logical 'AND' operation between two columns of the join bitmap. By applying logical operations to the join bitmap, the ground truth for 'Q' can be derived. Comparing the actual execution results of 'Q' on the target DBMS with its ground truth enables the detection of any potential logic bugs. It is important to highlight that after each random walk, the edge weights of the schema graph are updated. This reduces the likelihood of generating duplicate queries and enhances the efficiency of detecting logic bugs.

\subsection{Executor Testing}
The executor is responsible for translating a query plan into a combination of physical operators and executing them to obtain the query result. These physical operators are specific instructions for performing query operations, such as Index Scan, Hash Join, Sort, etc. During the development process of DBMS, any modification of the operator must ensure its correctness in various combinations. Therefore, fuzzing needs to be performed under different query plans that contain various combinations of physical operators.


GARan~\cite{bati2007genetic} utilizes a mutation-based generator with genetic algorithms to focus on checking executor logic bugs. By setting a fitness function that evaluates the quality of mutated SQL statements, GARan can generate statements that produce specific physical operators, such as Hash Join. Then, based on the idea of differential testing, it compares the execution results of the same statement on different DBMSs.

\section{Benchmarks and Comparison}\label{sec:exp}
As can be seen from the previous sections, there are quite a number of DBMS fuzzers and their structures are quite complex. In order to objectively evaluate the performance of each fuzzer, we present a comprehensive open-source toolkit, OpenDBFuzz~\footnote{https://github.com/Reverie4u/OpenDBFuzz.}. Based on this toolkit, we conduct fair experimental comparisons using the same configurations and focus on answering the following questions:
\begin{enumerate}
\item[\textbf{Q1:}] (Between fuzzers) For various database bugs, which fuzzer performs best under the same test environment? 
\item[\textbf{Q2:}] (Fuzzers themselves) Does each fuzzer perform as claimed, exhibiting superior performance?
\item[\textbf{Q3:}] (Inside fuzzers) Which module approach performs best among the various steps of fuzzing?
\end{enumerate}

Therefore, in this section, we first introduce the database instances, all the open-source fuzzers, and the evaluation metrics used in the experiments. Next, we thoroughly tested and evaluated existing fuzzers from the perspective of logic bugs, crashes, and performance bugs.

\subsection{Database Instances and Test Cases}
Some generation-based fuzzers start from scratch to generate database instances, while others, such as SQLsmith~\cite{git2023sqlsmith}, require existing database instances. For the latter experiments, we use TPC-H as the database instance. The mutation-based fuzzers generate diverse test cases by mutating seed statements. We use official unit test cases provided by DBMS vendors as our seed corpus, including SQLite TCL test scripts~\cite{sqlite2023testing}, PostgreSQL official test infrastructure~\cite{postgresql2023test}, and MySQL unit test samples~\cite{mysql2023unittest}.

\subsection{Open-source Fuzzers}
Source code is crucial for researchers to reproduce and compare the fuzzing methods. We have summarized the code of all open-source fuzzers and the DBMSs they support in Table~\ref{tab:open-source tools}. The following comparative experiments will be conducted on these fuzzers.

\begin{table}[ht]
\caption{Source Code of Related Fuzzers}
\label{tab:open-source tools}
\begin{tabular}{|c|c|l|}
\hline
\textbf{Fuzzer} & \textbf{\makecell[c]{Supported DBMS}} & \textbf{\makecell[c]{Link}}\\ \hline
\makecell[c]{SQLancer\cite{rigger2020testing}\\(PQS\cite{rigger2020testing}, NoREC\cite{Rigger_2020},\\TLP\cite{rigger2020finding}, QPG\cite{ba2023testing})} & \makecell[c]{SQLite, MySQL, TiDB\\ MariaDB, CockroachDB, OceanBase} & https://github.com/sqlancer/sqlancer\\ \hline

DQE\cite{song2023testing} & \makecell[c]{SQLite, MySQL, MariaDB\\TiDB, CockroachDB} & https://github.com/tcse-iscas/dqetool\\ \hline

SQLRight\cite{liang2022detecting} & \makecell[c]{SQLite, PostgreSQL, MySQL} & https: //github.com/psu-security-universe/sqlright\\ \hline

SQLsmith\cite{git2023sqlsmith} & \makecell[c]{SQLite, MonetDB, PostgreSQL} & https://github.com/anse1/sqlsmith\\ \hline
Go-Randgen\cite{git2023gorandgen}  & \makecell[c]{MySQL, TiDB} & https://github.com/pingcap/go-randgen\\ \hline

Squirrel\cite{zhong2020squirrel} & \makecell[c]{SQLite, PostgreSQL, MySQL, MariaDB} &

https://github.com/s3team/Squirrel\\ \hline
DT2\cite{cui2022differentially} &  \makecell[c]{MySQL, MariaDB, TiDB} & https://github.com/tcse-iscas/Troc\\ \hline
Troc\cite{dou2023detecting}  & \makecell[c]{MySQL, MariaDB, TiDB} & https://github.com/tcse-iscas/Troc\\ \hline

APOLLO\cite{jung2019apollo} & \makecell[c]{SQLite, PostgreSQL} & https://github.com/sslab-gatech/apollo\\ \hline

AMOEBA\cite{liu2022automatic} & \makecell[c]{PostgreSQL, CockroachDB} & https://bit.ly/3I995jL\\  \hline
\end{tabular}
\end{table}

\subsection{Evaluation Metrics}
We adopt the standard metrics for evaluating fuzzing performance:
\begin{itemize}
    \item \textbf{Number of bugs}: The number of bugs can directly quantify the bug detection capabilities of different fuzzers. Some studies submit the detected bugs to the open-source database community to obtain the number of community confirmed bugs. However, to make comparative experiments more discriminative, we need to conduct subsequent experiments on older versions of the DBMSs, so this method cannot be used. To avoid overrating methods that output highly redundant bugs, we used expert filtering to remove the obviously repetitive bugs, thereby improving the rationality of this metric.
    \item \textbf{Validity}: Validity refers to the proportion of test statements that are correct in both syntax and semantics. Only valid statements can trigger the underlying logic of the DBMS, making it more likely to detect bugs.
    \item \textbf{Valid Cases Per Second}: Valid cases per second are the number of valid test cases executed by the fuzzer within a given time range. It measures the overall efficiency of generation, execution, and comparison in a fuzzer.
    \item \textbf{New Edges}: New edges refer to the number of code branches explored during fuzzing, indicating the breadth of the testing scope.
    \item \textbf{Coverage}: Coverage indicates the proportion of code branches that have been explored. It is calculated using $new\ edges / total\ edges$, where $total\ edges$ represents the number of all code branches in the DBMS.
\end{itemize}

\subsection{Logic Bugs Detection Comparison}
\begin{figure}[htp]
    \centering
    \subfloat[Number of bugs (SQLite)]{\includegraphics[width=0.30\textwidth]{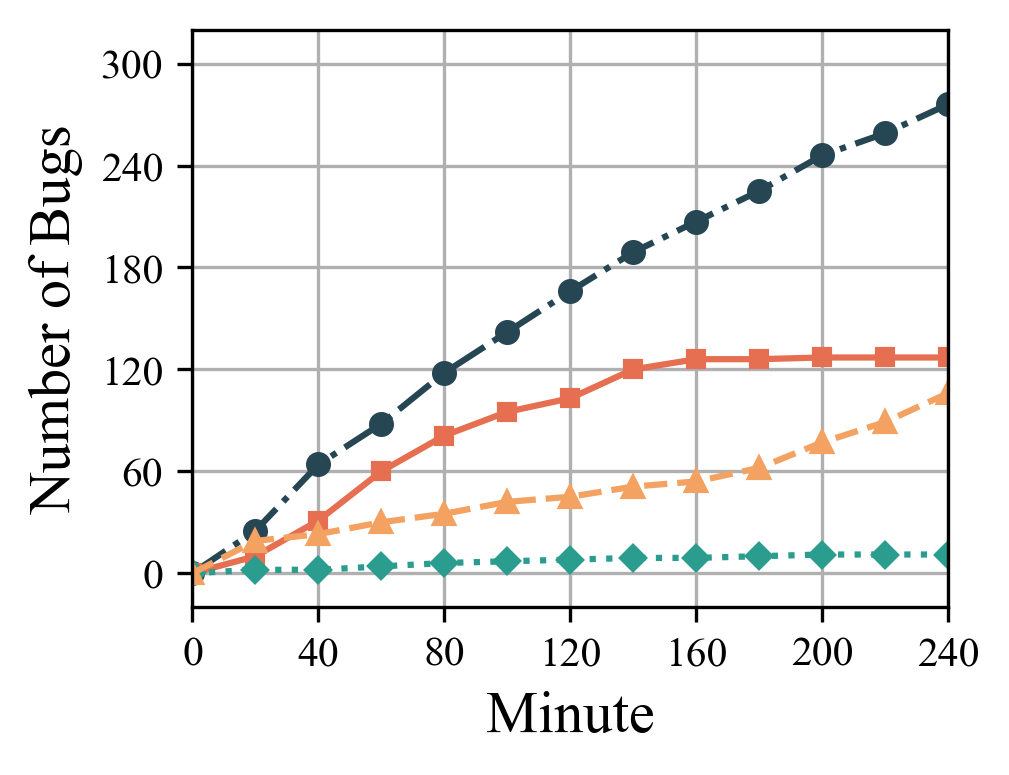}} \hfill
    \subfloat[Validity (SQLite)]{\includegraphics[width=0.30\textwidth]{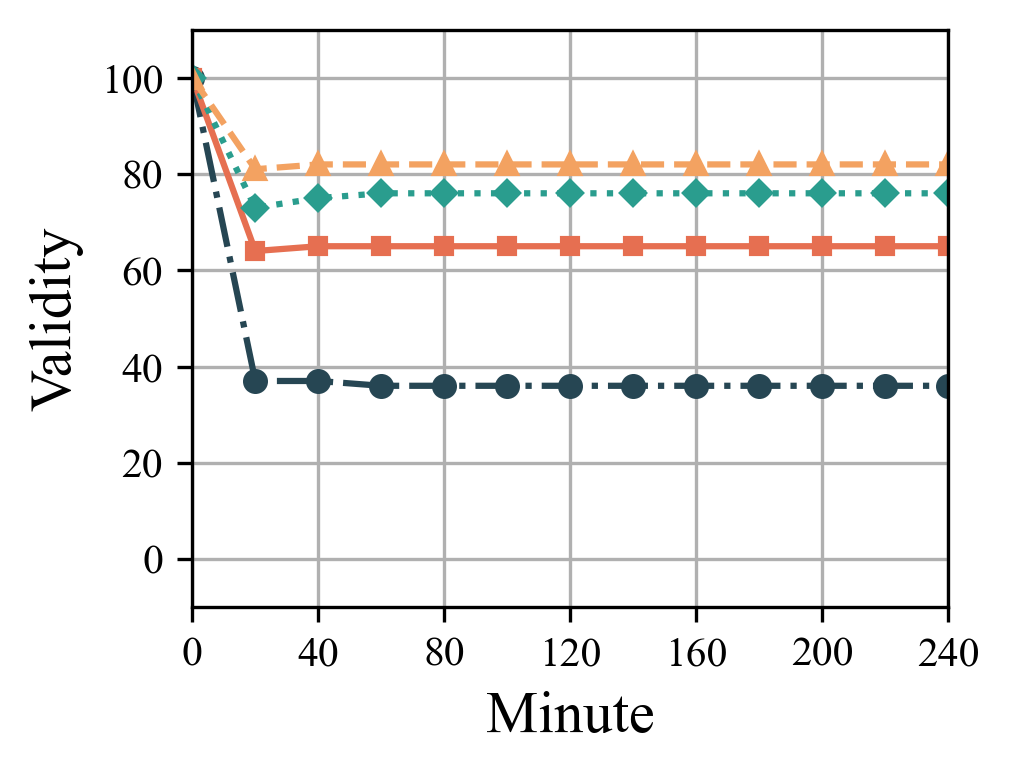}} \hfill
    \subfloat[Valid Cases/s (SQLite)]{\includegraphics[width=0.30\textwidth]{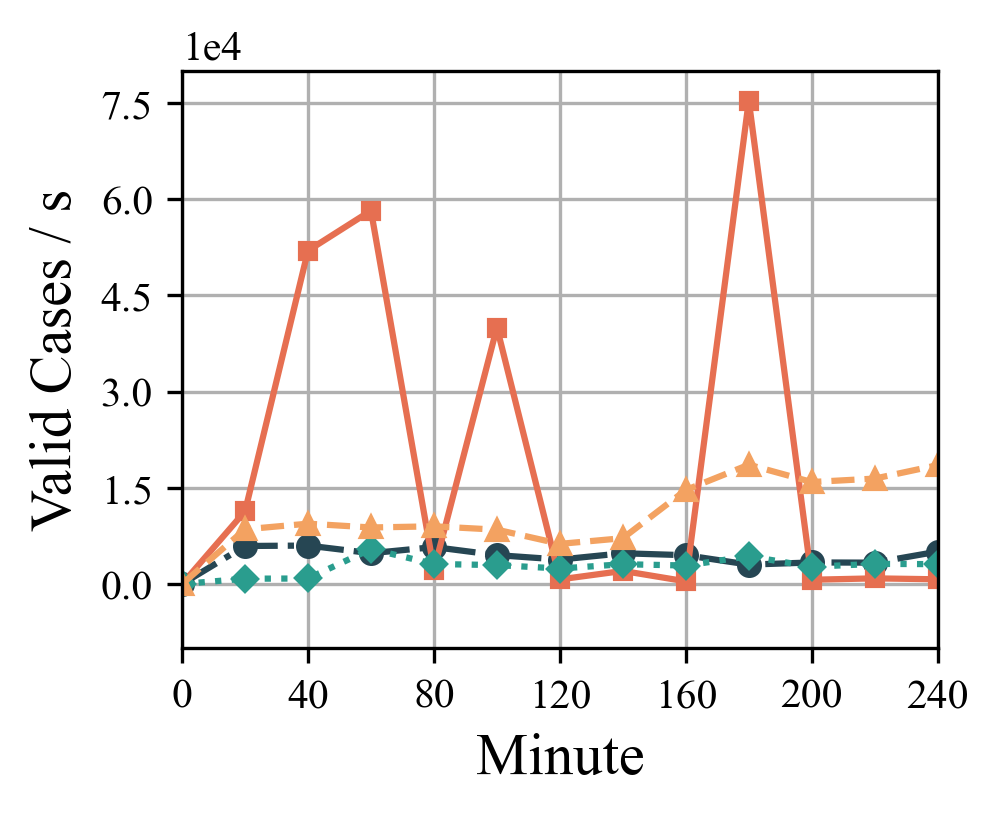}}
    
    \subfloat[Number of bugs (MySQL)]
    {\includegraphics[width=0.30\textwidth]{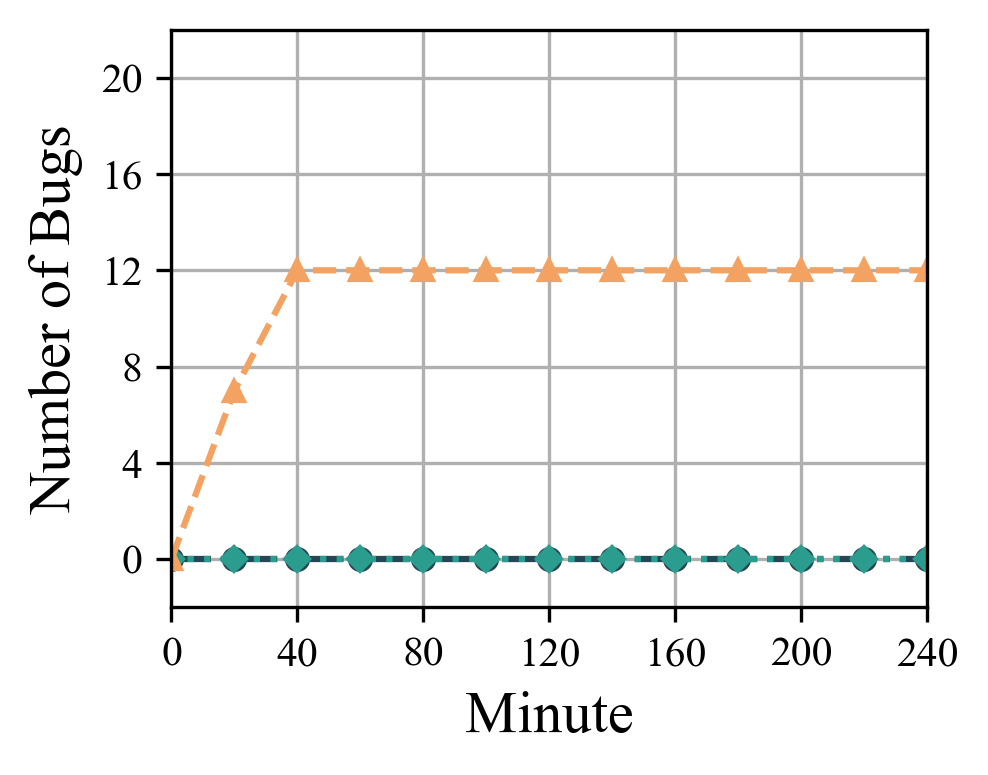}} \hfill
    \subfloat[Validity (MySQL)]{\includegraphics[width=0.30\textwidth]{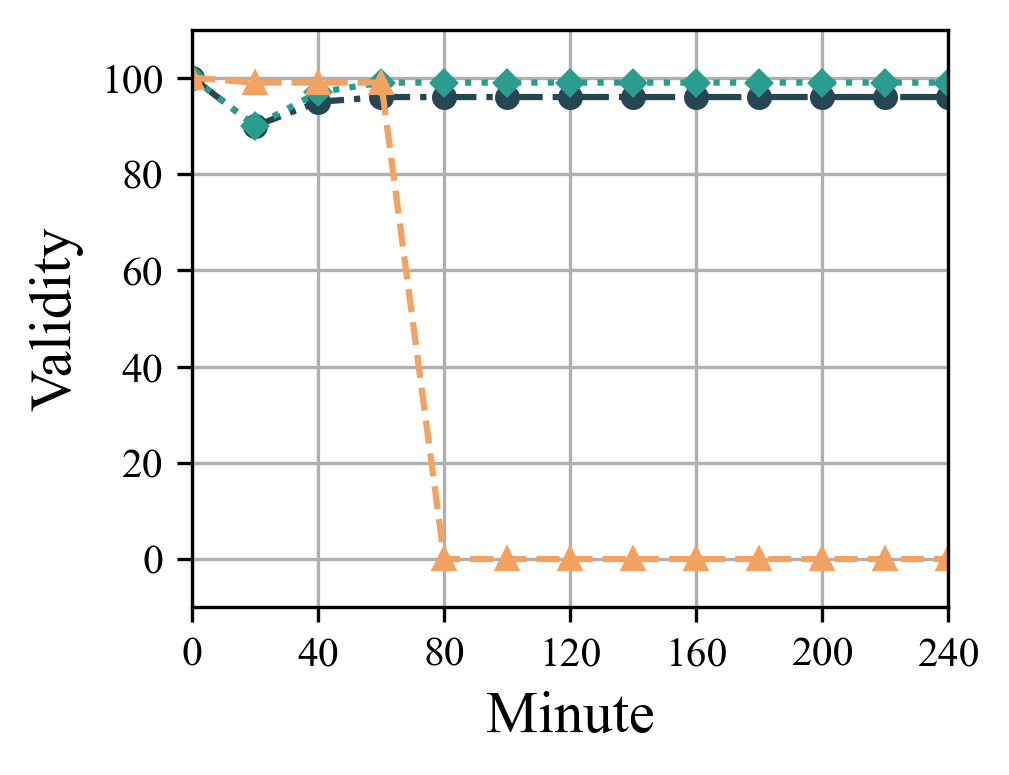}} \hfill
    \subfloat[Valid Cases/s (MySQL)]{\includegraphics[width=0.30\textwidth]{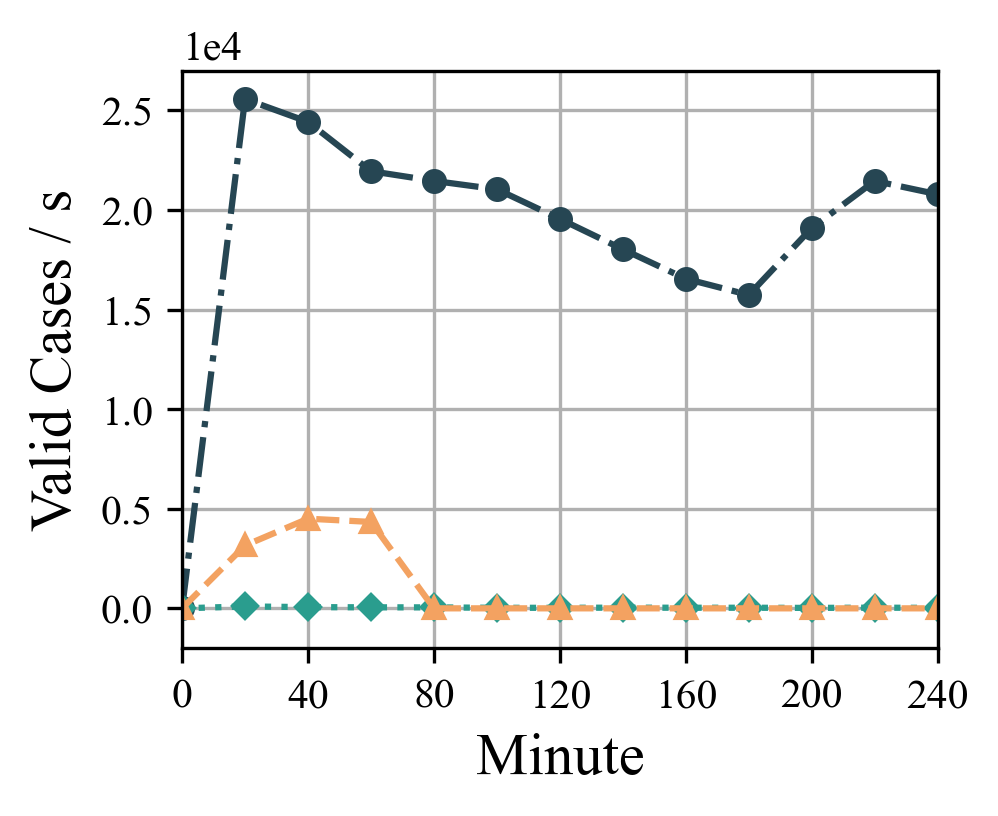}}
    
    \subfloat{\centering \includegraphics[width=0.9\textwidth]{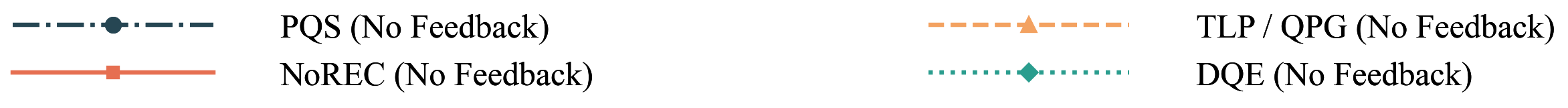}}
    \caption{Comparision of oracles for detecting logic bugs}
    \label{figure:Comparision of oracles for detecting logic bugs}
\end{figure}

Since the feedback module is relatively pluggable, we first tested the fuzzers without feedback to eliminate its impact on the evaluation of other modules and then tested the feedback module independently. For logic bug detection, we conducted comparative experiments on five fuzzers, including PQS, NoREC, TLP, QPG, and DQE. Among them, TLP and QPG only differ in feedback, and the other modules are consistent. Experiments were conducted on two popular DBMSs: MySQL 8.0.16 and SQLite 3.28.0. The number of bugs, the validity, and the valid cases per second are used as evaluation metrics. 


As shown in Figure~\ref{figure:Comparision of oracles for detecting logic bugs}, the experimental results indicate that these methods have detected more bugs in SQLite compared to MySQL, possibly due to the greater maturity of MySQL. Moreover, the validity of these methods on SQLite is relatively low, indicating that the SQL syntax of SQLite is not suitable for these fuzzers' generators. However, the valid cases per second on SQLite are actually higher, which could be attributed to its faster query performance on small data volumes. 

As can be seen from Figure~\ref{figure:Comparision of oracles for detecting logic bugs} (a), (b) and (c), in SQLite, PQS exhibits the lowest semantic validity and valid cases per second, but it demonstrated the highest bug detection efficiency, finding 276 bugs in only 240 minutes. The other fuzzers employ metamorphic testing strategies and have similar bug detection capabilities, stabilizing at around 100 bugs within 240 minutes. While NoREC stands out by using a strategy that transforms one statement into another, as opposed to the other methods that require transforming one statement into multiple ones~\cite{song2023testing}. Consequently, NoRec executes more valid cases per second, resulting in slightly faster bug detection efficiency as well. In summary, constraint solving is not limited by metamorphic applicability. Although the number of test cases per second is smaller, the overall bug detection efficiency is higher, and ultimately twice the number of bugs can be found compared to metamorphic testing methods.

Figure~\ref{figure:Comparision of oracles for detecting logic bugs} (d), (e) and (f) show the execution results on MySQL. Since NoREC does not support MySQL, we only compared the other tools. It can be observed that TLP and QPG are quite exceptional, as they quickly detected bugs in MySQL. They discovered 12 bugs in just 40 minutes, after which the system crashed (crash detected), leading to both validity and valid cases per second becoming zero. In contrast, DQE, which also falls under metamorphic testing, consistently failed to detect any logic bugs, indicating a more limited scope in its oracle. Further investigation revealed that in DQE, statements and their equivalent statements produced the same bug results, leading to missed detections. It is worth noting that PQS does not perform as well on MySQL as it does on SQLite. There are two main reasons for this. Firstly, MySQL has fewer inherent bugs, and our experiments did not run for months to detect these bugs. Secondly, the architecture of PQS speeds up the validation of each test case, but it is not a full comparison, only checking one of its query results. This is, in fact, sacrificing accuracy for speed. This approach may be more effective in environments with more bugs, but in the case of MySQL, i.e., with fewer bugs, the overall efficiency is lower.

\begin{figure}[htp]
    \centering
    \subfloat[Number of bugs (SQLite)]{\includegraphics[width=0.30\textwidth]{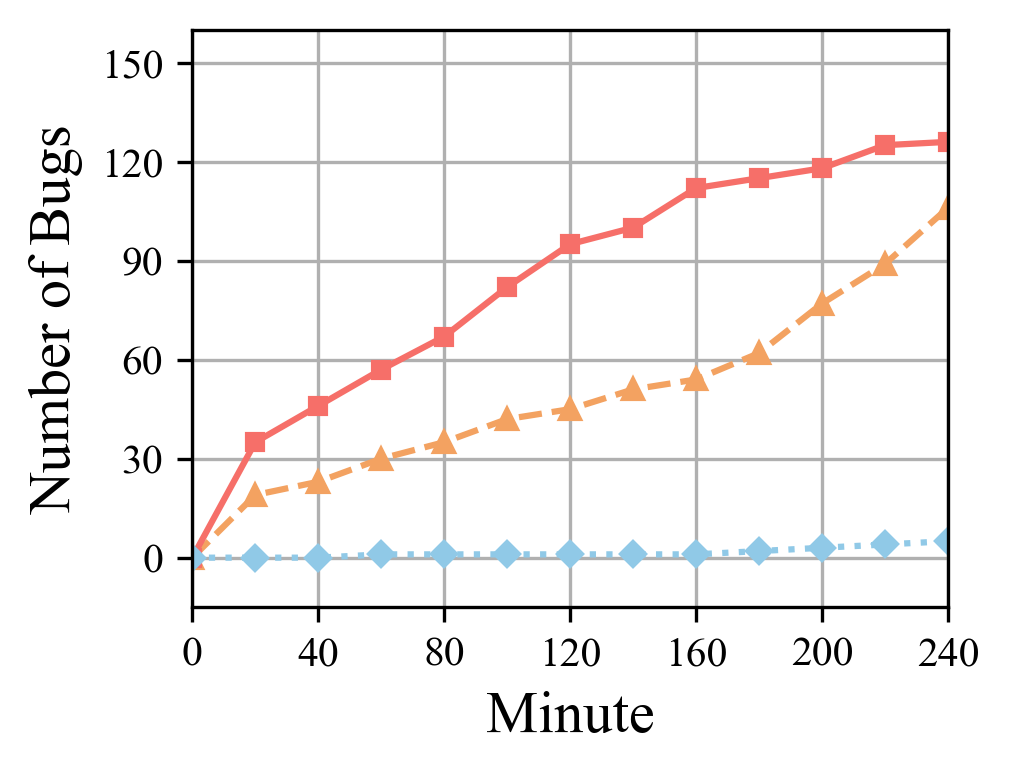}} \hfill
    \subfloat[Validity (SQLite)]{\includegraphics[width=0.30\textwidth]{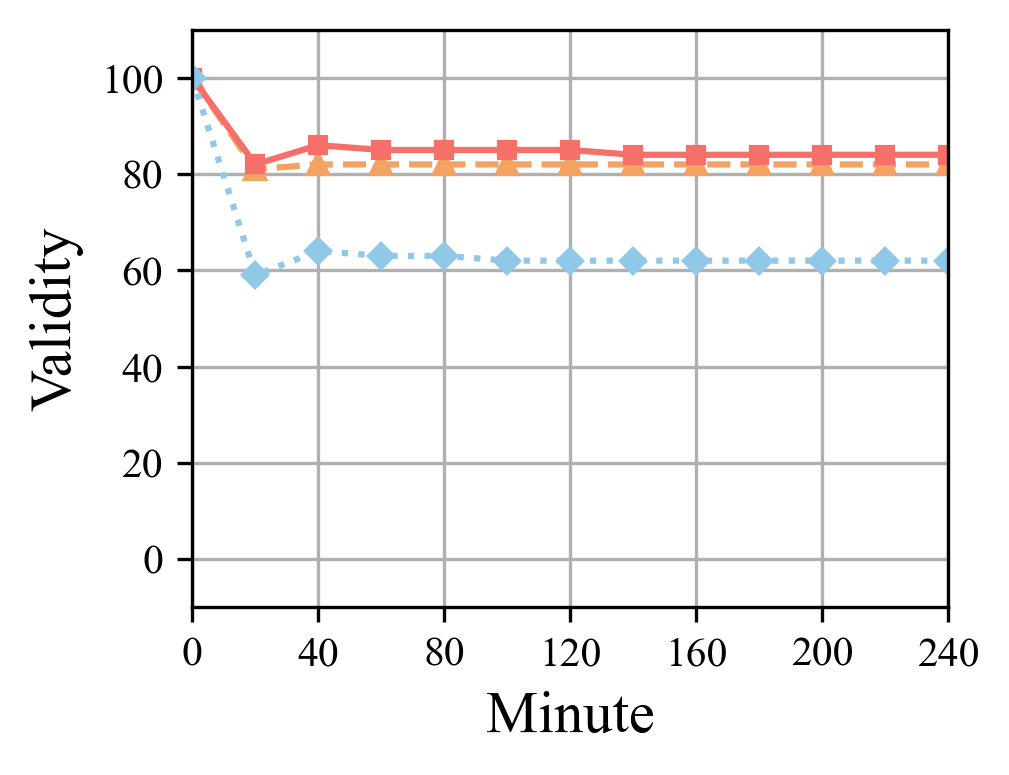}} \hfill
    \subfloat[Valid Cases/s (SQLite)]{\includegraphics[width=0.30\textwidth]{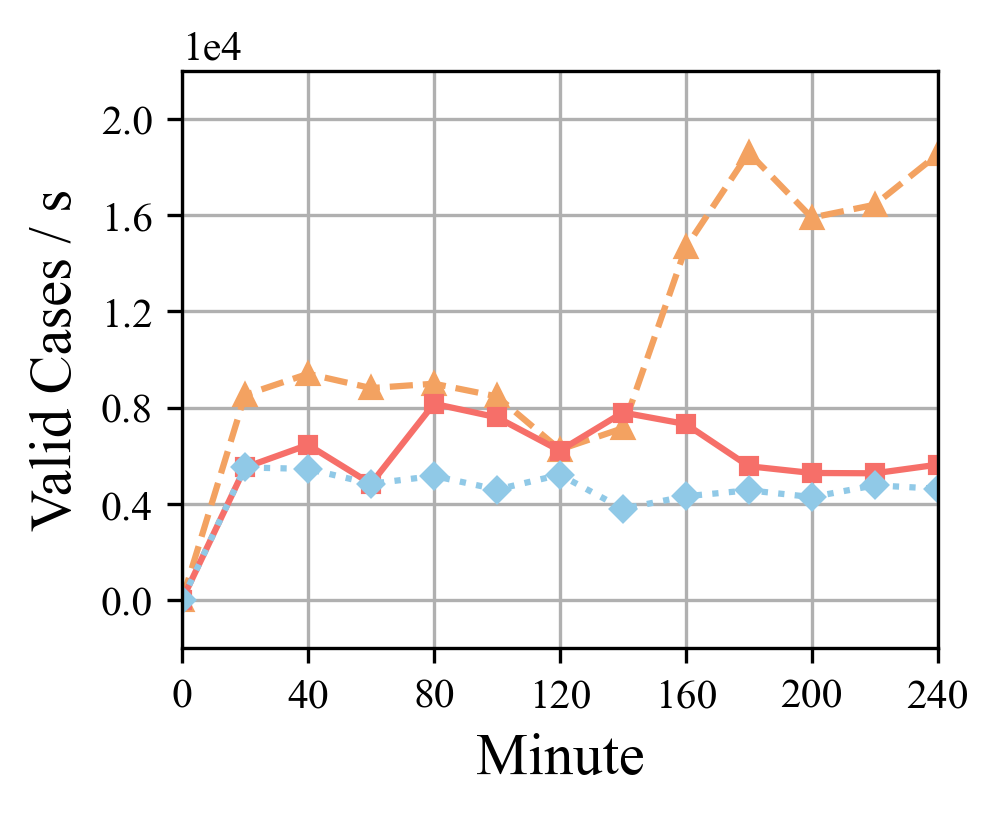}}

    \subfloat{\includegraphics[width=0.9\textwidth]{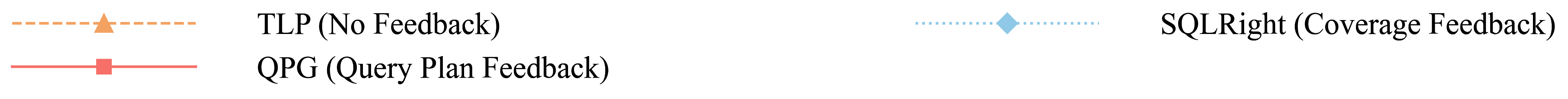}}
    \caption{Comparision of feedback ways for detecting logic bugs}
    \label{figure:Comparison of feedback methods for detecting logic bugs}
\end{figure}

Feedback currently applied to logic bug detection includes coverage feedback and query plan feedback. We conducted comparative experiments in SQLite using TLP, QPG, and SQLRight. They all use the same oracle, but TLP does not have feedback, QPG uses query plan feedback, and SQLRight utilizes coverage feedback. Among them, TLP and QPG employ the same generation-based generator. As can be seen in Figure~\ref{figure:Comparison of feedback methods for detecting logic bugs}, QPG detected bugs more efficiently than TLP within 240 minutes, and the number of bugs gradually equalized in the later period. This indicates that query plan feedback indeed contributes to improving the efficiency of logic bug detection. Furthermore, SQLRight only detected 5 bugs within 240 minutes, but this cannot be entirely attributed to coverage feedback. One important reason is that the mutation-based generator used by SQLRight is not effective, and its validity is obviously lower. The impact of coverage feedback on logic bug detection remains to be studied further. Also, as expected, the number of valid cases per second decreases slightly when feedback is added.

\subsection{Crash Detection Comparison}

\begin{figure}[htp]
    \centering
    \subfloat[Number of Crashes (SQLite)]{\includegraphics[width=0.30\textwidth]{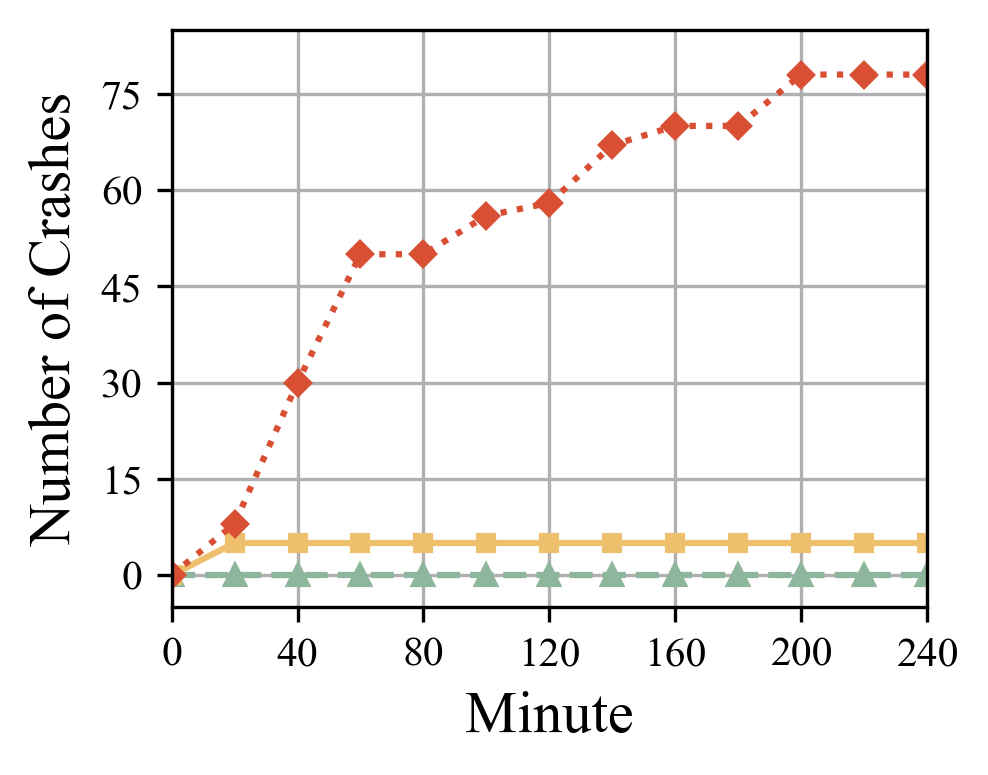}} \hfill
    \subfloat[New Edges (SQLite)]{\includegraphics[width=0.30\textwidth]{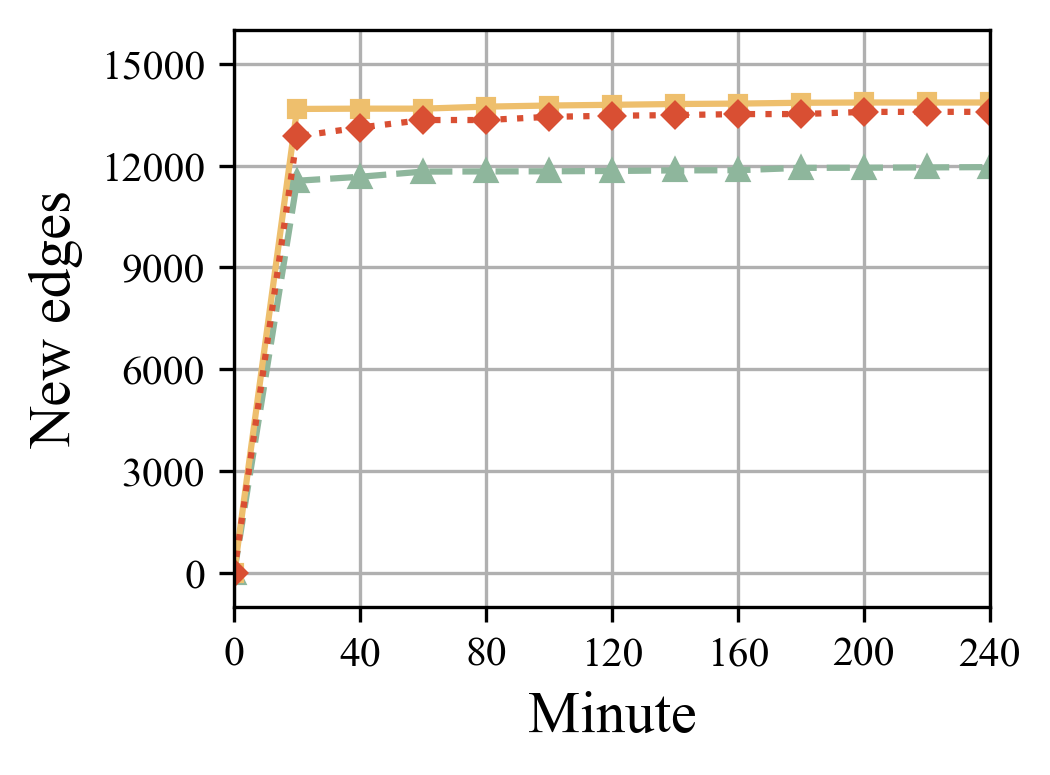}} \hfill
    \subfloat[Coverage (SQLite)]{\includegraphics[width=0.30\textwidth]{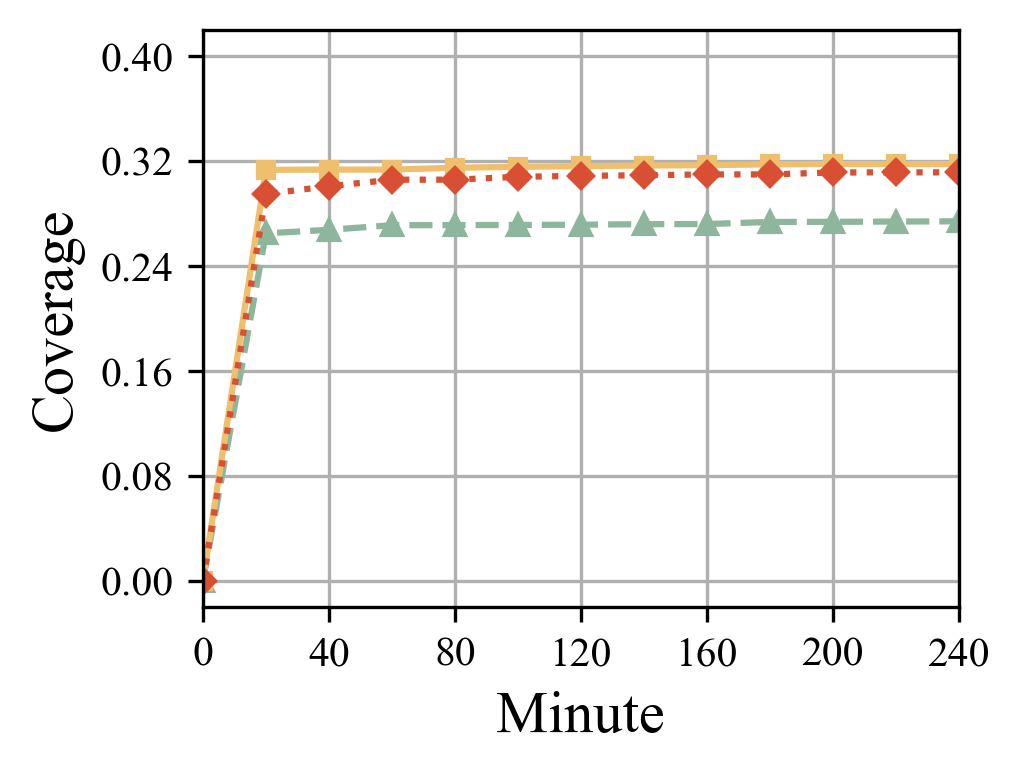}}
    
    \subfloat{\includegraphics[width=0.9\textwidth]
    {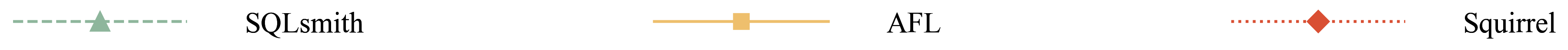}}
    \caption{Comparision of fuzzers for detecting crashes}
    \label{figure:Comparision of fuzzers for detecting crashes}
\end{figure}

We tested the performance of Squirrel, AFL, and SQLsmith in detecting crashes. Among them, Squirrel is a variant of AFL with improved semantics, and they both utilize coverage feedback. In contrast, sqlsmith does not have a feedback module. As shown in Figure~\ref{figure:Comparision of fuzzers for detecting crashes}, Squirrel can detect the highest number of crashes in the given time range. This indicates that Squirrel's IR-based mutation method is effective in improving crash detection. However, Squirrel does not show a significant improvement in coverage, probably because the semantically incorrect statements generated by AFL also cover a large portion of the DBMS code branches. In summary, increasing coverage does not necessarily improve crash detection efficiency, as only semantically correct statements contribute to crash detection.

\subsection{Performance Bugs Detection Evaluation}
\begin{figure}[htp]
    \centering
    \subfloat[Threshold Comparasion (Apollo)]{\includegraphics[width=0.30\textwidth]{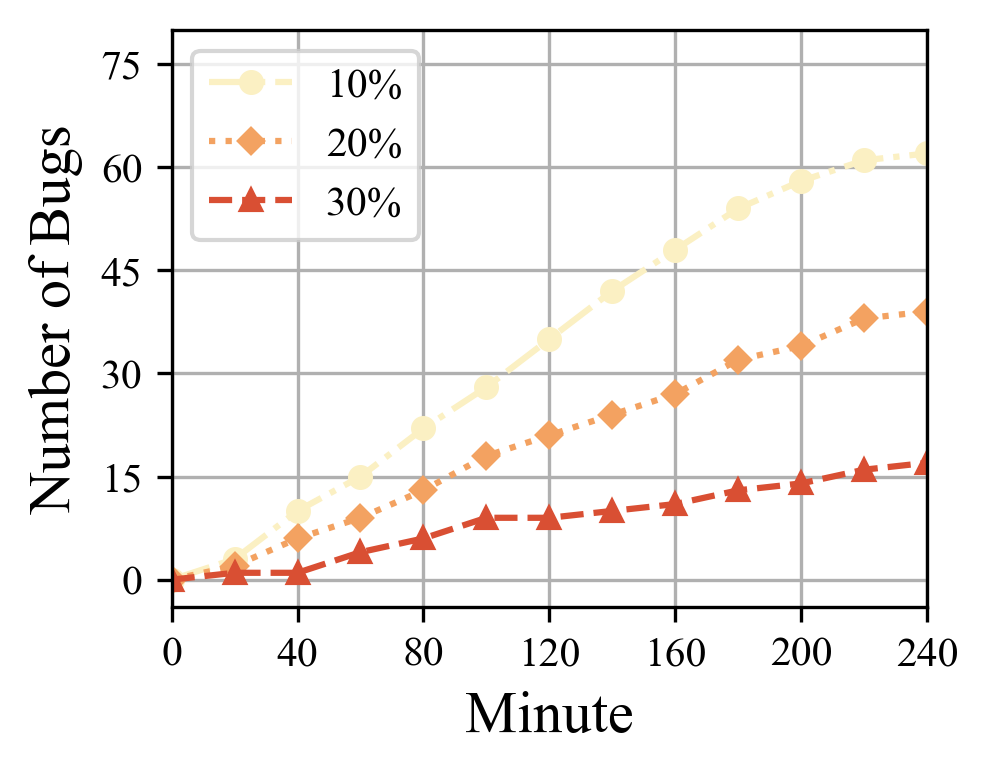}} \hfill
    \subfloat[Threshold Comparasion (AMOEBA)]{\includegraphics[width=0.30\textwidth]{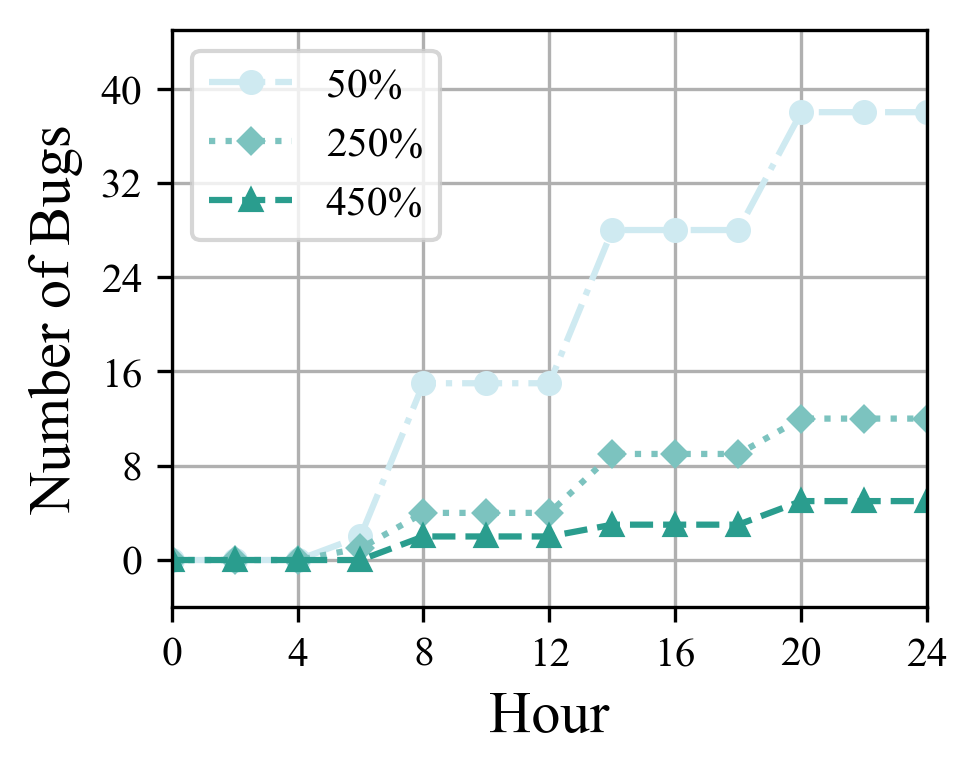}} \hfill
    \subfloat[Feedback Comparasion (AMOEBA)]{\includegraphics[width=0.30\textwidth]{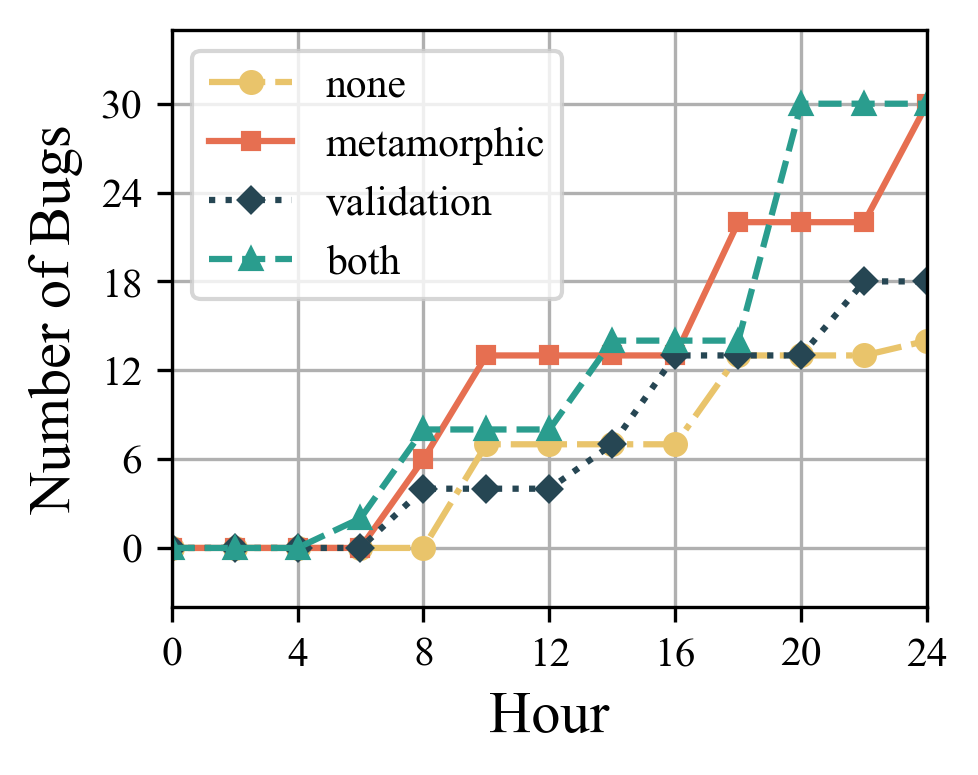}} 
    \caption{Evaluation of fuzzers for detecting performance bugs}
    \label{figure:Comparision of fuzzers for detecting performance bugs}
\end{figure}

Although both Apollo and AMOEBA can detect performance bugs, their focuses are different. Apollo focuses on detecting performance degradation caused by version updates, while AMOEBA is used to detect performance bugs caused by insufficient statement optimization. Therefore, we chose two PostgreSQL versions, 9.6 and 11.2, to evaluate Apollo and tested AMOEBA on the newer of the two versions, PostgreSQL 11.2.

In Figure~\ref{figure:Comparision of fuzzers for detecting performance bugs}, the experimental results show that both tools detected certain performance bugs in PostgreSQL. It is evident from Figure~\ref{figure:Comparision of fuzzers for detecting performance bugs} (a) that in Apollo’s experiment, the performance gap between the two versions did not exceed 30\% for most test cases, accounting for approximately 63.9\%. Figure~\ref{figure:Comparision of fuzzers for detecting performance bugs} (b) shows that in AMOEBA, the gap between most equivalent statements is less than 450\%, accounting for approximately 86.8\%. We analyzed the most severe performance bugs detected by both tools. The results reveal that the performance gap for the most severe bug detected by Apollo is 3.7 times, while for AMOEBA, it is 55.8 times. In general, performance issues caused by statement optimization are more serious than those caused by version updates. Figure~\ref{figure:Comparision of fuzzers for detecting performance bugs} (c) shows that metamorphic feedback enhances the efficiency of performance bug detection, as it guides the generator to produce more equivalent statements. Compared to no feedback, validation feedback slightly enhances detection capability, but its performance weakens within 10 to 14 minutes. Simultaneously, the combination of metamorphic feedback and validation feedback is less effective than using metamorphic feedback alone. Therefore, it is rational to suspect that validation feedback may not be as valuable as originally expected for the detection of performance bugs. This may be because, although validation feedback can help the generator learn the statement characteristics that trigger bugs, it also tends to generate repeated cases.

\section{Challenges and Future Directions}\label{sec:cha}
\subsection{Improved Constraint Solving}
The constraint-solving oracle is the only one that can acquire the ground truth. However, the low efficiency of SAT solvers prevents its widespread application. In addition, existing research only employs differential testing and metamorphic testing methods for the detection of transaction bugs, without utilizing constraint solving methods. Future research directions include:
\begin{itemize}
    \item \textbf{More Efficient Solver}: SAT solver is a general constraint solving method that requires evaluating whether each tuple satisfies predicate constraints, leading to inefficiency. However, there are already explorations like TQS~\cite{tang2023detecting}, which employs logical operations on join bitmaps to quickly acquire the ground truth for filter-free multi-table join statements without the necessity to iterate through each tuple. Therefore, for specific SQL statements, simplifying the solution process based on SQL characteristics is a focus of future research.
    \item \textbf{Constraint Solving For Transactions}: The closest existing work in this direction is Troc~\cite{dou2023detecting}, which employs external multi-version concurrency control to determine the visible data versions of statements. However, it still requires one to execute each statement within the DBMS to obtain the expected results, which can lead to false positives. In the future, this issue could potentially be addressed by incorporating constraint-solving methods.
\end{itemize}

\subsection{Component-oriented Fuzzing}
During the development phase of a DBMS, each update typically involves only a small portion of the code, making overall fuzzing less efficient. Component-oriented fuzzing can make testing more focused and efficient, and research directions include:
\begin{itemize}
    \item \textbf{Component-specific Fuzzing}: Most components in a DBMS can be regarded as standalone objects that take input and produce output. Therefore, we can perform separate fuzzing for these specific components, such as directly generating \textsf{PUT}, \textsf{GET} and \textsf{DELETE} operations to test bugs in key-value (KV) storage. Although there are some testing methods for the DBMS optimizer~\cite{leis2015good,li2016optmark,Gu_2012,giakoumakis2008testing}, they are not fuzzing methods and can only evaluate the quality of the query plans, not their correctness.
    \item \textbf{White-box Component Fuzzing}: Recent research~\cite{rigger2022intramorphic} has introduced white-box fuzzing in software testing. That method involves replacing components in a predictable manner to evaluate whether there are bugs in them. In theory, this method can be extended to DBMSs for component fuzzing. The primary challenge lies in achieving cost-effective replacement of DBMS components to improve its practicality. 
\end{itemize}

\subsection{Fuzzing of Modern Database Systems}
With the emergence of new computing technologies and application fields, traditional relational databases face new challenges, driving the development of databases in new directions. Emerging databases include distributed databases\cite{corbett2013spanner,huang2020tidb,taft2020cockroachdb,ajmani2022demonstration}, multi-model databases\cite{lu2019multi,benchmark2021,zhang2019unibench,yuan2021storing,multicategory2021}, autonomous databases\cite{li2021opengauss,li2019qtune,marcus2021bao,pavlo2017self,diagnosing2020}, and more. However, there is currently a lack of fuzzing methods for these new DBMSs.
\begin{itemize}
    \item \textbf{Distributed DBMS Fuzzing}: Compared to single-node databases, distributed databases involve more complex operations, such as data distribution, master-slave replication, distributed transactions, network communication, and more. This requires the fuzzers to further expand their exploration space to meet these specific needs. Jepsen~\cite{git2023jepsen} and MALLORY~\cite{meng2023greybox} were early attempts at fuzzing on distributed systems. Jepsen is a black-box fuzzer that detects consistency bugs by randomly generating network faults (network partitions, node removals). MALLORY, on the other hand, introduced gray-box feedback to generate faults that trigger unseen system behaviors. In distributed DBMSs, there are special components that support data management and querying across multiple nodes, but fuzzing on these components is currently far from sufficient.
    \item \textbf{Multi-Model DBMS Fuzzing}: Compared to single-model databases, multi-model databases support multiple data models within a unified backend, such as document, graph, relational, and KV models. Early research in NoSQL database fuzzing includes GDsmith~\cite{lin2022gdsmith}, Grand~\cite{zheng2022finding}, and UNICORN~\cite{wu2022unicorn}. GDsmith and Grand can identify bugs in graph database systems, while UNICORN is designed for time series databases. Nevertheless, there is currently a lack of fuzzing techniques for cross-model queries in multi-model DBMSs.
    \item \textbf{Autonomous DBMS Fuzzing}: Autonomous databases introduce autonomous features such as parameter self-adaptation and index recommendations. Verifying the correctness of autonomous tuning recommendations is a significant challenge.
\end{itemize}

\subsection{Improved Space Exploration Capabilities}
Currently, coverage feedback is often used in DBMS fuzzing to increase the coverage of code branches. However, this can lead to an ineffective inflation in the number of test cases, significantly increasing the time and computational resources required for testing. Therefore, potential future improvements may include:
\begin{itemize}
    \item \textbf{Balancing Full Coverage and Specific Coverage}: Current fuzzers based on coverage feedback treat all code branches equally~\cite{wang2020not}. While this approach is reasonable from a software testing perspective, it is not efficient in discovering DBMS bugs. In practice, a balance should be struck between full branch coverage and branch specific coverage. For example, it might be more efficient to prioritize testing the branches near code boundaries where exceptions or security vulnerabilities are more likely to occur.
    \item \textbf{Semantic Dimension Space Exploration}: Coverage feedback usually focuses on maximizing space exploration at the code level and ignores the semantic level. This means that test cases may trigger code branches but do not necessarily detect semantically related bugs. Existing work~\cite{ba2023testing} uses query plan feedback to maximize exploration of the query space, which is not achievable at the code level alone. Furthermore, maximizing space exploration in other dimensions, such as transaction space, is crucial.
\end{itemize}


\section{Conclusion}\label{sec:con}
DBMS is the fundamental software for managing and organizing data, and its internal structure is extremely complex. The development of commercial DBMSs often produces a large number of unexpected bugs (more than 10,000 bugs have been discovered and resolved in the MySQL forum~\cite{bugs_mysql}), which can easily lead to huge economic losses. Fuzzing is a method that detects bugs by automatically generating, mutating, and executing test cases. It has a history of nearly 50 years, but existing research mainly focuses on \textit{general-purpose} software. DBMS contains unique modules and functionalities, but there is no review to deeply study and compare these special DBMS fuzzing methods.

This paper offers a comprehensive review of existing DBMS fuzzing methods. It begins by providing a summary of essential concepts, establishing a systematic definition of the general fuzzing procedure, and categorizing methods horizontally. In addition, it vertically categorizes these methods based on testing components and provides detailed descriptions of representative techniques. The paper makes an effort to collect and organize available open-source DBMS fuzzers and conducts comparisons using the same experimental settings to help researchers understand the strengths, weaknesses, and suitability of various methods. Finally, we discuss future directions in fuzzing, including constraint solving, component orientation, modern databases, and query spatial exploration, emphasizing the importance of further exploration in these areas.

\begin{acks}
This work is supported by the National Natural Science Foundation of China under Grant No.62302370, 62002274, 62272369, 62372352, the Natural Science Basic Research Program of Shaanxi under Grant No.2023-JC-QN-0648 and the Fundamental Research Funds for the Central Universities under Grant No. XJS220301.
\end{acks}


%
\bibliographystyle{ACM-Reference-Format}
\bibliography{bibliography}


\begin{thebibliography}{110}


\ifx \showCODEN    \undefined \def \showCODEN     #1{\unskip}     \fi
\ifx \showDOI      \undefined \def \showDOI       #1{#1}\fi
\ifx \showISBNx    \undefined \def \showISBNx     #1{\unskip}     \fi
\ifx \showISBNxiii \undefined \def \showISBNxiii  #1{\unskip}     \fi
\ifx \showISSN     \undefined \def \showISSN      #1{\unskip}     \fi
\ifx \showLCCN     \undefined \def \showLCCN      #1{\unskip}     \fi
\ifx \shownote     \undefined \def \shownote      #1{#1}          \fi
\ifx \showarticletitle \undefined \def \showarticletitle #1{#1}   \fi
\ifx \showURL      \undefined \def \showURL       {\relax}        \fi
\providecommand\bibfield[2]{#2}
\providecommand\bibinfo[2]{#2}
\providecommand\natexlab[1]{#1}
\providecommand\showeprint[2][]{arXiv:#2}

\bibitem[tpc(1996)]%
        {tpc}
 \bibinfo{year}{1996}\natexlab{}.
\newblock \bibinfo{title}{TPC-H}.
\newblock \bibinfo{howpublished}{\url{https://www.tpc.org}}.
\newblock


\bibitem[fay(2011)]%
        {fayyad2011data}
 \bibinfo{year}{2011}\natexlab{}.
\newblock \bibinfo{title}{Data science revealed: A data-driven glimpse into the
  burgeoning new field}.
\newblock
  \bibinfo{howpublished}{\url{https://www.ndm.net/datawarehouse/pdf/EMC-Data_Science_Study_White_Paper.pdf
  }}.
\newblock


\bibitem[afl(2015)]%
        {afl}
 \bibinfo{year}{2015}\natexlab{}.
\newblock \bibinfo{title}{AFL: American Fuzzy Lop}.
\newblock \bibinfo{howpublished}{\url{ http://lcamtuf.coredump.cx/afl/}}.
\newblock


\bibitem[git(2015a)]%
        {git2023jepsen}
 \bibinfo{year}{2015}\natexlab{a}.
\newblock \bibinfo{title}{Jepsen}.
\newblock \bibinfo{howpublished}{\url{https://github.com/jepsen-io/jepsen}}.
\newblock


\bibitem[git(2015b)]%
        {git2023sqlsmith}
 \bibinfo{year}{2015}\natexlab{b}.
\newblock \bibinfo{title}{SQLsmith}.
\newblock \bibinfo{howpublished}{\url{https://github.com/anse1/sqlsmith}}.
\newblock


\bibitem[g20(2016)]%
        {g2016}
 \bibinfo{year}{2016}\natexlab{}.
\newblock \bibinfo{title}{Honggfuzz}.
\newblock \bibinfo{howpublished}{\url{ https://google.github.io/honggfuzz/}}.
\newblock


\bibitem[oss(2016)]%
        {oss}
 \bibinfo{year}{2016}\natexlab{}.
\newblock \bibinfo{title}{OSS-Fuzz: Continuous Fuzzing for Open Source
  Software}.
\newblock \bibinfo{howpublished}{\url{ https://github.com/google/oss-fuzz}}.
\newblock


\bibitem[llv(2017)]%
        {llvm}
 \bibinfo{year}{2017}\natexlab{}.
\newblock \bibinfo{title}{LibFuzzer - A Library For Coverage-guided Fuzz
  Testing}.
\newblock \bibinfo{howpublished}{\url{ http://llvm.org/docs/LibFuzzer.html}}.
\newblock


\bibitem[go(2019)]%
        {go}
 \bibinfo{year}{2019}\natexlab{}.
\newblock \bibinfo{title}{American Fuzzy Lop}.
\newblock \bibinfo{howpublished}{\url{ http://lcamtuf.coredump.cx/afl}}.
\newblock


\bibitem[ALL(2019)]%
        {ALLoy01}
 \bibinfo{year}{2019}\natexlab{}.
\newblock \bibinfo{title}{Documentation of Alloy SAT solver}.
\newblock \bibinfo{howpublished}{\url{
  https://alloytools.org/documentation.html}}.
\newblock


\bibitem[fb(2019)]%
        {fb}
 \bibinfo{year}{2019}\natexlab{}.
\newblock \bibinfo{title}{Technical "Whitepaper" For Afl-fuzz}.
\newblock \bibinfo{howpublished}{\url{
  http://lcamtuf.coredump.cx/afl/technical_details.txt}}.
\newblock


\bibitem[c20(2021)]%
        {c2021}
 \bibinfo{year}{2021}\natexlab{}.
\newblock \bibinfo{title}{CockroachDB Performance Bug Reports}.
\newblock \bibinfo{howpublished}{\url{
  https://github.com/cockroachdb/cockroach/issues?q=performance}}.
\newblock


\bibitem[p20(2021)]%
        {p2021}
 \bibinfo{year}{2021}\natexlab{}.
\newblock \bibinfo{title}{PostgreSQL Performance Bug Reports}.
\newblock \bibinfo{howpublished}{\url{
  https://www.postgresql.org/search/?m=1&q=performance&l=8&d=-1&s=r}}.
\newblock


\bibitem[sco(2021)]%
        {scott}
 \bibinfo{year}{2021}\natexlab{}.
\newblock \bibinfo{title}{SCOTT schema}.
\newblock \bibinfo{howpublished}{\url{https://www.orafaq.com/wiki/SCOTT}}.
\newblock


\bibitem[bug(2023)]%
        {bugs_mysql}
 \bibinfo{year}{2023}\natexlab{}.
\newblock \bibinfo{title}{MySQL Bugs Home}.
\newblock \bibinfo{howpublished}{\url{https://bugs.mysql.com/}}.
\newblock


\bibitem[Abdul~Khalek and Khurshid(2010)]%
        {abdul2010automated}
\bibfield{author}{\bibinfo{person}{Shadi Abdul~Khalek} {and}
  \bibinfo{person}{Sarfraz Khurshid}.} \bibinfo{year}{2010}\natexlab{}.
\newblock \showarticletitle{Automated SQL Query Generation for Systematic
  Testing of Database Engines}. In \bibinfo{booktitle}{\emph{International
  Conference on Automated Software Engineering}}. \bibinfo{publisher}{ACM},
  \bibinfo{pages}{329--332}.
\newblock


\bibitem[Ajmani et~al\mbox{.}(2022)]%
        {ajmani2022demonstration}
\bibfield{author}{\bibinfo{person}{Arul Ajmani}, \bibinfo{person}{Aayush Shah},
  \bibinfo{person}{Alexander Shraer}, \bibinfo{person}{Adam Storm},
  \bibinfo{person}{Rebecca Taft}, \bibinfo{person}{Oliver Tan}, {and}
  \bibinfo{person}{Nathan VanBenschoten}.} \bibinfo{year}{2022}\natexlab{}.
\newblock \showarticletitle{A Demonstration of Multi-Region CockroachDB}.
\newblock \bibinfo{journal}{\emph{Very Large Data Bases Endowment}}
  \bibinfo{volume}{15}, \bibinfo{number}{12} (\bibinfo{year}{2022}),
  \bibinfo{pages}{3610--3613}.
\newblock


\bibitem[Ba and Rigger(2023)]%
        {ba2023testing}
\bibfield{author}{\bibinfo{person}{Jinsheng Ba} {and} \bibinfo{person}{Manuel
  Rigger}.} \bibinfo{year}{2023}\natexlab{}.
\newblock \showarticletitle{Testing Database Engines via Query Plan Guidance}.
  In \bibinfo{booktitle}{\emph{International Conference on Software
  Engineering}}. \bibinfo{publisher}{ACM}, \bibinfo{pages}{2060–2071}.
\newblock


\bibitem[Bati et~al\mbox{.}(2007)]%
        {bati2007genetic}
\bibfield{author}{\bibinfo{person}{Hardik Bati}, \bibinfo{person}{Leo
  Giakoumakis}, \bibinfo{person}{Steve Herbert}, {and}
  \bibinfo{person}{Aleksandras Surna}.} \bibinfo{year}{2007}\natexlab{}.
\newblock \showarticletitle{A Genetic Approach for Random Testing of Database
  Systems}. In \bibinfo{booktitle}{\emph{International Conference on Very Large
  Data Bases}}. \bibinfo{publisher}{VLDB Endowment},
  \bibinfo{pages}{1243--1251}.
\newblock


\bibitem[Beizer(1992)]%
        {software1992}
\bibfield{author}{\bibinfo{person}{B. Beizer}.}
  \bibinfo{year}{1992}\natexlab{}.
\newblock \showarticletitle{Software testing techniques}.
\newblock \bibinfo{journal}{\emph{Software Testing Verification and
  Reliability}} \bibinfo{volume}{2}, \bibinfo{number}{4}
  (\bibinfo{year}{1992}), \bibinfo{pages}{215--216}.
\newblock


\bibitem[Berry and Fristedt(1985)]%
        {berry1985bandit}
\bibfield{author}{\bibinfo{person}{Donald~A Berry} {and} \bibinfo{person}{Bert
  Fristedt}.} \bibinfo{year}{1985}\natexlab{}.
\newblock \showarticletitle{Bandit problems: sequential allocation of
  experiments (Monographs on statistics and applied probability)}.
\newblock \bibinfo{journal}{\emph{London: Chapman and Hall}}
  \bibinfo{volume}{5}, \bibinfo{number}{71-87} (\bibinfo{year}{1985}),
  \bibinfo{pages}{7--7}.
\newblock


\bibitem[Blazytko et~al\mbox{.}(2019)]%
        {blazytko2019grimoire}
\bibfield{author}{\bibinfo{person}{Tim Blazytko}, \bibinfo{person}{Matt
  Bishop}, \bibinfo{person}{Cornelius Aschermann}, \bibinfo{person}{Justin
  Cappos}, \bibinfo{person}{Moritz Schl{\"o}gel}, \bibinfo{person}{Nadia
  Korshun}, \bibinfo{person}{Ali Abbasi}, \bibinfo{person}{Marco
  Schweighauser}, \bibinfo{person}{Sebastian Schinzel}, \bibinfo{person}{Sergej
  Schumilo}, {et~al\mbox{.}}} \bibinfo{year}{2019}\natexlab{}.
\newblock \showarticletitle{GRIMOIRE: Synthesizing Structure While Fuzzing}. In
  \bibinfo{booktitle}{\emph{USENIX Conference on Security Symposium}}.
  \bibinfo{publisher}{USENIX Association}, \bibinfo{pages}{1985--2002}.
\newblock


\bibitem[Bruno and Chaudhuri(2005)]%
        {bruno2005flexible}
\bibfield{author}{\bibinfo{person}{Nicolas Bruno} {and}
  \bibinfo{person}{Surajit Chaudhuri}.} \bibinfo{year}{2005}\natexlab{}.
\newblock \showarticletitle{Flexible Database Generators}. In
  \bibinfo{booktitle}{\emph{International Conference on Very Large Data
  Bases}}. \bibinfo{publisher}{VLDB Endowment}, \bibinfo{pages}{1097--1107}.
\newblock


\bibitem[Bruno et~al\mbox{.}(2006)]%
        {Bruno_2006}
\bibfield{author}{\bibinfo{person}{N. Bruno}, \bibinfo{person}{S. Chaudhuri},
  {and} \bibinfo{person}{D. Thomas}.} \bibinfo{year}{2006}\natexlab{}.
\newblock \showarticletitle{Generating Queries with Cardinality Constraints for
  DBMS Testing}.
\newblock \bibinfo{journal}{\emph{Transactions on Knowledge and Data
  Engineering}} \bibinfo{volume}{18}, \bibinfo{number}{12}
  (\bibinfo{year}{2006}), \bibinfo{pages}{1721--1725}.
\newblock


\bibitem[Böhme et~al\mbox{.}(2017)]%
        {B_hme_2017}
\bibfield{author}{\bibinfo{person}{Marcel Böhme}, \bibinfo{person}{Van-Thuan
  Pham}, \bibinfo{person}{Manh-Dung Nguyen}, {and} \bibinfo{person}{Abhik
  Roychoudhury}.} \bibinfo{year}{2017}\natexlab{}.
\newblock \showarticletitle{Directed Greybox Fuzzing}. In
  \bibinfo{booktitle}{\emph{Conference on Computer and Communications
  Security}}. \bibinfo{publisher}{ACM}, \bibinfo{pages}{2329–2344}.
\newblock


\bibitem[Böhme et~al\mbox{.}(2016)]%
        {B_hme_2016}
\bibfield{author}{\bibinfo{person}{Marcel Böhme}, \bibinfo{person}{Van-Thuan
  Pham}, {and} \bibinfo{person}{Abhik Roychoudhury}.}
  \bibinfo{year}{2016}\natexlab{}.
\newblock \showarticletitle{Coverage-based Greybox Fuzzing as Markov Chain}. In
  \bibinfo{booktitle}{\emph{Conference on Computer and Communications
  Security}}. \bibinfo{publisher}{ACM}, \bibinfo{pages}{1032–1043}.
\newblock


\bibitem[Chen et~al\mbox{.}(1998)]%
        {chen1998metamorphic}
\bibfield{author}{\bibinfo{person}{Tsong~Y Chen}, \bibinfo{person}{Shing~C
  Cheung}, {and} \bibinfo{person}{Shiu~Ming Yiu}.}
  \bibinfo{year}{1998}\natexlab{}.
\newblock \showarticletitle{Metamorphic Testing: A New Approach for Generating
  Next Test Cases}.
\newblock \bibinfo{journal}{\emph{arXiv preprint arXiv:2002.12543}}
  (\bibinfo{year}{1998}).
\newblock


\bibitem[Chen et~al\mbox{.}(2020)]%
        {chen2020testing}
\bibfield{author}{\bibinfo{person}{Xinyue Chen}, \bibinfo{person}{Chenglong
  Wang}, {and} \bibinfo{person}{Alvin Cheung}.}
  \bibinfo{year}{2020}\natexlab{}.
\newblock \showarticletitle{Testing Query Execution Engines with Mutations}. In
  \bibinfo{booktitle}{\emph{Workshop on Testing Database Systems}}.
  \bibinfo{publisher}{ACM}, \bibinfo{pages}{1--5}.
\newblock


\bibitem[Chu et~al\mbox{.}(2017)]%
        {chu2017cosette}
\bibfield{author}{\bibinfo{person}{Shumo Chu}, \bibinfo{person}{Chenglong
  Wang}, \bibinfo{person}{Konstantin Weitz}, {and} \bibinfo{person}{Alvin
  Cheung}.} \bibinfo{year}{2017}\natexlab{}.
\newblock \showarticletitle{Cosette: An Automated Prover for SQL}. In
  \bibinfo{booktitle}{\emph{Conference on Innovative Data Systems Research}}.
\newblock


\bibitem[Corbett et~al\mbox{.}(2013)]%
        {corbett2013spanner}
\bibfield{author}{\bibinfo{person}{James~C Corbett}, \bibinfo{person}{Jeffrey
  Dean}, \bibinfo{person}{Michael Epstein}, \bibinfo{person}{Andrew Fikes},
  \bibinfo{person}{Christopher Frost}, \bibinfo{person}{Jeffrey~John Furman},
  \bibinfo{person}{Sanjay Ghemawat}, \bibinfo{person}{Andrey Gubarev},
  \bibinfo{person}{Christopher Heiser}, \bibinfo{person}{Peter Hochschild},
  {et~al\mbox{.}}} \bibinfo{year}{2013}\natexlab{}.
\newblock \showarticletitle{Spanner: Google’s Globally Distributed Database}.
\newblock \bibinfo{journal}{\emph{ACM Transactions on Computer Systems}}
  \bibinfo{volume}{31}, \bibinfo{number}{3} (\bibinfo{year}{2013}),
  \bibinfo{pages}{1--22}.
\newblock


\bibitem[Cui et~al\mbox{.}(2022)]%
        {cui2022differentially}
\bibfield{author}{\bibinfo{person}{Ziyu Cui}, \bibinfo{person}{Wensheng Dou},
  \bibinfo{person}{Qianwang Dai}, \bibinfo{person}{Jiansen Song},
  \bibinfo{person}{Wei Wang}, \bibinfo{person}{Jun Wei}, {and}
  \bibinfo{person}{Dan Ye}.} \bibinfo{year}{2022}\natexlab{}.
\newblock \showarticletitle{Differentially Testing Database Transactions for
  Fun and Profit}. In \bibinfo{booktitle}{\emph{International Conference on
  Automated Software Engineering}}. \bibinfo{publisher}{ACM},
  \bibinfo{pages}{1--12}.
\newblock


\bibitem[Dou et~al\mbox{.}(2023)]%
        {dou2023detecting}
\bibfield{author}{\bibinfo{person}{Wensheng Dou}, \bibinfo{person}{Ziyu Cui},
  \bibinfo{person}{Qianwang Dai}, \bibinfo{person}{Jiansen Song},
  \bibinfo{person}{Dong Wang}, \bibinfo{person}{Yu Gao}, \bibinfo{person}{Wei
  Wang}, \bibinfo{person}{Jun Wei}, \bibinfo{person}{Lei Chen},
  \bibinfo{person}{Hanmo Wang}, {et~al\mbox{.}}}
  \bibinfo{year}{2023}\natexlab{}.
\newblock \showarticletitle{Detecting Isolation Bugs via Transaction Oracle
  Construction}. In \bibinfo{booktitle}{\emph{International Conference on
  Software Engineering}}. \bibinfo{publisher}{IEEE}.
\newblock


\bibitem[Fu et~al\mbox{.}(2022)]%
        {fu2022griffin}
\bibfield{author}{\bibinfo{person}{Jingzhou Fu}, \bibinfo{person}{Jie Liang},
  \bibinfo{person}{Zhiyong Wu}, \bibinfo{person}{Mingzhe Wang}, {and}
  \bibinfo{person}{Yu Jiang}.} \bibinfo{year}{2022}\natexlab{}.
\newblock \showarticletitle{Griffin: Grammar-Free DBMS Fuzzing}. In
  \bibinfo{booktitle}{\emph{International Conference on Automated Software
  Engineering}}. \bibinfo{publisher}{ACM}, \bibinfo{pages}{1--12}.
\newblock


\bibitem[Galindo-Legaria et~al\mbox{.}(2004)]%
        {Galindo_Legaria_2004}
\bibfield{author}{\bibinfo{person}{C\'{e}sar~A. Galindo-Legaria},
  \bibinfo{person}{Stefano Stefani}, {and} \bibinfo{person}{Florian Waas}.}
  \bibinfo{year}{2004}\natexlab{}.
\newblock \showarticletitle{Query Processing for SQL Updates}. In
  \bibinfo{booktitle}{\emph{International Conference on Management of Data}}.
  \bibinfo{publisher}{ACM}, \bibinfo{pages}{844–849}.
\newblock


\bibitem[Giakoumakis and Galindo-Legaria(2008)]%
        {giakoumakis2008testing}
\bibfield{author}{\bibinfo{person}{Leo Giakoumakis} {and}
  \bibinfo{person}{C{\'e}sar~A Galindo-Legaria}.}
  \bibinfo{year}{2008}\natexlab{}.
\newblock \showarticletitle{Testing SQL Server's Query Optimizer: Challenges,
  Techniques and Experiences}.
\newblock \bibinfo{journal}{\emph{Data Engineering Bulletin}}
  \bibinfo{volume}{31}, \bibinfo{number}{1} (\bibinfo{year}{2008}),
  \bibinfo{pages}{36--43}.
\newblock


\bibitem[Godefroid et~al\mbox{.}(2008)]%
        {Godefroid_2008}
\bibfield{author}{\bibinfo{person}{Patrice Godefroid}, \bibinfo{person}{Adam
  Kiezun}, {and} \bibinfo{person}{Michael~Y. Levin}.}
  \bibinfo{year}{2008}\natexlab{}.
\newblock \showarticletitle{Grammar-based whitebox fuzzing}. In
  \bibinfo{booktitle}{\emph{Conference on Programming Language Design and
  Implementation}}. \bibinfo{publisher}{ACM}, \bibinfo{pages}{206–215}.
\newblock


\bibitem[Goldberg and Novikov(2007)]%
        {goldberg2007berkmin}
\bibfield{author}{\bibinfo{person}{Eugene Goldberg} {and}
  \bibinfo{person}{Yakov Novikov}.} \bibinfo{year}{2007}\natexlab{}.
\newblock \showarticletitle{BerkMin: A Fast and Robust SAT-solver}.
\newblock \bibinfo{journal}{\emph{Discrete Applied Mathematics}}
  \bibinfo{volume}{155}, \bibinfo{number}{12} (\bibinfo{year}{2007}),
  \bibinfo{pages}{1549--1561}.
\newblock


\bibitem[Graefe(1993)]%
        {Graefe_1993}
\bibfield{author}{\bibinfo{person}{Goetz Graefe}.}
  \bibinfo{year}{1993}\natexlab{}.
\newblock \showarticletitle{Query Evaluation Techniques for Large Databases}.
\newblock \bibinfo{journal}{\emph{Computing Surveys}} \bibinfo{volume}{25},
  \bibinfo{number}{2} (\bibinfo{year}{1993}), \bibinfo{pages}{73--169}.
\newblock


\bibitem[Graefe(1995)]%
        {graefe1995cascades}
\bibfield{author}{\bibinfo{person}{Goetz Graefe}.}
  \bibinfo{year}{1995}\natexlab{}.
\newblock \showarticletitle{The Cascades Framework for Query Optimization}.
\newblock \bibinfo{journal}{\emph{Data Engineering Bulletin}}
  \bibinfo{volume}{18}, \bibinfo{number}{3} (\bibinfo{year}{1995}),
  \bibinfo{pages}{19--29}.
\newblock


\bibitem[Gray et~al\mbox{.}(1994)]%
        {gray1994quickly}
\bibfield{author}{\bibinfo{person}{Jim Gray}, \bibinfo{person}{Prakash
  Sundaresan}, \bibinfo{person}{Susanne Englert}, \bibinfo{person}{Ken
  Baclawski}, {and} \bibinfo{person}{Peter~J Weinberger}.}
  \bibinfo{year}{1994}\natexlab{}.
\newblock \showarticletitle{Quickly Generating Billion-Record Synthetic
  Databases}. In \bibinfo{booktitle}{\emph{International Conference on
  Management of Data}}. \bibinfo{publisher}{ACM}, \bibinfo{pages}{243--252}.
\newblock


\bibitem[Gu et~al\mbox{.}(2012)]%
        {Gu_2012}
\bibfield{author}{\bibinfo{person}{Zhongxian Gu}, \bibinfo{person}{Mohamed~A.
  Soliman}, {and} \bibinfo{person}{Florian~M Waas}.}
  \bibinfo{year}{2012}\natexlab{}.
\newblock \showarticletitle{Testing the Accuracy of Query Optimizers}. In
  \bibinfo{booktitle}{\emph{Workshop on Testing Database Systems}}.
  \bibinfo{publisher}{ACM}, \bibinfo{pages}{1--6}.
\newblock


\bibitem[Houkj{\ae}r et~al\mbox{.}(2006)]%
        {houkjaer2006simple}
\bibfield{author}{\bibinfo{person}{Kenneth Houkj{\ae}r},
  \bibinfo{person}{Kristian Torp}, {and} \bibinfo{person}{Rico Wind}.}
  \bibinfo{year}{2006}\natexlab{}.
\newblock \showarticletitle{Simple and Realistic Data Generation}. In
  \bibinfo{booktitle}{\emph{International Conference on Very Large Data
  Bases}}. \bibinfo{publisher}{VLDB Endowment}, \bibinfo{pages}{1243--1246}.
\newblock


\bibitem[Howden(1978)]%
        {howden1978theoretical}
\bibfield{author}{\bibinfo{person}{William~E Howden}.}
  \bibinfo{year}{1978}\natexlab{}.
\newblock \showarticletitle{Theoretical and empirical studies of program
  testing}.
\newblock \bibinfo{journal}{\emph{Transactions on Software Engineering}}
  \bibinfo{volume}{SE-4}, \bibinfo{number}{4} (\bibinfo{year}{1978}),
  \bibinfo{pages}{293--298}.
\newblock


\bibitem[Huang et~al\mbox{.}(2020)]%
        {huang2020tidb}
\bibfield{author}{\bibinfo{person}{Dongxu Huang}, \bibinfo{person}{Qi Liu},
  \bibinfo{person}{Qiu Cui}, \bibinfo{person}{Zhuhe Fang},
  \bibinfo{person}{Xiaoyu Ma}, \bibinfo{person}{Fei Xu}, \bibinfo{person}{Li
  Shen}, \bibinfo{person}{Liu Tang}, \bibinfo{person}{Yuxing Zhou},
  \bibinfo{person}{Menglong Huang}, {et~al\mbox{.}}}
  \bibinfo{year}{2020}\natexlab{}.
\newblock \showarticletitle{TiDB: A Raft-Based HTAP Database}.
\newblock \bibinfo{journal}{\emph{Very Large Data Bases Endowment}}
  \bibinfo{volume}{13}, \bibinfo{number}{12} (\bibinfo{year}{2020}),
  \bibinfo{pages}{3072--3084}.
\newblock


\bibitem[Hunt(2020)]%
        {db733}
\bibfield{author}{\bibinfo{person}{T. Hunt}.} \bibinfo{year}{2020}\natexlab{}.
\newblock \bibinfo{title}{The 773 Million Record "Collection 1" Data Breach}.
\newblock
  \bibinfo{howpublished}{\url{https://www.troyhunt.com/the-773-million-record-collection-1-data-reach/,
  January 2020}}.
\newblock


\bibitem[Ibaraki and Kameda(1984)]%
        {ibaraki1984optimal}
\bibfield{author}{\bibinfo{person}{Toshihide Ibaraki} {and}
  \bibinfo{person}{Tiko Kameda}.} \bibinfo{year}{1984}\natexlab{}.
\newblock \showarticletitle{On the Optimal Nesting Order for Computing
  N-Relational Joins}.
\newblock \bibinfo{journal}{\emph{Transactions on Database Systems}}
  \bibinfo{volume}{9}, \bibinfo{number}{3} (\bibinfo{year}{1984}),
  \bibinfo{pages}{482--502}.
\newblock


\bibitem[Jackson(2002)]%
        {jackson2002alloy}
\bibfield{author}{\bibinfo{person}{Daniel Jackson}.}
  \bibinfo{year}{2002}\natexlab{}.
\newblock \showarticletitle{Alloy: A Lightweight Object Modelling Notation}.
\newblock \bibinfo{journal}{\emph{Transactions on Software Engineering and
  Methodology}} \bibinfo{volume}{11}, \bibinfo{number}{2}
  (\bibinfo{year}{2002}), \bibinfo{pages}{256--290}.
\newblock


\bibitem[Jayakumar and Abran(2013)]%
        {Jayakumar_2013}
\bibfield{author}{\bibinfo{person}{Kamala~Ramasubramani Jayakumar} {and}
  \bibinfo{person}{Alain Abran}.} \bibinfo{year}{2013}\natexlab{}.
\newblock \showarticletitle{A Survey of Software Test Estimation Techniques}.
\newblock \bibinfo{journal}{\emph{Journal of Software Engineering and
  Applications}} \bibinfo{volume}{6}, \bibinfo{number}{10}
  (\bibinfo{year}{2013}), \bibinfo{pages}{47--52}.
\newblock


\bibitem[Jiang et~al\mbox{.}(2023)]%
        {jiang2023dynsql}
\bibfield{author}{\bibinfo{person}{Zu-Ming Jiang}, \bibinfo{person}{Jia-Ju
  Bai}, {and} \bibinfo{person}{Zhendong Su}.} \bibinfo{year}{2023}\natexlab{}.
\newblock \showarticletitle{DynSQL: Stateful Fuzzing for Database Management
  Systems with Complex and Valid SQL Query Generation}. In
  \bibinfo{booktitle}{\emph{USENIX Security Symposium}}.
  \bibinfo{publisher}{USENIX Association}, \bibinfo{pages}{4949--4965}.
\newblock


\bibitem[Jibson(2019)]%
        {sc2019}
\bibfield{author}{\bibinfo{person}{Matt Jibson}.}
  \bibinfo{year}{2019}\natexlab{}.
\newblock \bibinfo{title}{SQLsmith: Randomized sql testing in cockroachdb}.
\newblock
  \bibinfo{howpublished}{\url{https://www.cockroachlabs.com/blog/sqlsmith-randomized-sql-testing/}}.
\newblock


\bibitem[Jung et~al\mbox{.}(2019)]%
        {jung2019apollo}
\bibfield{author}{\bibinfo{person}{Jinho Jung}, \bibinfo{person}{Hong Hu},
  \bibinfo{person}{Joy Arulraj}, \bibinfo{person}{Taesoo Kim}, {and}
  \bibinfo{person}{Woonhak Kang}.} \bibinfo{year}{2019}\natexlab{}.
\newblock \showarticletitle{APOLLO: Automatic Detection and Diagnosis of
  Performance Regressions in Database Systems}.
\newblock \bibinfo{journal}{\emph{Very Large Data Bases Endowment}}
  \bibinfo{volume}{13}, \bibinfo{number}{1} (\bibinfo{year}{2019}),
  \bibinfo{pages}{57--70}.
\newblock


\bibitem[Khurshid and Marinov(2004)]%
        {khurshid2004testera}
\bibfield{author}{\bibinfo{person}{Sarfraz Khurshid} {and}
  \bibinfo{person}{Darko Marinov}.} \bibinfo{year}{2004}\natexlab{}.
\newblock \showarticletitle{TestEra: Specification-based testing of Java
  programs using SAT}.
\newblock \bibinfo{journal}{\emph{Automated Software Engineering}}
  \bibinfo{volume}{11} (\bibinfo{year}{2004}), \bibinfo{pages}{403--434}.
\newblock


\bibitem[Lee et~al\mbox{.}(2011)]%
        {lee2011performance}
\bibfield{author}{\bibinfo{person}{Donghun Lee}, \bibinfo{person}{Sang~K Cha},
  {and} \bibinfo{person}{Arthur~H Lee}.} \bibinfo{year}{2011}\natexlab{}.
\newblock \showarticletitle{A Performance Anomaly Detection and Analysis
  Framework for DBMS Development}.
\newblock \bibinfo{journal}{\emph{Transactions on Knowledge and Data
  Engineering}} \bibinfo{volume}{24}, \bibinfo{number}{8}
  (\bibinfo{year}{2011}), \bibinfo{pages}{1345--1360}.
\newblock


\bibitem[Leis et~al\mbox{.}(2015)]%
        {leis2015good}
\bibfield{author}{\bibinfo{person}{Viktor Leis}, \bibinfo{person}{Andrey
  Gubichev}, \bibinfo{person}{Atanas Mirchev}, \bibinfo{person}{Peter Boncz},
  \bibinfo{person}{Alfons Kemper}, {and} \bibinfo{person}{Thomas Neumann}.}
  \bibinfo{year}{2015}\natexlab{}.
\newblock \showarticletitle{How Good Are Query Optimizers, Really?}
\newblock \bibinfo{journal}{\emph{Very Large Data Bases Endowment}}
  \bibinfo{volume}{9}, \bibinfo{number}{3} (\bibinfo{year}{2015}),
  \bibinfo{pages}{204--215}.
\newblock


\bibitem[Lemieux and Sen(2018)]%
        {Lemieux_2018}
\bibfield{author}{\bibinfo{person}{Caroline Lemieux} {and}
  \bibinfo{person}{Koushik Sen}.} \bibinfo{year}{2018}\natexlab{}.
\newblock \showarticletitle{FairFuzz: A Targeted Mutation Strategy for
  Increasing Greybox Fuzz Testing Coverage}. In
  \bibinfo{booktitle}{\emph{International Conference on Automated Software
  Engineering}}. \bibinfo{publisher}{ACM}, \bibinfo{pages}{475--485}.
\newblock


\bibitem[Li et~al\mbox{.}(2019)]%
        {li2019qtune}
\bibfield{author}{\bibinfo{person}{Guoliang Li}, \bibinfo{person}{Xuanhe Zhou},
  \bibinfo{person}{Shifu Li}, {and} \bibinfo{person}{Bo Gao}.}
  \bibinfo{year}{2019}\natexlab{}.
\newblock \showarticletitle{QTune: A Query-Aware Database Tuning System with
  Deep Reinforcement Learning}.
\newblock \bibinfo{journal}{\emph{Very Large Data Bases Endowment}}
  \bibinfo{volume}{12}, \bibinfo{number}{12} (\bibinfo{year}{2019}),
  \bibinfo{pages}{2118--2130}.
\newblock


\bibitem[Li et~al\mbox{.}(2021)]%
        {li2021opengauss}
\bibfield{author}{\bibinfo{person}{Guoliang Li}, \bibinfo{person}{Xuanhe Zhou},
  \bibinfo{person}{Ji Sun}, \bibinfo{person}{Xiang Yu}, \bibinfo{person}{Yue
  Han}, \bibinfo{person}{Lianyuan Jin}, \bibinfo{person}{Wenbo Li},
  \bibinfo{person}{Tianqing Wang}, {and} \bibinfo{person}{Shifu Li}.}
  \bibinfo{year}{2021}\natexlab{}.
\newblock \showarticletitle{OpenGauss: An Autonomous Database System}.
\newblock \bibinfo{journal}{\emph{Very Large Data Bases Endowment}}
  \bibinfo{volume}{14}, \bibinfo{number}{12} (\bibinfo{year}{2021}),
  \bibinfo{pages}{3028--3042}.
\newblock


\bibitem[Li et~al\mbox{.}(2016)]%
        {li2016optmark}
\bibfield{author}{\bibinfo{person}{Zhan Li}, \bibinfo{person}{Olga
  Papaemmanouil}, {and} \bibinfo{person}{Mitch Cherniack}.}
  \bibinfo{year}{2016}\natexlab{}.
\newblock \showarticletitle{Optmark: A Toolkit for Benchmarking Query
  Optimizers}. In \bibinfo{booktitle}{\emph{Information and Knowledge
  Management}}. \bibinfo{publisher}{ACM}, \bibinfo{pages}{2155--2160}.
\newblock


\bibitem[Liang et~al\mbox{.}(2023)]%
        {liang2023sequence}
\bibfield{author}{\bibinfo{person}{Jie Liang}, \bibinfo{person}{Yaoguang Chen},
  \bibinfo{person}{Zhiyong Wu}, \bibinfo{person}{Jingzhou Fu},
  \bibinfo{person}{Mingzhe Wang}, \bibinfo{person}{Yu Jiang},
  \bibinfo{person}{Xiangdong Huang}, \bibinfo{person}{Ting Chen},
  \bibinfo{person}{Jiashui Wang}, {and} \bibinfo{person}{Jiajia Li}.}
  \bibinfo{year}{2023}\natexlab{}.
\newblock \showarticletitle{Sequence-Oriented DBMS Fuzzing}. In
  \bibinfo{booktitle}{\emph{International Conference on Data Engineering}}.
  \bibinfo{publisher}{IEEE}, \bibinfo{pages}{668--681}.
\newblock


\bibitem[Liang et~al\mbox{.}(2022)]%
        {liang2022detecting}
\bibfield{author}{\bibinfo{person}{Yu Liang}, \bibinfo{person}{Song Liu}, {and}
  \bibinfo{person}{Hong Hu}.} \bibinfo{year}{2022}\natexlab{}.
\newblock \showarticletitle{Detecting Logical Bugs of DBMS with Coverage-based
  Guidance}. In \bibinfo{booktitle}{\emph{USENIX Security Symposium}}.
  \bibinfo{publisher}{USENIX Association}, \bibinfo{pages}{4309--4326}.
\newblock


\bibitem[Lin et~al\mbox{.}(2022)]%
        {lin2022gdsmith}
\bibfield{author}{\bibinfo{person}{Wei Lin}, \bibinfo{person}{Ziyue Hua},
  \bibinfo{person}{Luyao Ren}, \bibinfo{person}{Zongyang Li},
  \bibinfo{person}{Lu Zhang}, {and} \bibinfo{person}{Tao Xie}.}
  \bibinfo{year}{2022}\natexlab{}.
\newblock \showarticletitle{GDsmith: Detecting Bugs in Graph Database Engines}.
\newblock \bibinfo{journal}{\emph{arXiv preprint arXiv:2206.08530}}
  (\bibinfo{year}{2022}).
\newblock


\bibitem[Liu et~al\mbox{.}(2022)]%
        {liu2022automatic}
\bibfield{author}{\bibinfo{person}{Xinyu Liu}, \bibinfo{person}{Qi Zhou},
  \bibinfo{person}{Joy Arulraj}, {and} \bibinfo{person}{Alessandro Orso}.}
  \bibinfo{year}{2022}\natexlab{}.
\newblock \showarticletitle{Automatic Detection of Performance Bugs in Database
  Systems Using Equivalent Queries}. In \bibinfo{booktitle}{\emph{International
  Conference on Software Engineering}}. \bibinfo{publisher}{ACM},
  \bibinfo{pages}{225--236}.
\newblock


\bibitem[Lo et~al\mbox{.}(2010)]%
        {lo2010framework}
\bibfield{author}{\bibinfo{person}{Eric Lo}, \bibinfo{person}{Carsten Binnig},
  \bibinfo{person}{Donald Kossmann}, \bibinfo{person}{M Tamer~{\"O}zsu}, {and}
  \bibinfo{person}{Wing-Kai Hon}.} \bibinfo{year}{2010}\natexlab{}.
\newblock \showarticletitle{A Framework for Testing DBMS Features}.
\newblock \bibinfo{journal}{\emph{Very Large Data Bases Endowment}}
  \bibinfo{volume}{19} (\bibinfo{year}{2010}), \bibinfo{pages}{203--230}.
\newblock


\bibitem[Lu and Holubov{\'a}(2019)]%
        {lu2019multi}
\bibfield{author}{\bibinfo{person}{Jiaheng Lu} {and} \bibinfo{person}{Irena
  Holubov{\'a}}.} \bibinfo{year}{2019}\natexlab{}.
\newblock \showarticletitle{Multi-Model Databases: A New Journey to Handle the
  Variety of Data}.
\newblock \bibinfo{journal}{\emph{Comput. Surveys}} \bibinfo{volume}{52},
  \bibinfo{number}{3} (\bibinfo{year}{2019}), \bibinfo{pages}{1--38}.
\newblock


\bibitem[Ma et~al\mbox{.}(2020)]%
        {diagnosing2020}
\bibfield{author}{\bibinfo{person}{Minghua Ma}, \bibinfo{person}{Zheng Yin},
  \bibinfo{person}{Shenglin Zhang}, \bibinfo{person}{Sheng Wang},
  \bibinfo{person}{Christopher Zheng}, \bibinfo{person}{Xinhao Jiang},
  \bibinfo{person}{Hanwen Hu}, \bibinfo{person}{Cheng Luo},
  \bibinfo{person}{Yilin Li}, \bibinfo{person}{Nengjun Qiu},
  \bibinfo{person}{Feifei Li}, \bibinfo{person}{Changcheng Chen}, {and}
  \bibinfo{person}{Dan Pei}.} \bibinfo{year}{2020}\natexlab{}.
\newblock \showarticletitle{Diagnosing Root Causes of Intermittent Slow Queries
  in Cloud Databases}.
\newblock \bibinfo{journal}{\emph{Very Large Data Bases Endowment}}
  \bibinfo{volume}{13}, \bibinfo{number}{8} (\bibinfo{year}{2020}),
  \bibinfo{pages}{1176–1189}.
\newblock


\bibitem[Marcus et~al\mbox{.}(2021)]%
        {marcus2021bao}
\bibfield{author}{\bibinfo{person}{Ryan Marcus}, \bibinfo{person}{Parimarjan
  Negi}, \bibinfo{person}{Hongzi Mao}, \bibinfo{person}{Nesime Tatbul},
  \bibinfo{person}{Mohammad Alizadeh}, {and} \bibinfo{person}{Tim Kraska}.}
  \bibinfo{year}{2021}\natexlab{}.
\newblock \showarticletitle{Bao: Making Learned Query Optimization Practical}.
  In \bibinfo{booktitle}{\emph{International Conference on Management of
  Data}}. \bibinfo{publisher}{ACM}, \bibinfo{pages}{1275--1288}.
\newblock


\bibitem[Marinov(2010)]%
        {Marinov_2010}
\bibfield{author}{\bibinfo{person}{Darko Marinov}.}
  \bibinfo{year}{2010}\natexlab{}.
\newblock \showarticletitle{Session Details: Technical Session 8: Concurrency
  and Differential Testing}. In \bibinfo{booktitle}{\emph{International
  Symposium on Software Testing and Analysis}}. \bibinfo{publisher}{ACM}.
\newblock


\bibitem[McKeeman(1998)]%
        {mckeeman1998differential}
\bibfield{author}{\bibinfo{person}{William~M McKeeman}.}
  \bibinfo{year}{1998}\natexlab{}.
\newblock \showarticletitle{Differential Testing for Software}.
\newblock \bibinfo{journal}{\emph{Digital Technical Journal}}
  \bibinfo{volume}{10}, \bibinfo{number}{1} (\bibinfo{year}{1998}),
  \bibinfo{pages}{100--107}.
\newblock


\bibitem[Mcminn et~al\mbox{.}(2015)]%
        {Mcminn_2015}
\bibfield{author}{\bibinfo{person}{Phil Mcminn}, \bibinfo{person}{Chris~J.
  Wright}, {and} \bibinfo{person}{Gregory~M. Kapfhammer}.}
  \bibinfo{year}{2015}\natexlab{}.
\newblock \showarticletitle{The Effectiveness of Test Coverage Criteria for
  Relational Database Schema Integrity Constraints}.
\newblock \bibinfo{journal}{\emph{Transactions on Software Engineering and
  Methodology}} \bibinfo{volume}{25}, \bibinfo{number}{1}
  (\bibinfo{year}{2015}), \bibinfo{pages}{1--49}.
\newblock


\bibitem[Meng et~al\mbox{.}(2023)]%
        {meng2023greybox}
\bibfield{author}{\bibinfo{person}{Ruijie Meng}, \bibinfo{person}{George
  Pîrlea}, \bibinfo{person}{Abhik Roychoudhury}, {and} \bibinfo{person}{Ilya
  Sergey}.} \bibinfo{year}{2023}\natexlab{}.
\newblock \showarticletitle{Greybox Fuzzing of Distributed Systems}.
\newblock \bibinfo{journal}{\emph{arXiv preprint arXiv:2305.02601}}
  (\bibinfo{year}{2023}).
\newblock


\bibitem[Mi et~al\mbox{.}(2021)]%
        {mi2021artemis}
\bibfield{author}{\bibinfo{person}{Kaiming Mi}, \bibinfo{person}{Chunxi Zhang},
  \bibinfo{person}{Weining Qian}, {and} \bibinfo{person}{Rong Zhang}.}
  \bibinfo{year}{2021}\natexlab{}.
\newblock \showarticletitle{Artemis: An Automatic Test Suite Generator for
  Large Scale OLAP Database}. In \bibinfo{booktitle}{\emph{Benchmarking,
  Measuring, and Optimizing: Third BenchCouncil International Symposium}}.
  \bibinfo{publisher}{Springer}, \bibinfo{pages}{74--89}.
\newblock


\bibitem[Miller et~al\mbox{.}(1990)]%
        {Miller_1990}
\bibfield{author}{\bibinfo{person}{Barton~P. Miller}, \bibinfo{person}{Lars
  Fredriksen}, {and} \bibinfo{person}{Bryan So}.}
  \bibinfo{year}{1990}\natexlab{}.
\newblock \showarticletitle{An Empirical Study of the Reliability of UNIX
  Utilities}.
\newblock \bibinfo{journal}{\emph{Commun. ACM}} \bibinfo{volume}{33},
  \bibinfo{number}{12} (\bibinfo{year}{1990}), \bibinfo{pages}{32--44}.
\newblock


\bibitem[Mishra et~al\mbox{.}(2008)]%
        {Mishra_2008}
\bibfield{author}{\bibinfo{person}{Chaitanya Mishra}, \bibinfo{person}{Nick
  Koudas}, {and} \bibinfo{person}{Calisto Zuzarte}.}
  \bibinfo{year}{2008}\natexlab{}.
\newblock \showarticletitle{Generating targeted queries for database testing}.
  In \bibinfo{booktitle}{\emph{International Conference on Management of
  Data}}. \bibinfo{publisher}{ACM}, \bibinfo{pages}{499--510}.
\newblock


\bibitem[Mohan et~al\mbox{.}(2018)]%
        {mohan2018finding}
\bibfield{author}{\bibinfo{person}{Jayashree Mohan}, \bibinfo{person}{Ashlie
  Martinez}, \bibinfo{person}{Soujanya Ponnapalli}, \bibinfo{person}{Pandian
  Raju}, {and} \bibinfo{person}{Vijay Chidambaram}.}
  \bibinfo{year}{2018}\natexlab{}.
\newblock \showarticletitle{Finding {Crash-Consistency} Bugs with Bounded
  {Black-Box} Crash Testing}. In \bibinfo{booktitle}{\emph{USENIX Symposium on
  Operating Systems Design and Implementations}}. \bibinfo{publisher}{USENIX
  Association}, \bibinfo{pages}{33--50}.
\newblock


\bibitem[MySQL(2023)]%
        {mysql2023unittest}
\bibfield{author}{\bibinfo{person}{MySQL}.} \bibinfo{year}{2023}\natexlab{}.
\newblock \bibinfo{title}{The MySQL Test Framework}.
\newblock
  \bibinfo{howpublished}{\url{https://dev.mysql.com/doc/dev/mysql-server/latest/PAGE_MYSQL_TEST_RUN.html}}.
\newblock


\bibitem[Neufeld et~al\mbox{.}(1993)]%
        {neufeld1993generating}
\bibfield{author}{\bibinfo{person}{Andrea Neufeld}, \bibinfo{person}{Guido
  Moerkotte}, {and} \bibinfo{person}{Peter~C Loekemann}.}
  \bibinfo{year}{1993}\natexlab{}.
\newblock \showarticletitle{Generating Consistent Test Data: Restricting the
  Search Space by A Generator Formula}.
\newblock \bibinfo{journal}{\emph{Very Large Data Bases Endowment}}
  \bibinfo{volume}{2} (\bibinfo{year}{1993}), \bibinfo{pages}{173--213}.
\newblock


\bibitem[on~Knowledge et~al\mbox{.}(2008)]%
        {Abdul_Khalek_2008}
\bibfield{author}{\bibinfo{person}{ShaTransactions on Knowledge},
  \bibinfo{person}{Data Engineeringdi~Abdul Khalek}, \bibinfo{person}{Bassem
  Elkarablieh}, \bibinfo{person}{Yai~O. Laleye}, {and} \bibinfo{person}{Sarfraz
  Khurshid}.} \bibinfo{year}{2008}\natexlab{}.
\newblock \showarticletitle{Query-Aware Test Generation Using a Relational
  Constraint Solver}. In \bibinfo{booktitle}{\emph{International Conference on
  Automated Software Engineering}}. \bibinfo{publisher}{IEEE},
  \bibinfo{pages}{238--247}.
\newblock


\bibitem[Osman and Knottenbelt(2012)]%
        {Osman_2012}
\bibfield{author}{\bibinfo{person}{Rasha Osman} {and}
  \bibinfo{person}{William~J. Knottenbelt}.} \bibinfo{year}{2012}\natexlab{}.
\newblock \showarticletitle{Database System Performance Evaluation Models: A
  Survey}.
\newblock \bibinfo{journal}{\emph{Performance Evaluation}}
  \bibinfo{volume}{69}, \bibinfo{number}{10} (\bibinfo{year}{2012}),
  \bibinfo{pages}{471--493}.
\newblock


\bibitem[Pacheco et~al\mbox{.}(2007)]%
        {Pacheco_2007}
\bibfield{author}{\bibinfo{person}{Carlos Pacheco},
  \bibinfo{person}{Shuvendu~K. Lahiri}, \bibinfo{person}{Michael~D. Ernst},
  {and} \bibinfo{person}{Thomas Ball}.} \bibinfo{year}{2007}\natexlab{}.
\newblock \showarticletitle{Feedback-Directed Random Test Generation}. In
  \bibinfo{booktitle}{\emph{International Conference on Software Engineering}}.
  \bibinfo{publisher}{IEEE}, \bibinfo{pages}{75--84}.
\newblock


\bibitem[Pavlo et~al\mbox{.}(2017)]%
        {pavlo2017self}
\bibfield{author}{\bibinfo{person}{Andrew Pavlo}, \bibinfo{person}{Gustavo
  Angulo}, \bibinfo{person}{Joy Arulraj}, \bibinfo{person}{Haibin Lin},
  \bibinfo{person}{Jiexi Lin}, \bibinfo{person}{Lin Ma},
  \bibinfo{person}{Prashanth Menon}, \bibinfo{person}{Todd~C Mowry},
  \bibinfo{person}{Matthew Perron}, \bibinfo{person}{Ian Quah},
  {et~al\mbox{.}}} \bibinfo{year}{2017}\natexlab{}.
\newblock \showarticletitle{Self-Driving Database Management Systems}. In
  \bibinfo{booktitle}{\emph{Conference on Innovative Data Systems Research}}.
  \bibinfo{pages}{1}.
\newblock


\bibitem[PingCap(2023)]%
        {git2023gorandgen}
\bibfield{author}{\bibinfo{person}{PingCap}.} \bibinfo{year}{2023}\natexlab{}.
\newblock \bibinfo{title}{go randgen}.
\newblock \bibinfo{howpublished}{\url{https://github.com/pingcap/go-randgen}}.
\newblock


\bibitem[Poess and Stephens(2004)]%
        {Poess_2004}
\bibfield{author}{\bibinfo{person}{Meikel Poess} {and} \bibinfo{person}{John~M.
  Stephens}.} \bibinfo{year}{2004}\natexlab{}.
\newblock \showarticletitle{Generating Thousand Benchmark Queries in Seconds}.
\newblock In \bibinfo{booktitle}{\emph{International Conference on Very Large
  Data Bases}}. \bibinfo{publisher}{VLDB Endowment},
  \bibinfo{pages}{1045--1053}.
\newblock


\bibitem[PostgreSQL(2023)]%
        {postgresql2023test}
\bibfield{author}{\bibinfo{person}{PostgreSQL}.}
  \bibinfo{year}{2023}\natexlab{}.
\newblock \bibinfo{title}{Regression Tests}.
\newblock
  \bibinfo{howpublished}{\url{https://www.postgresql.org/docs/current/regress.html}}.
\newblock


\bibitem[Rigger and Su(2020a)]%
        {Rigger_2020}
\bibfield{author}{\bibinfo{person}{Manuel Rigger} {and}
  \bibinfo{person}{Zhendong Su}.} \bibinfo{year}{2020}\natexlab{a}.
\newblock \showarticletitle{Detecting Optimization Bugs in Database Engines via
  Non-Optimizing Reference Engine Construction}. In
  \bibinfo{booktitle}{\emph{Joint Meeting on European Software Engineering
  Conference and Symposium on the Foundations of Software Engineering}}.
  \bibinfo{publisher}{ACM}.
\newblock


\bibitem[Rigger and Su(2020b)]%
        {rigger2020finding}
\bibfield{author}{\bibinfo{person}{Manuel Rigger} {and}
  \bibinfo{person}{Zhendong Su}.} \bibinfo{year}{2020}\natexlab{b}.
\newblock \showarticletitle{Finding Bugs in Database Systems via Query
  Partitioning}.
\newblock \bibinfo{journal}{\emph{Programming Languages}} \bibinfo{volume}{4},
  \bibinfo{number}{OOPSLA} (\bibinfo{year}{2020}), \bibinfo{pages}{1--30}.
\newblock


\bibitem[Rigger and Su(2020c)]%
        {rigger2020testing}
\bibfield{author}{\bibinfo{person}{Manuel Rigger} {and}
  \bibinfo{person}{Zhendong Su}.} \bibinfo{year}{2020}\natexlab{c}.
\newblock \showarticletitle{Testing Database Engines via Pivoted Query
  Synthesis}. In \bibinfo{booktitle}{\emph{USENIX Symposium on Operating
  Systems Design and Implementation}}. \bibinfo{publisher}{USENIX Association},
  \bibinfo{pages}{667--682}.
\newblock


\bibitem[Rigger and Su(2022)]%
        {rigger2022intramorphic}
\bibfield{author}{\bibinfo{person}{Manuel Rigger} {and}
  \bibinfo{person}{Zhendong Su}.} \bibinfo{year}{2022}\natexlab{}.
\newblock \showarticletitle{Intramorphic Testing: A New Approach to the Test
  Oracle Problem}. In \bibinfo{booktitle}{\emph{International Symposium on New
  Ideas, New Paradigms, and Reflections on Programming and Software}}.
  \bibinfo{publisher}{ACM}, \bibinfo{pages}{128--136}.
\newblock


\bibitem[Shah et~al\mbox{.}(2011)]%
        {Shah_2011}
\bibfield{author}{\bibinfo{person}{Shetal Shah}, \bibinfo{person}{S.
  Sudarshan}, \bibinfo{person}{Suhas Kajbaje}, \bibinfo{person}{Sandeep
  Patidar}, \bibinfo{person}{Bhanu~Pratap Gupta}, {and} \bibinfo{person}{Devang
  Vira}.} \bibinfo{year}{2011}\natexlab{}.
\newblock \showarticletitle{Generating Test Data for Killing SQL Mutants: A
  Constraint-Based Approach}. In \bibinfo{booktitle}{\emph{International
  Conference on Data Engineering}}. \bibinfo{publisher}{IEEE},
  \bibinfo{pages}{1175--1186}.
\newblock


\bibitem[Slutz(1998)]%
        {slutz1998massive}
\bibfield{author}{\bibinfo{person}{Donald~R Slutz}.}
  \bibinfo{year}{1998}\natexlab{}.
\newblock \showarticletitle{Massive Stochastic Testing of SQL}. In
  \bibinfo{booktitle}{\emph{International Conference on Very Large Data
  Bases}}. \bibinfo{publisher}{VLDB Endowment}, \bibinfo{pages}{618--622}.
\newblock


\bibitem[Song et~al\mbox{.}(2023)]%
        {song2023testing}
\bibfield{author}{\bibinfo{person}{Jiansen Song}, \bibinfo{person}{Wensheng
  Dou}, \bibinfo{person}{Ziyu Cui}, \bibinfo{person}{Qianwang Dai},
  \bibinfo{person}{Wei Wang}, \bibinfo{person}{Jun Wei}, \bibinfo{person}{Hua
  Zhong}, {and} \bibinfo{person}{Tao Huang}.} \bibinfo{year}{2023}\natexlab{}.
\newblock \showarticletitle{Testing Database Systems via Differential Query
  Execution}. In \bibinfo{booktitle}{\emph{International Conference on Software
  Engineering}}. \bibinfo{publisher}{IEEE}, \bibinfo{pages}{2072–2084}.
\newblock


\bibitem[SQLite(2023)]%
        {sqlite2023testing}
\bibfield{author}{\bibinfo{person}{SQLite}.} \bibinfo{year}{2023}\natexlab{}.
\newblock \bibinfo{title}{How SQLite Is Tested}.
\newblock \bibinfo{howpublished}{\url{https://sqlite.org/testing.html}}.
\newblock


\bibitem[Stobie(2005)]%
        {Stobie_2005}
\bibfield{author}{\bibinfo{person}{Keith Stobie}.}
  \bibinfo{year}{2005}\natexlab{}.
\newblock \showarticletitle{Too Darned Big to Test: Testing Large Systems is a
  Daunting Task, but There Are Steps We Can Take to Ease the Pain}.
\newblock \bibinfo{journal}{\emph{Queue}} \bibinfo{volume}{3},
  \bibinfo{number}{1} (\bibinfo{year}{2005}), \bibinfo{pages}{30--37}.
\newblock


\bibitem[Taft et~al\mbox{.}(2020)]%
        {taft2020cockroachdb}
\bibfield{author}{\bibinfo{person}{Rebecca Taft}, \bibinfo{person}{Irfan
  Sharif}, \bibinfo{person}{Andrei Matei}, \bibinfo{person}{Nathan
  VanBenschoten}, \bibinfo{person}{Jordan Lewis}, \bibinfo{person}{Tobias
  Grieger}, \bibinfo{person}{Kai Niemi}, \bibinfo{person}{Andy Woods},
  \bibinfo{person}{Anne Birzin}, \bibinfo{person}{Raphael Poss},
  {et~al\mbox{.}}} \bibinfo{year}{2020}\natexlab{}.
\newblock \showarticletitle{CockroachDB: The Resilient Geo-Distributed SQL
  Database}. In \bibinfo{booktitle}{\emph{International Conference on
  Management of Data}}. \bibinfo{publisher}{ACM}, \bibinfo{pages}{1493--1509}.
\newblock


\bibitem[Tang et~al\mbox{.}(2023)]%
        {tang2023detecting}
\bibfield{author}{\bibinfo{person}{Xiu Tang}, \bibinfo{person}{Sai Wu},
  \bibinfo{person}{Dongxiang Zhang}, \bibinfo{person}{Feifei Li}, {and}
  \bibinfo{person}{Gang Chen}.} \bibinfo{year}{2023}\natexlab{}.
\newblock \showarticletitle{Detecting Logic Bugs of Join Optimizations in
  DBMS}.
\newblock \bibinfo{journal}{\emph{International Conference on Management of
  Data}} \bibinfo{volume}{1}, \bibinfo{number}{1} (\bibinfo{year}{2023}),
  \bibinfo{pages}{1--26}.
\newblock


\bibitem[Th\'{e}venod-Fosse and Waeselynck(1993)]%
        {Th_venod_Fosse_1993}
\bibfield{author}{\bibinfo{person}{P. Th\'{e}venod-Fosse} {and}
  \bibinfo{person}{H. Waeselynck}.} \bibinfo{year}{1993}\natexlab{}.
\newblock \showarticletitle{STATEMATE Applied to Statistical Software Testing}.
  In \bibinfo{booktitle}{\emph{International Symposium on Software Testing and
  Analysis}}. \bibinfo{publisher}{ACM}, \bibinfo{pages}{99–109}.
\newblock


\bibitem[Uotila et~al\mbox{.}(2021)]%
        {multicategory2021}
\bibfield{author}{\bibinfo{person}{Valter Uotila}, \bibinfo{person}{Jiaheng
  Lu}, \bibinfo{person}{Dieter Gawlick}, \bibinfo{person}{Zhen~Hua Liu},
  \bibinfo{person}{Souripriya Das}, {and} \bibinfo{person}{Gregory
  Pogossiants}.} \bibinfo{year}{2021}\natexlab{}.
\newblock \showarticletitle{MultiCategory: Multi-Model Query Processing Meets
  Category Theory and Functional Programming}.
\newblock \bibinfo{journal}{\emph{Very Large Data Bases Endowment}}
  \bibinfo{volume}{14}, \bibinfo{number}{12} (\bibinfo{year}{2021}),
  \bibinfo{pages}{2663–2666}.
\newblock


\bibitem[Valduriez and Danforth(1992)]%
        {Valduriez_1992}
\bibfield{author}{\bibinfo{person}{Patrick Valduriez} {and}
  \bibinfo{person}{Scott Danforth}.} \bibinfo{year}{1992}\natexlab{}.
\newblock \showarticletitle{Functional SQL (FSQL), an SQL Upward-Compatible
  Database Programming Language}.
\newblock \bibinfo{journal}{\emph{Information Sciences}} \bibinfo{volume}{62},
  \bibinfo{number}{3} (\bibinfo{year}{1992}), \bibinfo{pages}{183--203}.
\newblock


\bibitem[Wang et~al\mbox{.}(2019)]%
        {Wang_2019}
\bibfield{author}{\bibinfo{person}{Junjie Wang}, \bibinfo{person}{Bihuan Chen},
  \bibinfo{person}{Lei Wei}, {and} \bibinfo{person}{Yang Liu}.}
  \bibinfo{year}{2019}\natexlab{}.
\newblock \showarticletitle{Superion: Grammar-Aware Greybox Fuzzing}. In
  \bibinfo{booktitle}{\emph{International Conference on Software Engineering}}.
  \bibinfo{publisher}{IEEE}, \bibinfo{pages}{724--735}.
\newblock


\bibitem[Wang et~al\mbox{.}(2021)]%
        {Wang_2021}
\bibfield{author}{\bibinfo{person}{Mingzhe Wang}, \bibinfo{person}{Zhiyong Wu},
  \bibinfo{person}{Xinyi Xu}, \bibinfo{person}{Jie Liang},
  \bibinfo{person}{Chijin Zhou}, \bibinfo{person}{Huafeng Zhang}, {and}
  \bibinfo{person}{Yu Jiang}.} \bibinfo{year}{2021}\natexlab{}.
\newblock \showarticletitle{Industry Practice of Coverage-Guided
  Enterprise-Level DBMS Fuzzing}. In \bibinfo{booktitle}{\emph{International
  Conference on Software Engineering: Software Engineering in Practice}}.
  \bibinfo{publisher}{IEEE}, \bibinfo{pages}{328--337}.
\newblock


\bibitem[Wang et~al\mbox{.}(2020)]%
        {wang2020not}
\bibfield{author}{\bibinfo{person}{Yanhao Wang}, \bibinfo{person}{Xiangkun
  Jia}, \bibinfo{person}{Yuwei Liu}, \bibinfo{person}{Kyle Zeng},
  \bibinfo{person}{Tiffany Bao}, \bibinfo{person}{Dinghao Wu}, {and}
  \bibinfo{person}{Purui Su}.} \bibinfo{year}{2020}\natexlab{}.
\newblock \showarticletitle{Not All Coverage Measurements Are Equal: Fuzzing by
  Coverage Accounting for Input Prioritization}. In
  \bibinfo{booktitle}{\emph{Network and Distributed System Security
  Symposium}}.
\newblock


\bibitem[Wen et~al\mbox{.}(2023)]%
        {wen2023squill}
\bibfield{author}{\bibinfo{person}{Shihao Wen}, \bibinfo{person}{Peng Jia},
  \bibinfo{person}{Pin Yang}, {and} \bibinfo{person}{Chi Hu}.}
  \bibinfo{year}{2023}\natexlab{}.
\newblock \showarticletitle{Squill: Testing DBMS with Correctness Feedback and
  Accurate Instantiation}.
\newblock \bibinfo{journal}{\emph{Applied Sciences}} \bibinfo{volume}{13},
  \bibinfo{number}{4} (\bibinfo{year}{2023}), \bibinfo{pages}{2519}.
\newblock


\bibitem[Wu et~al\mbox{.}(2022)]%
        {wu2022unicorn}
\bibfield{author}{\bibinfo{person}{Zhiyong Wu}, \bibinfo{person}{Jie Liang},
  \bibinfo{person}{Mingzhe Wang}, \bibinfo{person}{Chijin Zhou}, {and}
  \bibinfo{person}{Yu Jiang}.} \bibinfo{year}{2022}\natexlab{}.
\newblock \showarticletitle{Unicorn: Detect Runtime Errors in Time-Series
  Databases with Hybrid Input Synthesis}. In
  \bibinfo{booktitle}{\emph{International Symposium on Software Testing and
  Analysis}}. \bibinfo{publisher}{ACM}, \bibinfo{pages}{251--262}.
\newblock


\bibitem[Yan et~al\mbox{.}(2018)]%
        {Yan_2018}
\bibfield{author}{\bibinfo{person}{Jiaqi Yan}, \bibinfo{person}{Qiuye Jin},
  \bibinfo{person}{Shrainik Jain}, \bibinfo{person}{Stratis~D. Viglas}, {and}
  \bibinfo{person}{Allison Lee}.} \bibinfo{year}{2018}\natexlab{}.
\newblock \showarticletitle{Snowtrail: Testing with Production Queries on a
  Cloud Database}. In \bibinfo{booktitle}{\emph{Workshop on Testing Database
  Systems}}. \bibinfo{publisher}{ACM}, \bibinfo{pages}{1--6}.
\newblock


\bibitem[Yang et~al\mbox{.}(2021)]%
        {benchmark2021}
\bibfield{author}{\bibinfo{person}{Man Yang}, \bibinfo{person}{Mingwang Zhang},
  \bibinfo{person}{Tao Leng}, \bibinfo{person}{Weize Dong}, {and}
  \bibinfo{person}{Yu Yan}.} \bibinfo{year}{2021}\natexlab{}.
\newblock \showarticletitle{Design and Implementation of Benchmarks for
  Multimodal Databases}. In \bibinfo{booktitle}{\emph{International Conference
  on Industrial Application of Artificial Intelligence}}.
  \bibinfo{publisher}{IEEE}, \bibinfo{pages}{490--494}.
\newblock


\bibitem[Yuan et~al\mbox{.}(2021)]%
        {yuan2021storing}
\bibfield{author}{\bibinfo{person}{Gongsheng Yuan}, \bibinfo{person}{Jiaheng
  Lu}, \bibinfo{person}{Shuxun Zhang}, {and} \bibinfo{person}{Zhengtong Yan}.}
  \bibinfo{year}{2021}\natexlab{}.
\newblock \showarticletitle{Storing Multi-Model Data in RDBMSs Based on
  Reinforcement Learning}. In \bibinfo{booktitle}{\emph{International
  Conference on Information \& Knowledge Management}}.
  \bibinfo{publisher}{ACM}, \bibinfo{pages}{3608--3611}.
\newblock


\bibitem[Zhang et~al\mbox{.}(2019)]%
        {zhang2019unibench}
\bibfield{author}{\bibinfo{person}{Chao Zhang}, \bibinfo{person}{Jiaheng Lu},
  \bibinfo{person}{Pengfei Xu}, {and} \bibinfo{person}{Yuxing Chen}.}
  \bibinfo{year}{2019}\natexlab{}.
\newblock \showarticletitle{UniBench: A Benchmark for Multi-model Database
  Management Systems}. In \bibinfo{booktitle}{\emph{Performance Evaluation and
  Benchmarking for the Era of Artificial Intelligence}}.
  \bibinfo{publisher}{Springer}, \bibinfo{pages}{7--23}.
\newblock


\bibitem[Zhang et~al\mbox{.}(2021)]%
        {zhang2021duplicate}
\bibfield{author}{\bibinfo{person}{Yushan Zhang}, \bibinfo{person}{Peisen Yao},
  \bibinfo{person}{Rongxin Wu}, {and} \bibinfo{person}{Charles Zhang}.}
  \bibinfo{year}{2021}\natexlab{}.
\newblock \showarticletitle{Duplicate-sensitivity Guided Transformation
  Synthesis for DBMS Correctness Bug Detection}.
\newblock \bibinfo{journal}{\emph{arXiv preprint arXiv:2107.03660}}
  (\bibinfo{year}{2021}).
\newblock


\bibitem[Zheng et~al\mbox{.}(2022)]%
        {zheng2022finding}
\bibfield{author}{\bibinfo{person}{Yingying Zheng}, \bibinfo{person}{Wensheng
  Dou}, \bibinfo{person}{Yicheng Wang}, \bibinfo{person}{Zheng Qin},
  \bibinfo{person}{Lei Tang}, \bibinfo{person}{Yu Gao}, \bibinfo{person}{Dong
  Wang}, \bibinfo{person}{Wei Wang}, {and} \bibinfo{person}{Jun Wei}.}
  \bibinfo{year}{2022}\natexlab{}.
\newblock \showarticletitle{Finding Bugs in Gremlin-Based Graph Database
  Systems via Randomized Differential Testing}. In
  \bibinfo{booktitle}{\emph{International Symposium on Software Testing and
  Analysis}}. \bibinfo{publisher}{ACM}, \bibinfo{pages}{302--313}.
\newblock


\bibitem[Zhong et~al\mbox{.}(2020)]%
        {zhong2020squirrel}
\bibfield{author}{\bibinfo{person}{Rui Zhong}, \bibinfo{person}{Yongheng Chen},
  \bibinfo{person}{Hong Hu}, \bibinfo{person}{Hangfan Zhang},
  \bibinfo{person}{Wenke Lee}, {and} \bibinfo{person}{Dinghao Wu}.}
  \bibinfo{year}{2020}\natexlab{}.
\newblock \showarticletitle{SQUIRREL: Testing Database Management Systems with
  Language Validity and Coverage Feedback}. In
  \bibinfo{booktitle}{\emph{Conference on Computer and Communications
  Security}}. \bibinfo{publisher}{ACM}, \bibinfo{pages}{955--970}.
\newblock


\bibitem[Zhou et~al\mbox{.}(2021)]%
        {Zhou_2021}
\bibfield{author}{\bibinfo{person}{Zenghui Zhou}, \bibinfo{person}{Zheng
  Zheng}, \bibinfo{person}{Tsong~Yueh Chen}, \bibinfo{person}{Jinyi Zhou},
  {and} \bibinfo{person}{Kun Qiu}.} \bibinfo{year}{2021}\natexlab{}.
\newblock \showarticletitle{Follow-up Test Cases are Better Than Source Test
  Cases in Metamorphic Testing: A Preliminary Study}. In
  \bibinfo{booktitle}{\emph{International Workshop on Metamorphic Testing}}.
  \bibinfo{publisher}{IEEE}, \bibinfo{pages}{69--74}.
\newblock


\end{thebibliography}

\end{document}